\documentclass[aps,prd,twocolumn,superscriptaddress,floatfix,longbibliography,preprintnumbers,nofootinbib]{revtex4-1} 
\usepackage{natbib}
\usepackage{ifthen,amsmath,amssymb, bm}
\usepackage{graphicx}
\usepackage{color}
\usepackage[normalem]{ulem}
\usepackage{float}
\usepackage{xcolor}
\usepackage{aas_macros}
\usepackage{soul}
\usepackage[caption=false]{subfig}
\usepackage{mathtools}
\newcommand{\vk} {{\bm{k}}}

\newcommand{\vl} { {\bm{\ell}} }
\newcommand{\vL} { {\bm{L}} }
\newcommand{\vx} {\bm{x}}
\newcommand{\fN} {\hat{N}}
\newcommand{\vd} {{\bm{d}}}
\newcommand{\vi} {{\bm{i}}}
\newcommand{\vj} {{\bm{j}}}
\newcommand{\vll} {{\bm{l}}}

\newcommand{\va} {{\bm{a}}}
\newcommand{\vb} {{\bm{b}}}
\newcommand{\vc} {{\bm{c}}}

\newcommand{\vJ} {{\bm{J}}}

\newcommand{\Tz} {T^{(0)}}

\newcommand{\To} {T^{(1)}}
\newcommand{\Tt} {T^{(2)}}
\newcommand{\vlo}{{\vl_{1}}}
\newcommand{\vlt}{{\vl_{2}}}
\newcommand{\vlth}{{\vl_{3}}}
\newcommand{\vlf}{{\vl_{4}}}
\DeclareRobustCommand{\Sec}[1]{Sec.~\ref{sec:#1}}
\DeclareRobustCommand{\Secs}[2]{Secs.~\ref{sec:#1} and \ref{sec:#2}}
\DeclareRobustCommand{\App}[1]{App.~\ref{app:#1}}

\DeclareRobustCommand{\Fig}[1]{Fig.~\ref{fig:#1}}

\DeclareRobustCommand{\Eq}[1]{Eq.~(\ref{eq:#1})}
\DeclareRobustCommand{\Eqs}[2]{Eqs.~(\ref{eq:#1}) and (\ref{eq:#2})}

\DeclareRobustCommand{\Reff}[1]{Ref.~\cite{#1}}
\DeclareRobustCommand{\Refs}[1]{Refs.~\cite{#1}}

\newcommand{\beq} {\begin{equation}}
\newcommand{\eeq} {\end{equation}}
\newcommand{\bal} {\begin{aligned}}
\newcommand{\eal} {\end{aligned}}

\definecolor{darkgreen}{rgb}{0,0.7,0}

\definecolor{green}{RGB}{64,160,64}
\definecolor{c1}{RGB}{249,65,68} 
\definecolor{lg}{RGB}{211,211,211}
\definecolor{c4}{RGB}{255,111,114} 
\definecolor{c2}{RGB}{0,168,50} 
\definecolor{c3}{RGB}{39,125,161} 
\definecolor{c5}{RGB}{157,111,255} 
\definecolor{c6}{RGB}{251,105,255} 

\usepackage{hyperref}
\usepackage{cleveref}

\begin{document}

\title{CMB lensing power spectrum without noise bias 
}

\author{Delon Shen}
\email{delon@stanford.edu}
\affiliation{Department of Physics, Stanford University, Stanford, CA, USA 94305-4085}
\affiliation{Kavli Institute for Particle Astrophysics and Cosmology, 382 Via Pueblo Mall Stanford, CA 94305-4060, USA}
\affiliation{SLAC National Accelerator Laboratory 2575 Sand Hill Road Menlo Park, California 94025, USA}
\author{Emmanuel Schaan}
\affiliation{Kavli Institute for Particle Astrophysics and Cosmology,
382 Via Pueblo Mall Stanford, CA 94305-4060, USA}
\affiliation{SLAC National Accelerator Laboratory 2575 Sand Hill Road Menlo Park, California 94025, USA}
\author{Simone Ferraro}
\affiliation{Physics Division, Lawrence Berkeley National Laboratory, Berkeley, CA 94720, USA}
\affiliation{Berkeley Center for Cosmological Physics, Department of Physics, University of California, Berkeley, CA 94720, USA}

\begin{abstract}

Upcoming surveys will measure the cosmic microwave background (CMB) weak lensing power spectrum in exquisite detail, allowing for strong constraints on the sum of neutrino masses among other cosmological parameters.
%
Standard CMB lensing power spectrum estimators aim to extract the connected non-Gaussian trispectrum of CMB temperature maps.
However, they are generically dominated by a large disconnected, or Gaussian, noise bias, which thus needs to be subtracted at high accuracy.
This is currently done with realistic map simulations of the CMB and noise, whose finite accuracy currently limits our ability to recover {CMB lensing on small-scales}.
%
In this paper, we propose a novel estimator which instead avoids this large Gaussian bias.
This estimator relies only on the data and avoids the need for bias subtraction with simulations.
Thus our bias avoidance method is (1) insensitive to misestimates in simulated CMB and noise models and (2) avoids the large computational cost of standard simulation-based methods like ``realization-dependent $N^{(0)}$'' (${\rm RDN}^{(0)}$).
We show that our estimator is as robust as standard methods in the presence of realistic inhomogeneous noise (e.g. from scan strategy) and masking.
Moreover, our method can be combined with split-based methods, making it completely insensitive to mode coupling from inhomogeneous atmospheric and detector noise.
We derive the corresponding expressions for our estimator when estimating lensing from CMB temperature and polarization.
%
Although we specifically consider CMB weak lensing power spectrum estimation {in this paper}, we illuminate the relation between our new estimator, ${\rm RDN}^{(0)}$ subtraction, and general optimal trispectrum estimation.
Through this discussion we conclude that our estimator is applicable to analogous problems in other fields which rely on estimating connected trispectra/four-point functions like large-scale structure. 
We release the code implementing our proposed estimator and numerical experiments {\href{https://github.com/DelonShen/LensQuEst}{here}}.

\end{abstract}

\maketitle

\section{Introduction}

Cosmic microwave background (CMB) photons are gravitationally lensed by large scale structure as they propagate through the Universe. 
This lensing distorts our images of CMB anisotropies and imprints onto the CMB four-point {correlation} function\footnote{{In what follows, we use $n$-point correlation function and $n$-point function interchangeably.}} 
a distinct non-Gaussian component \cite{Lewis:2006fu}.
Measurements of this non-Gaussian component, the CMB weak lensing power spectrum $\langle\kappa\kappa\rangle$, provides us with a wealth of information on the growth of structure in our Universe. 
Thus a measurement of $\langle\kappa\kappa\rangle$ can provide us with some of the strongest constraints on the properties of neutrinos \cite{Allison:2015qca}, primordial non-Gaussianity \cite{Schmittfull:2017ffw}, dark matter \cite{Li:2018zdm} and dark energy \cite{2011PhRvL.107b1302S}. 
The first detection of CMB lensing was reported with data from the Wilkinson Microwave Anisotropy Probe (WMAP) in \Reff{Smith:2007rg} and CMB lensing power spectrum in \Refs{2011PhRvL.107b1302S, 2011PhRvL.107b1301D}. 
Since then, many additional detections of CMB lensing spectra have been reported \cite{van_Engelen_2012,polarbear,Story_2015,BICEP2016,PhysRevD.95.123529,Omori_2017,Wu_2019,SPT:2019fqo, ACT:2023dou, SPT:2023jql}.

In order to robustly and accurately estimate $\langle\kappa\kappa\rangle$, we must navigate the problem of extracting a non-Gaussian signal, the connected trispectrum, from the four-point function which is generically dominated by a Gaussian bias $N^{(0)}$ that is typically several orders of magnitude larger than the non-Gaussian signal. 
In typical analyses, this $N^{(0)}$ is estimated from realistic map simulations of the CMB and noise and then subtracted off. 
However, since $N^{(0)}$ is orders of magnitude larger than our signal CMB lensing power spectrum, even a $0.1\%$ misestimate of $N^{(0)}$ can lead to a $10\%$ bias in our estimate of the lensing power spectrum on small scales. 
The level of accuracy needed for robust estimates of $\langle\kappa\kappa\rangle$ on small scales is difficult to achieve with traditional methods since any mismodeling of the maps that are input to traditional methods propagates to errors in estimates of $N^{(0)}$. 
Two particular concerns is (1) that the complex and inhomogeneous noise structure arising from ground-based CMB experiments is difficult to simulate accurately and (2) uncertainties on the cosmological parameters both of which can bias our estimates of $\langle\kappa\kappa\rangle$. 

In this paper we propose an estimator of the CMB weak lensing power spectrum which avoids the large $N^{(0)}$ Gaussian bias. 
This is achieved by examining exactly which terms contribute to the $N^{(0)}$ bias and constructing a scheme which estimates $\langle\kappa\kappa\rangle$ while avoiding these terms. 
It turns out this leads to a negligible reduction in signal-to-noise. 
In addition, this method relies only on data, completely avoiding the need for simulations to avoid $N^{(0)}$ making it (1) insensitive to inaccuracies in the modeling of the CMB and noise and (2) significantly more computationally efficient compared to standard ${\rm RDN}^{(0)}$ methods. 
We also demonstrate that this estimator is as robust as standard methods in the presence of realistic inhomogeneous noise, such as the noise pattern that may arise from scan strategy, and masking.
In addition we describe how our estimator may be combined with split-based methods proposed in \Reff{Madhavacheril:2020ido} which make it insensitive to spurious mode couplings from inhomogeneous atmospheric and detector noise. 
This new estimator was hinted at in \Reff{2010arXiv1011.4510S} but not implemented or studied there. Instead they focused on an alternative estimator.
Crucially, this estimator can be implemented efficiently (at least in the flat-sky limit) with the use of Fast Fourier Transforms (FFTs), often required in a real analysis.

The rest of this paper is organized as follows. 
In \Sec{standard} we briefly review how standard CMB lensing and lensing spectrum estimation is done. 
In \Sec{Nhat} we outline our proposed estimator for the lensing power spectrum which avoids $N^{(0)}$ bias. 
Following this we compare our method with the standard ${\rm RDN}^{(0)}$ subtraction method in \Sec{RDN0}.
In \Sec{cov} we comment on the effect of $N^{(0)}$ on the covariance of CMB lensing spectra as well as how our method and the standard ${\rm RDN}^{(0)}$ prescription for handling $N^{(0)}$ removes covariances introduced by $N^{(0)}$. 
We explore an illuminating toy model for optimal trispectrum estimating first presented in \Reff{Smith:2015uia} in \Sec{toy} which allows us to see how our estimator, the standard ${\rm RDN}^{(0)}$ estimator, and optimal trispectrum estimation are related and builds intuition for the various features of each method. 
In \Sec{aniso-noise} we study the robustness of our proposed estimator in the presence of realistic anisotropic noise which lead to additional complications in CMB weak lensing power spectrum estimation.
This motivates a discussion of how one might combine our method with a split-based method in \Sec{splits}.
In \Sec{masking} we study the robustness of our proposed estimator in the presence of masking which also leads to additional complications in CMB weak lensing power spectrum estimation. 
Throughout this paper we focus mostly on estimating CMB lensing using only temperature anisotropies, but in \Sec{polarization} we spell out the general form of our proposed estimator when estimating CMB lensing using both temperature and polarization anisotropies. 
Finally we conclude in \Sec{conclusion}.

\section{Standard CMB lensing power spectrum estimation}
\label{sec:standard}

\begin{figure*}
    \centering
    \includegraphics[width=\linewidth]{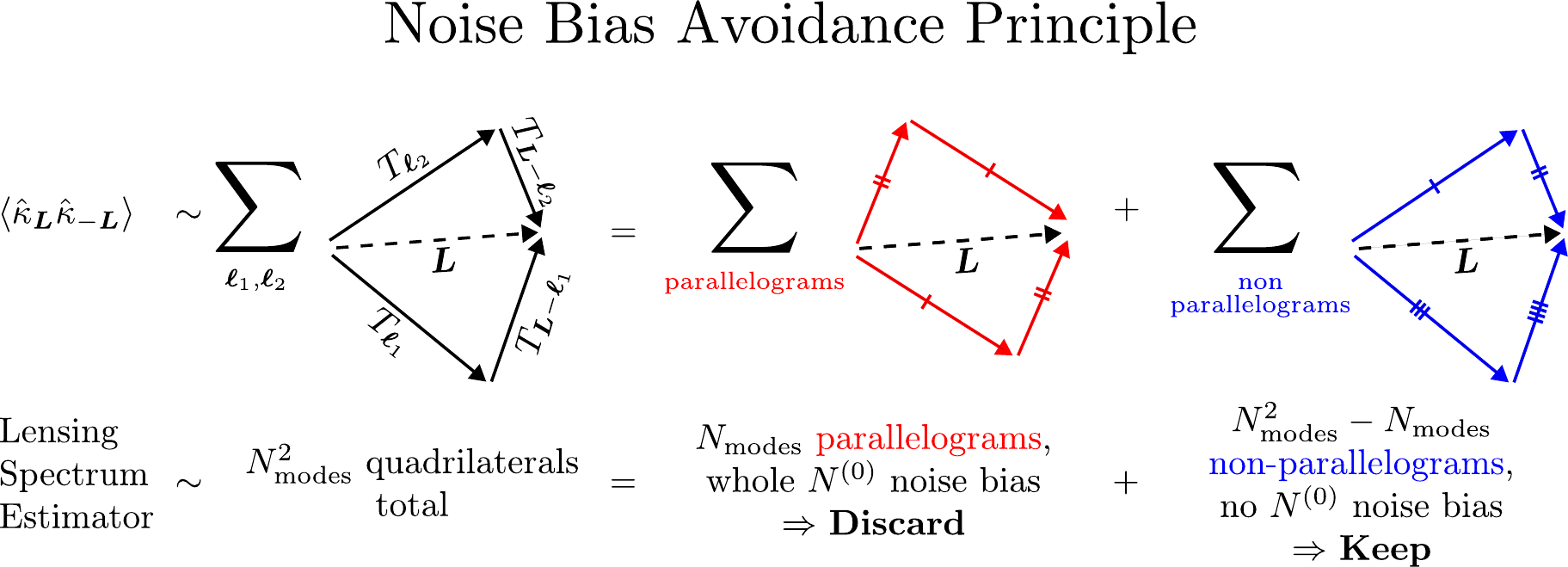}
    \caption{
    A visualization of our proposed $\hat N$ subtraction method. One can schematically think of the estimator for the CMB lensing power spectrum, \Eq{kk}, as a sum over quadrilaterals where each side of the quadrilateral corresponds to a mode of the temperature map. All quadrilaterals contain information about the lensing power spectrum but \textbf{only the parallelograms contain information about the disconnected Gaussian bias}. We show this explicitly in \Eq{Nhat_explicit}. So if we neglect  parallelograms in computing $\langle\hat\kappa\hat\kappa\rangle$ we can remove the large disconnected Gaussian bias. We expand on this more in \Sec{Nhat}}
    \label{fig:summary}
\end{figure*}

The effect of lensing on the CMB temperature maps {at the position $\vx$} can be expressed as 
\begin{equation}
    T(\vx) = T^0(\vx + \vd(\vx)) = T^0(\vx) + \vd \cdot \nabla T^0(\vx) + O(\vd^2).
    \label{eq:Tx}
\end{equation}
One can then do a Helmholtz decomposition of the displacement field
$\vd = \nabla\psi$,
where we have neglected the field rotation component since it is negligible for current sensitivities \cite{Hirata:2003ka}. 
From this we can also define the convergence
$\kappa \equiv -\nabla^2 \psi / 2.$
{Throughout this paper we will refer to lensing potential $\psi$ and convergence $\kappa$ interchangeably} but will primarily work with the lensing convergence $\kappa$.
With this we can rewrite \Eq{Tx} in Fourier space:
\begin{align}
\nonumber    T_\vl &= T^0_\vl  - \int \frac{d^2 \vl'}{(2\pi)^2}  \vl'\cdot(\vl - \vl') \frac{2 \kappa_{\vl - \vl'}}{(\vl - \vl')^2} T^0_{\vl'} + O(\kappa^2)\\
    &\equiv T^{(0)}_\vl  + T^{(1)}_\vl + O(\kappa^2).\label{eq:T}
\end{align}
From statistical homogeneity and isotropy one can show that for the unlensed CMB 
\begin{equation}
    \langle T^0_{\vl_1} T^{0}_{\vl_2}\rangle = (2\pi)^2 \delta^{(D)}(\vl_1 +\vl_2) C_{\vl_1}^{TT}
    \label{eq:T0T0}
\end{equation}
At fixed realization of the lensing convergence $\kappa_\vL$, 
lensing introduces non-trivial off-diagonal correlations  
\begin{align}
\nonumber
    \langle T_{\vl_1} T_{\vl_2}\rangle &= (2\pi)^2 \delta(\vl_1 + \vl_2) C_{\vl_1}^{TT} \\
&+ \kappa_{\vl_1 +\vl_2} f^\kappa_{\vl_1, \vl_2}+O(\kappa^2).
    \label{eq:TT}    
\end{align}
where we defined $f^\kappa_{\vl_1,\vl_2}$ in the second line with the relation
\begin{align}
\nonumber f^\kappa_{\vl_1,\vl_2} {\kappa_{\vl_1+\vl_2}}&\equiv   \langle T^{(0)}_{\vl_1} T^{(1)}_{\vl_2} \rangle +  \langle T^{(1)}_{\vl_1} T^{(0)}_{\vl_2}\rangle \\
\Rightarrow    f^\kappa_{\vl_1,\vl_2}&=\frac{2 (\vl_1 + \vl_2)}{(\vl_1+\vl_2)^2} \cdot [\vl_1 C_{\vl_1}^{TT} + \vl_2 C_{\vl_2}^{TT}].\label{eq:fK}
\end{align}
This means that CMB lensing breaks statistical homogeneity and isotropy. 
It is particularly useful to consider
\beq
\langle T_{\bm{\ell}} T_{\bm{L}-\bm{\ell}} \rangle
=
f^\kappa_{\bm{\ell},\bm{L}-\bm{\ell}}\kappa_{\bm{L}}
+
\mathcal{O}\left( \kappa^2 \right). 
\label{eq:TTspecial}
\eeq
From this we can intuit a quadratic estimator (QE) for the lensing potential:
\begin{align}
\nonumber\hat\kappa_\vL &= N_\vL^\kappa \int \frac{d^2 \vl}{(2\pi)^2} F^\kappa_{\vl,\vL-\vl} T_\vl T_{\vL-\vl}\\
&\sim    \vcenter{\hbox{\includegraphics{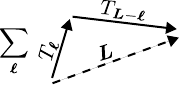}}} .\label{eq:kappa}
\end{align}

The normalization for this estimator is derived from asserting the estimator is unbiased, $\langle\hat\kappa_\vL\rangle = \kappa_\vL$, which leads to
\beq
\left(N^\kappa_{\bm{L}}\right)^{-1} = \int \frac{d^2\bm{\ell}}{(2\pi)^2} F^\kappa_{\bm{\ell},\bm{L}-\bm{\ell}} f^\kappa_{\bm{\ell},\bm{L}-\bm{\ell}}.
\label{eq:NkappaL}
\eeq
Let  $\tilde{C}^{TT}_{\vl}$ be the \textit{total observed} temperature power spectrum, i.e. the power spectra of the \textit{lensed} temperature field which includes instrument noise and foregrounds. 
Expanding \Eq{kappa} to linear order in $\kappa$ allows one to derive the {minimum-variance} weights at this order:
\begin{equation}
F^\kappa_{\vl, \vL-\vl} = \frac{f^\kappa_{\vl,\vL-\vl}}{2 \tilde C_\vl^{\rm TT}\tilde C_{|\vL - \vl|}^{\rm TT}}.
\label{eq:weights}
\end{equation}

Intuitively, it makes sense we need two powers of the temperature field to build an estimator. 
This is because we assume the CMB is isotropic and homogeneous when emitted therefore independent modes of $T_\vl^0$ are statistically independent. 
However, as we saw in \Eq{TT}, weak lensing of CMB introduces correlations between previously independent modes. 
Thus it should be possible to extract the lensing potential that causes this lensing by considering pairs of modes $T_\vl, T_{\vL-\vl}$ and seeing what kind of correlations we have measured. 

Using this estimator for the lensing potential $\kappa_\vL$ we can now study the power spectrum of the lensing potential by first considering the two-point function of our estimator for the lensing potential:
\begin{align}
    \nonumber\langle \hat\kappa_\vL \hat\kappa_{\vL}^*\rangle = (N^\kappa_\vL)^2 \int &\frac{d^2\vl_1}{(2\pi)^2}\int\frac{d^2\vl_2}{(2\pi)^2} F^\kappa_{\vl_1,\vL-\vl_1}F^\kappa_{-\vl_2,-\vL+\vl_2}\\
    &\times \langle T_{\vl_1} T_{\vL-\vl_1} T_{-\vl_2}T_{-\vL+\vl_2}\rangle\label{eq:kk}.
\end{align}
This estimate for the lensing power spectrum has several significant noise biases which can helpfully be denoted by $N^{(i)}$ biases. 

\subsection{$N^{(i)}$ Noise Biases}
\label{sec:Ni}
There are several noise biases hindering our estimate of the true $\kappa$ power spectrum:

\begin{itemize}
    \item $N^{(0)}$: large disconnected Gaussian noise bias. 
    This bias is expected even for a Gaussian unlensed CMB map, or Gaussian noise and foregrounds.
    \item $N^{(1)}$: bias arising from an integral over one power of the lensing power spectrum $\iint\langle \kappa^2\rangle$
\end{itemize}
{In} general we have :
\begin{itemize}
    \item $N^{(i)}$: bias arising from an integral over {$i$ powers} of the lensing power spectrum
\end{itemize}
We spell out biases of these type in more detail within \App{Ni}.
Moreover, taking into account the non-Gaussianity of the lensing field, there exists a non-zero bispectrum which allows contraction that lead to other noise biases:
\begin{itemize}
    \item $N^{(i/2)}$: bias arising from intrinsic non-Gaussianity in the lensing potential, involving {the $i$-th power} of the lensing field \cite{Bohm:2016gzt,Bohm:2018omn,Beck:2018wud,Fabbian:2019tik}
\end{itemize}

The standard way to estimate the Gaussian bias $N^{(0)}$ is to compute what is called a realization-dependent $N^{(0)}$, ${\rm RDN}^{(0)}$ Monte Carlo (MC) correction as described in \Refs{Namikawa:2012pe, Planck:2018}. 
This estimator is constructed to be robust to misestimates of the underlying total power spectrum in comparison to a naive method where one subtracts $N_{\rm theory}\equiv N^{\kappa}$ from \Eq{NkappaL}.
We expand on ${\rm RDN}^{(0)}$ in \Sec{RDN0} and the $N_{\rm theory}$ subtraction in \App{lL2}

We can also estimate the first relevant higher order $N^{(1)}$ noise bias in two ways: analytic computation of the $N^{(1)}$ bias as described in \Refs{Kesden:2003cc, 2011PhRvD..83d3005H, Jenkins:2014hza, Jenkins:2014oja} which we review in \App{Ni} 
or numerical computation from simulations as described in \Reff{ACT:2023dou} which we will discuss in \App{N1}. In our numerical studies we utilize both of these methods to estimate $N^{(1)}$ as we will discuss in \App{N1} as well. 

Higher order biases such as the $N^{(2)}$ bias can still lead to significant misestimates of the lensing power spectrum at low $\vL$ \cite{2011PhRvD..83d3005H}. However in \Reff{2011JCAP...03..018L}, it is shown that one can do a non-perturbative treatment of the lensing potential power spectrum and derive that replacing $C_\vl^{TT}$ with $C_{\vl}^{T\nabla T}$ in \Eq{fK} significantly reduces the $N^{(2)}$ bias: 
\begin{equation}
f^{\kappa}_{\bm{\ell},\bm{L}-\bm{\ell}}=\frac{2\bm{L}}{L^2}\cdot\left[\bm{\ell}C^{T\nabla T}_\ell + (\bm{L}-\bm{\ell})C^{T\nabla T}_{|\bm{L}-\bm{\ell}|}\right].
\label{eq:fKappa}
\end{equation}
We make use of these $C_\vL^{T\nabla T}$ weights in this paper for our numerical studies.


\section{Exact noise bias avoidance}
\label{sec:Nhat}

Our proposed method to avoid the Gaussian $N^{(0)}$ noise bias, outlined in \Fig{summary}, boils down to isolating and subsequently discarding the set of 
$\{\vl_1, \vl_2 \}$ 
in \Eq{kk} which contain the entire $N^{(0)}$ Gaussian noise bias. 
Formally this is a measure zero set and in practice means ignoring $N_{\rm modes}$ out of the $N_{\rm modes}^2$  terms
$\langle{T_{\vl_1} T_{\vL-\vl_1}  T_{\vl_2}  T_{-\vL - \vl_2}  }\rangle$
used to estimate the lensing power spectrum.

This $N^{(0)}$ noise bias avoidance improves over traditional methods by 
(1) circumventing the need for extensive MC simulations to subtract $N^{(0)}$ and
(2) leads to a negligible fractional reduction in the SNR (of order $\sim 1/N_\text{modes} \sim 10^{-6}$). 
Also by nature of (1) our estimator is insensitive to inexact assumptions about the power spectrum $C_{\vl}^{TT}$, which is a source of misestimation for traditional MC-based corrections.

To implement this avoidance of the configurations which source the $N^{(0)}$ bias, we simply keep the usual CMB lensing power spectrum estimator (which contains all the configurations) and subtract from it the problematic configurations from the data.
As illustrated schematically in \Eq{kappa}, the CMB lensing quadratic estimator can be thought of as the sum of $\sim N_\text{modes}$ elementary quadratic estimators.
The noise bias only arises from the auto-spectra of these elementary estimators.
By keeping only the 
$N_\text{modes} \left( N_\text{modes}-1 \right)/2$ 
cross-spectra, we avoid the noise bias, while discarding only a fraction $\sim 1/N_\text{modes} \sim 10^{-6}$ of the signal to noise.
Our method thus amounts to estimating the CMB lensing auto-spectrum only from cross-spectra.
We denote these {problematic configurations} $\hat{N}_\vL$, defined as\footnote{In \Eq{Nhat} we are suppressing a factor of $[(2\pi)^2 \delta^{(D)}(0)]^{-1}$ that turns into a finite area correction when computing this in the discrete case. Namely {$\delta^{(D)}(0)$ becomes the area of the map in the discrete case}. See discussion around \Eq{deltaFinite}. \label{footnote:area}}
\beq
\hat{N}_\vL
=
2
\left(N^\kappa_{\bm{L}}\right)^{2}
\int \frac{d^2\vl}{(2\pi)^2}
F^\kappa_{\vl, \vL-\vl}
F^\kappa_{-\vl, -\vL+\vl}\
\left| T_\vl \right|^2
\left| T_{\vL - \vl} \right|^2.
\label{eq:Nhat}
\eeq
Note that due to the structure of the integral, this can be efficiently implemented with the use of FFTs as shown in our code \href{https://github.com/DelonShen/LensQuEst}{\texttt{LensQuEst}}.
We can thus form a new estimator for the CMB lensing auto-spectrum:
\beq
\label{eq:ClmNhat}
\hat{C}^{\kappa\kappa,\textrm{(no bias)}}_\vL
\equiv
 \hat\kappa_\vL\hat\kappa^*_\vL
-
\hat{N}_\vL.
\eeq
Crucially, just like the standard lensing quadratic estimator, this quantity can be computed efficiently with FFT on the flat sky (resp. spherical harmonics transforms on the curved sky).
As a result, it does does not increase the complexity or the computation time of the analysis.
This new estimator subtracts the Gaussian noise bias exactly on average.
Indeed, in the absence of lensing signal when $\langle\hat\kappa_\vL \hat\kappa_\vL^*\rangle_{\rm GRF} = N^{(0)}_\vL$ only\footnote{In \Eqs{ClmNhat}{Clnobias}, there is actually an additional $\vl = \vL/2$ contraction not included. 
However, we show in \App{lL2} that this additional contraction is smaller, again by a factor 
$\sim 1/N_\text{modes}\sim 10^{-6}$.\label{footnote:l2}},
\beq
\label{eq:Clnobias}
\langle \hat\kappa_\vL\hat\kappa^*_\vL\rangle_{\rm GRF}
-
\langle \hat{N}_\vL \rangle_{\rm GRF}
= 0.
\eeq

\begin{figure}
\includegraphics[width=\linewidth]{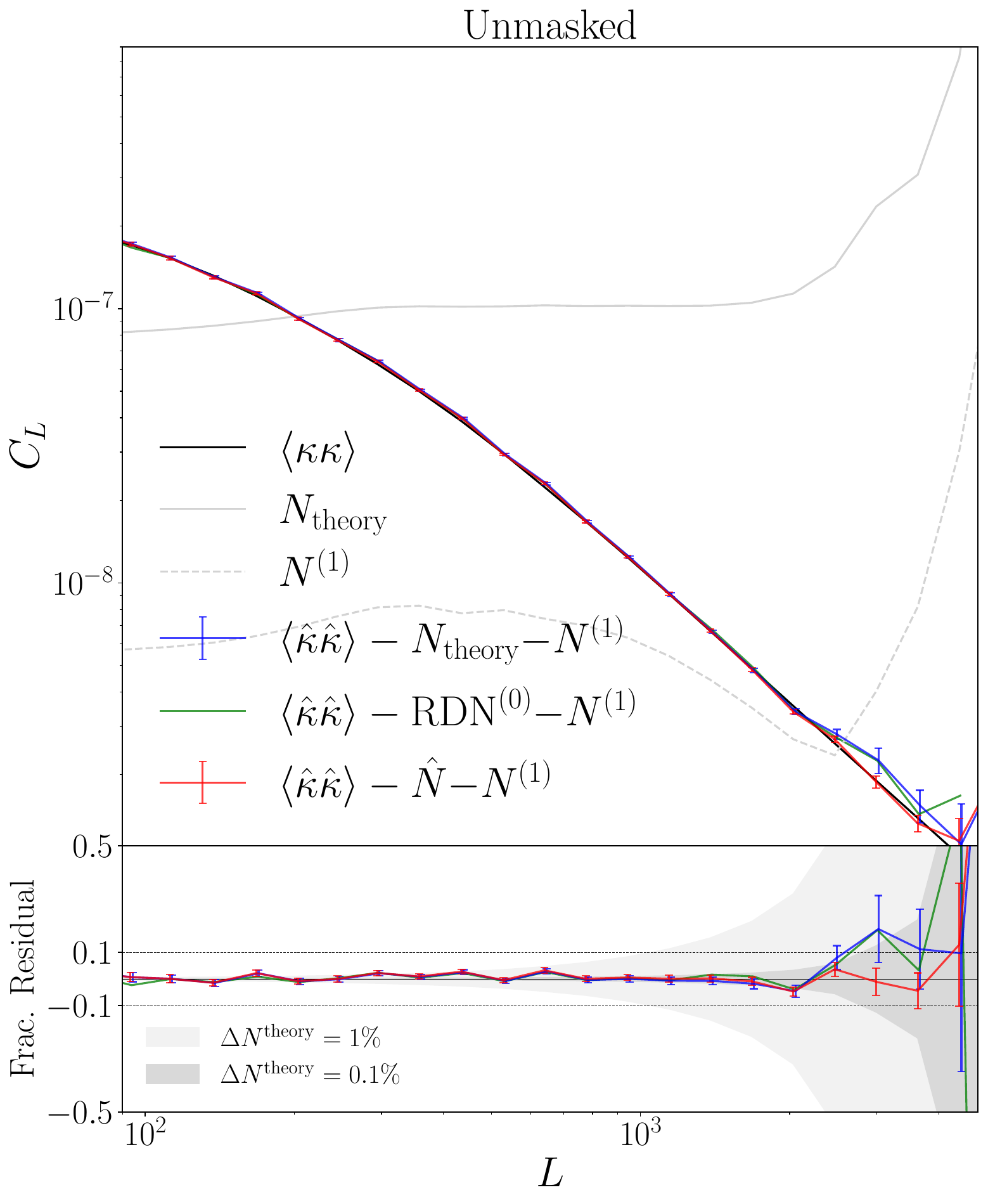}
\caption{
When estimating the lensing power spectrum $\langle \kappa\kappa\rangle$ (black) as a function of angular multipole {for a Simons Observatory (SO)-like survey}, the $N^{(0)}$ Gaussian bias ({gray solid}) dominates on small scales.
As a result, as the bottom residual plot shows, even a $0.1\%$ misestimate of $N^{(0)}$ at high $L$ leads to a $10\%$ bias in $\langle\kappa\kappa\rangle$.
When the input CMB and noise power spectra are perfectly known, standard methods like theory subtraction ({\color{blue}blue}) and ${\rm RDN}^{(0)}$ subtraction ({\color{green}green}) perform equally well as our new bias avoidance method ({\color{red}red}).
This new method continues to work when the input models are inaccurate as we will see in \Secs{aniso-noise}{masking}, and is computationally faster, by not relying on a set of simulations.
In fact, since we computed the ${\rm RDN}^{(0)}$ correction using common pool of simulations thus introducing non-trivial correlations, it is difficult, {though not impossible through semi-analytic methods \cite{Planck:2015mym,Planck:2018lbu}}, to estimate a meaningful error bar on it. 
{So for the {\color{green}${\rm RDN}^{(0)}$ subtracted spectra} we do not include error bars.
We expand on how we simulate our maps in \App{sims}.}
} 
\label{fig:Nhat_unmasked}
\end{figure}

In \Fig{Nhat_unmasked}, we show how our exact noise bias avoidance works on a lensed CMB temperature map with no masking and isotropic noise in  comparison to the standard method. 
In \Sec{aniso-noise} and \Sec{masking} we comment on the robustness of our estimator in the presence of realistic anisotropic noise and masking. 
In this plot we use a hybrid method to estimate the $N^{(1)}$ term which we describe in \App{N1}\footnote{{We also show in \App{Nigeq1} that our proposed $\hat N$ subtraction does not affect $N^{(1)}$ biases thus allowing us to use standard methods of computing $N^{(1)}$.}}. 
We see from this plot that in this case our method is able to estimate the $N^{(0)}$ bias to better than $0.1\%$, doing just as well as standard methods. 

\section{Comparison with the standard realization-dependent $N^{(0)}$ subtraction}
\label{sec:RDN0}

Currently, the standard way to estimate the Gaussian bias is called realization-dependent $N^{(0)}$ \Refs{Namikawa:2012pe, Planck:2018}. 
It makes use of the actual measured data $d$, along with two sets of simulations 
$\{s\},\{s'\}$ 
which are Gaussian random fields with power spectra that are equal to the total lensed CMB power spectrum $\tilde C_{\vl}^{TT}$:
\begin{align}
    \nonumber{\rm RDN}^{(0)}_\vL 
    \equiv
    \big<&C_{\vL} (\hat\kappa^{ds} ,\hat\kappa^{ds}) + C_{\vL}(\hat\kappa^{ds},\hat\kappa^{sd})\\
    \nonumber+&C_{\vL}(\hat\kappa^{sd},\hat\kappa^{ds}) + C_{\vL}(\hat\kappa^{sd},\hat\kappa^{sd})\\
    -&(C_{\vL}(\hat\kappa^{ss'},\hat\kappa^{ss'}) + C_{\vL}(\hat\kappa^{ss'},\hat\kappa^{s's}))\big>_{s,s'}.
    \label{eq:RDN0}
\end{align}
This estimator for the disconnected noise bias arises naturally from an Edgeworth expansion of the CMB likelihood which we show explicitly in \App{RDN0}. 
It has two advantages over simply evaluating $N_{\rm theory}$ (\Eq{NkappaL}) numerically.
First, while both methods require modeling the power spectrum of the observed data $d$, RD$N^{(0)}$ is parametrically less sensitive to inaccuracies in the power spectrum model.
Second, as we show below in \Fig{corr_mNhatvmRDN0}, it optimally suppresses the covariance of the lensing band powers.

Our method differs from the usual ${\rm RDN}^{(0)}$ subtraction in CMB lensing in several important ways. 
First, it does not require running the lensing estimator on Gaussian simulations with the same power spectrum as the data. 
Thus our estimator is insensitive to errors in the modelling of the total map power spectrum, which is a required input for those Gaussian simulations. 
Second, by not requiring to run our estimator on many simulations we are also able to significantly reduce the computational run time and memory cost of estimating the $N^{(0)}$ bias.

However, our method shares several limitations with the usual RD$N^{(0)}$ subtraction. 
Non-Gaussian foregrounds contribute a non-Gaussian bias to the lensing power spectrum estimator. 
This contribution is not subtracted automatically in our method, similarly to the other $N^{(0)}$ subtraction methods mentioned. 
Similarly, another bias to the lensing power spectrum can occur from inhomogeneous atmospheric noise, or detector noise from a non-uniform scan strategy, or due to mode coupling from the survey mask.
We study these in more detail in \Secs{aniso-noise}{masking}.

\section{Effect on the CMB Weak Lensing Power Spectrum Covariance}
\label{sec:cov}

In this section we argue that the bulk of the covariance in the CMB weak lensing power spectrum is due to the given $N^{(0)}$ realization. Correspondingly we should expect the bulk of the covariance to be removed if we remove the $N^{(0)}$ noise bias. 
This is desirable for our measured CMB weak lensing power spectra: making the covariance matrix more diagonal simplifies the problem of its estimation.

\begin{figure}
    \centering
    \includegraphics[width=\linewidth]{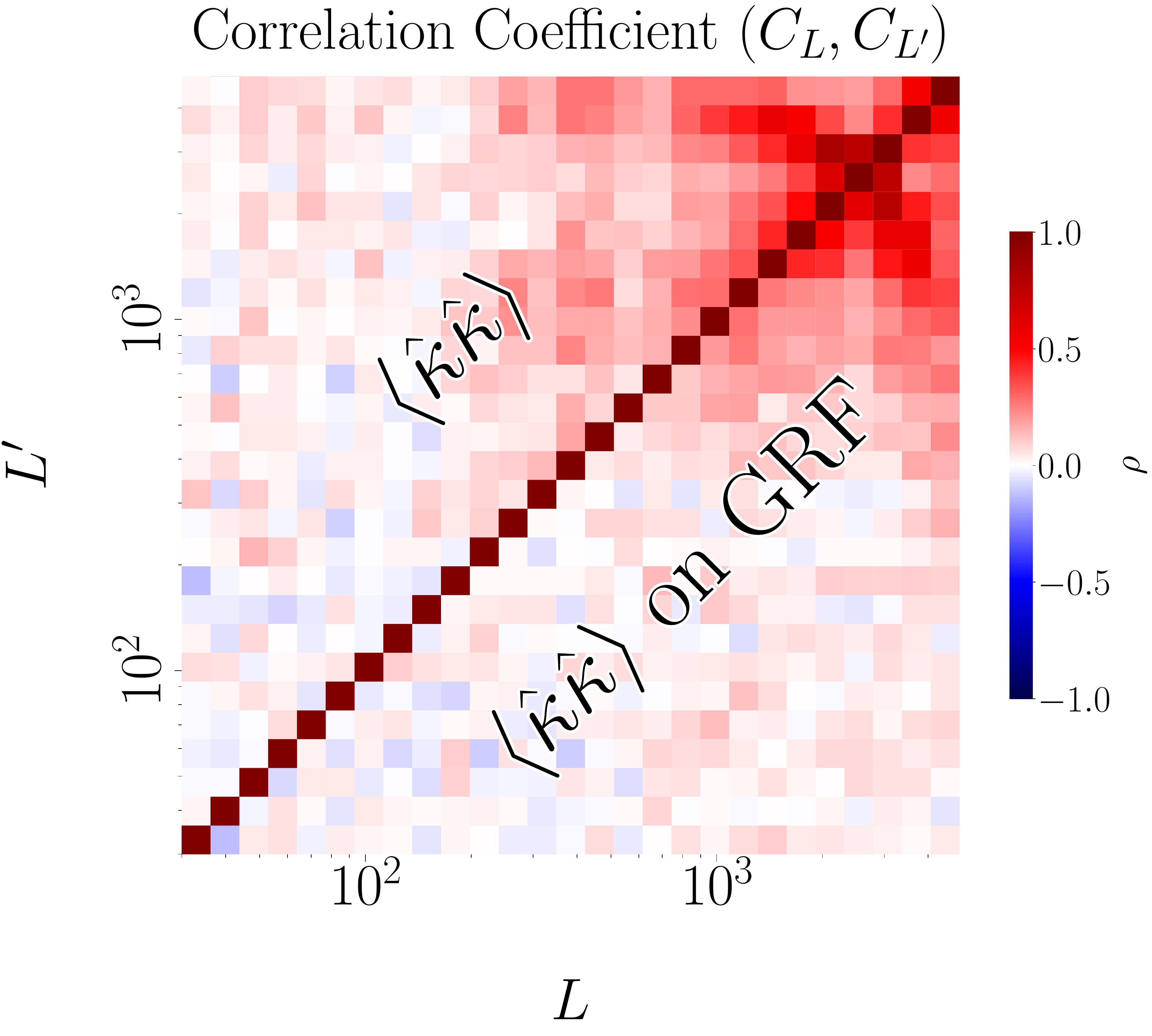}
    \caption{There is non-trivial covariance structure between different angular multipoles of $\langle\hat\kappa\hat\kappa\rangle \sim {\rm CMB\ lensing\ spectrum} + N^{(0)} + \dots$ as shown in the upper left where we plot covariance structure of $\langle\hat\kappa\hat\kappa\rangle$. 
    The dominant contribution to this covariance is from the $N^{(0)}$ bias. 
    This is shown in the lower right where we plot the covariance structure of $\langle\hat\kappa\hat\kappa\rangle$ run on GRFs with the same power spectrum as an actually lensed map. 
    Since in the bottom right plot we are only using Gaussian random fields, there is only the Gaussian $N^{(0)}$ contribution and no non-Gaussian lensing signal or higher order $N^{(i)}$ within $\langle\hat\kappa\hat\kappa\rangle_{\rm GRF}$. 
    Because the dominant covariance structure in the upper left also shows up in the bottom right we see that the dominant contribution to the covariance in CMB lensing spectrum is from $N^{(0)}$. 
    We spell out the origin of this covariance structure in \App{off_diagonal}.
    Since the dominant contribution to covariances in the CMB lensing spectrum are due to $N^{(0)}$, it should be expected that any prescription to handle the $N^{(0)}$ bias should also handle the covariance introduced by the $N^{(0)}$ bias. 
    We shall see in \Fig{corr_mNhatvmRDN0} that both our proposed method and the standard ${\rm RDN}^{(0)}$ do this. 
    }
    \label{fig:corr_QEQEvGRF}
\end{figure}

Indeed, \Fig{corr_QEQEvGRF} shows that the correlation structure of $C_L^{\hat\kappa\hat\kappa}$ is almost identical in two cases: (1) $\hat\kappa$ run on lensed temperature maps (upper left) and (2) $\hat\kappa$ run on an unlensed temperature map (Gaussian random fields) with the same power spectrum.
In (1), the resulting $C_L^{\hat\kappa\hat\kappa}$ includes $N^{(0)}$, the true lensing power spectrum, and higher order $N^{(i)}$ biases.
In (2), $C_L^{\hat\kappa\hat\kappa}$ includes only $N^{(0)}$.
{Since the dominant correlation structure in (1) also appears in (2) where there is only $N^{(0)}$, we see that the bulk of the covariance in the CMB weak lensing power spectrum is due to $N^{(0)}$.}
We analytically explain the origin of these off-diagonal covariance in $N^{(0)}$ in \App{off_diagonal}. 

\begin{figure}
\includegraphics[width=\columnwidth]{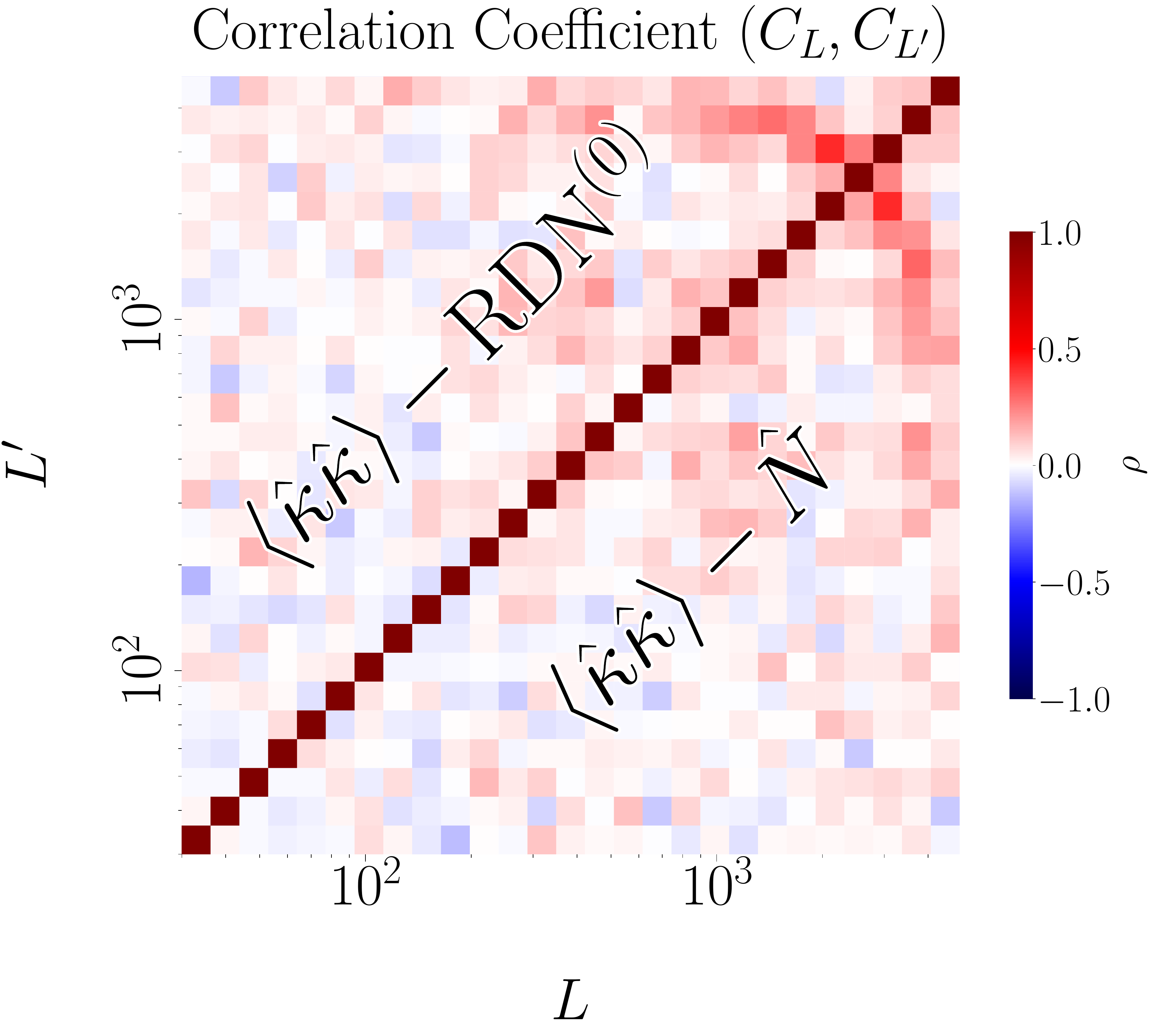}
\caption{The non-trivial covariance between different angular multipoles of $\langle\hat\kappa\hat\kappa\rangle \sim {\rm CMB\ lensing\ spectrum} + N^{(0)} + \dots$ shown in \Fig{corr_QEQEvGRF} which we argue are mostly due to $N^{(0)}$ are removed by both the standard ${\rm RDN}^{(0)}$ subtraction (upper left) and our proposed method of $N^{(0)}$ bias avoidance (lower right). Note that doing the naive $N^{\rm theory}$ subtraction would not remove the covariance since $N^{\rm theory}$ is a constant which cannot affect the covariance.
}
\label{fig:corr_mNhatvmRDN0}
\end{figure}

Subtracting the expected $N^{(0)}$, computed from the power spectrum of the input maps, would not change the correlation structure in \Fig{corr_QEQEvGRF}.
Indeed, this $N_\text{theory}$ subtraction only removes the mean $N^{(0)}$, not its exact realization.
On the other hand, both our $\hat N$ estimator and the ${\rm RDN}^{(0)}$ subtraction successfully suppress the off-diagonal covariances (\Fig{corr_mNhatvmRDN0}).
While both method are as effective in this respect, the $\hat N$ subtraction is dramatically cheaper computationally, requiring no simulations.

\section{Toy Model of Optimal Trispectrum Estimation}
\label{sec:toy}

\Reff{Smith:2015uia} presents an illuminating toy model for optimal trispectrum estimation which we connect to our method of $N^{(0)}$ subtraction, the standard realization-dependent $N^{(0)}$ subtraction, and the naive $N_{\rm theory}$ subtraction methods discussed in this paper. 
Consider a weakly non-Gaussian random variable $X$ with zero mean, known variance $\sigma^2$, and a small kurtosis $\mathcal K$ we wish to estimate:
\begin{align}
  \langle X^2\rangle = \sigma^2\label{eq:toyX2}\\ 
  \langle X^4 \rangle = 3 \sigma^4 + \mathcal K.\label{eq:toyX4}
\end{align}
Let $x_1,\dots, x_N$ be independent realization of this random variable $X$:
\begin{align}
\left<x_i x_j\right>&= \delta_{ij}\sigma^2 \label{eq:xixj}\\
\left<x_ix_jx_k x_l\right> &= \sigma^4(\delta_{ij}\delta_{kl} + \delta_{ik}\delta_{jl} + \delta_{il}\delta_{jk}) + \mathcal K \delta_{ij}\delta_{jk}\delta_{kl}.\label{eq:xixjxkxl}
\end{align}
In this toy model, $x_i$ is analogous to a specific Fourier mode of the CMB temperature anisotropy field $T_\vl$ and knowing $\sigma^2$ is analogous to knowing the total power spectrum of the temperature field.

From \Eqs{toyX4}{xixjxkxl} one might write down the naive estimator for $\mathcal K$:
\begin{equation}
    \hat{\mathcal K}_{\rm naive} = \frac 1 N \left(\sum_{i=1}^N x_i^4 \right) - 3 \sigma^4
\end{equation}
However if we are searching for a minimum variance estimator, this is suboptimal. 
Indeed, one can compute that
\begin{equation}
    {\rm Var}(\hat{\mathcal K}_{\rm naive}) = \frac{96 \sigma^8}{N}.\label{eq:varKnaive}
\end{equation}
Furthermore it suffers from a strong parametric dependence on our estimate of $\sigma^2$. 
Suppose we estimate $\sigma^2$ with some error:
\begin{equation}
\sigma_{\rm est}^2 = \sigma_{\rm true}^2 - \Delta \sigma^2   .\label{eq:robust}
\end{equation} 
Then the estimate of $\hat{\mathcal K}_{\rm naive}$ is biased as
\begin{equation}
    \langle \hat{\mathcal K}_{\rm naive}\rangle = \mathcal K - 6\sigma^2 (\Delta\sigma^2) +O(\Delta \sigma^2)^2.
    \label{eq:error_Knaive}
\end{equation}
This makes the naive estimator fragile to misestimates of $\sigma^2$ in comparison to other methods we will discuss shortly.

This naive method is analogous to the naive $N^{(0)}$ subtraction method we described earlier in this paper of subtracting off $N_{\rm theory}$. 
So we can gleam that the naive $N_{\rm theory}$ subtraction suffers the same problems as this naive kurtosis estimator. 
It is not the optimal minimum variance estimator of $C_L^{\kappa\kappa}$ and is vulnerable to misestimates of the total CMB temperature power spectrum (``$\sigma^2$'').

$\hat{\mathcal K}_{\rm opt}$, the minimum variance {unbiased} estimator of this kurtosis given that $\sigma^2$ is known, can be shown to be (\Reff{Smith:2015uia} and \App{Kopt_deriv})
\begin{align}
\hat {\mathcal K}_{\rm opt} &= \frac 1 N \sum_{i=1}^N x_i^4 - {\color{red}\underbrace{\frac{6\sigma^2 }{N} \sum_{i=1}^N x_i^2}_{{\rm (A)}}} + {\color{blue}\underbrace{{3\sigma^4 \vphantom{\sum_{i=1}^N}}}_{\rm (B)}}.\label{eq:Kopt}
\end{align}
Another computation will lead to the variance of this estimator:
\begin{equation}
    {\rm Var}(\hat{\mathcal K}_{\rm opt}) = \frac{24\sigma^8}N. \label{eq:varKopt}
\end{equation}
Comparing this to \Eq{varKnaive} we see that the naive estimator is significantly suboptimal. We can also perform a similar robustness analysis as in \Eq{robust} to find that if we misestimate $\sigma^2$ then
\begin{equation}
\langle\hat{\mathcal K}_{\rm opt}\rangle = \mathcal K + 3 (\Delta \sigma^2)^2
+...
\label{eq:error_Kopt}.
\end{equation}
In comparison to \Eq{error_Knaive} we see that the optimal estimator is also much more robust to misestimates of $\sigma^2$ by one power in the error. 
This is extremely attractive for problems where $\sigma^2$ is difficult to estimate.

The optimal kurtosis estimator $\hat{\mathcal K}_{\rm opt}$ is exactly analogous to the standard realization dependent $N^{(0)}$ subtraction method. 
{We show this rigorously in \App{Kopt_deriv} and \App{RDN0} by demonstrating that the derivation of ${\rm RDN}^{(0)}$ is the multivariate generalization of $\hat {\mathcal K}_{\rm opt}$. But this can also be seen schematically.
Indeed, terms {\color{red}(A)} and {\color{blue}(B)} in \Eq{Kopt} can be related to the corresponding terms in ${\rm RDN}^{(0)}_{L}$ \Eq{RDN0}
\begin{equation*}
\vcenter{\hbox{\includegraphics[width=0.85\linewidth]{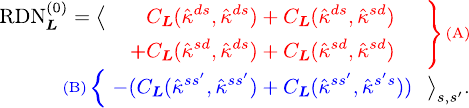}}} \tag{\ref{eq:RDN0}}
\end{equation*}
Namely:
\begin{itemize}
    \item  The {\color{red}(A) terms} of the form ${\color{red} C_{\mathbf L}(\hat\kappa^{ds}\hat\kappa^{ds})}$ are the same as being a product of the sample variance $\sum_i x_i^2 / N$ from the $dd$ contractions and the assumed ``true'' input variance $\sigma^2$ from the $s s$ contractions.
    \item The {\color{blue}(B) terms} of the form ${\color{blue} C_\vL (\hat\kappa^{ss'}, \hat\kappa^{ss'})}$ only containing information about the assumed ``true" variance $\sigma^2$ from the $ss$ and $s's'$ contractions.
\end{itemize}
}
 We put ``true'' in quotation marks since in reality, the underlying power spectra is not known making the optimal kurtosis (and thus trispectrum) estimator sensitive to how well we understand the underlying power spectrum as discussed in \Eq{error_Kopt}. 
 Thus from this toy model and correspondence between $\hat{\mathcal K}_{\rm opt}$ and ${\rm RDN}^{(0)}$ we can gleam the fact that ${\rm RDN}^{(0)}$ is more robust to misestimates of ``$\sigma^2$" compared to the naive method of $N^{\rm theory}$ as it was designed to be and can be thought of as a generalization of the optimal minimum-variance kurtosis estimator given that $\sigma^2$ is known perfectly.

In addition to the optimal estimator, \Reff{Smith:2015uia} presents an alternative near-optimal kurtosis estimator $\hat{\mathcal K}_{\rm alt}$ of the form
\begin{equation}
\hat{\mathcal K}_{\rm alt} = \frac 1 N \sum_i x_i^4 - {\color{c5}{\underbrace{\frac 3 {N(N-1)} \sum_{i\ne j} x_i^2 x_j^2}_{\rm (C)}}}. 
\label{eq:Kalt}
\end{equation}
As before, you can compute the variance of this estimator:
\begin{align}
{\rm Var}(\hat{\mathcal K}_{\rm alt}) &= \frac{24 \sigma^8}{N}\left(1 + \frac{3}{N-1}\right)  \\
&= {\rm Var}(\hat{\mathcal K}_{\rm opt}) + O\left( \frac 1 N\right)^2
\end{align}
From the last line where we compare to the variance of the optimal minimum variance estimator \Eq{varKopt} we see that this estimator is \textit{nearly} optimal. But a key difference is that this alternative estimator \textit{has no dependence on an assumed $\sigma^2$}. 
This means that unlike the optimal estimator $\hat{\mathcal K}_{\rm opt}$, this alternative estimator $\hat{\mathcal K}_{\rm alt}$ is completely insensitive to misestimates in $\sigma^2$. 

We can show that this alternative kurtosis estimator $\hat{\mathcal K}_{\rm alt}$ is analogous to our proposed method of subtracting the $N^{(0)}$ bias. 
Indeed, the {\color{c5} (C) term} in \Eq{Kalt} can be identified with the terms in our $\hat N$ subtraction scheme in \Eq{Nhat}:
\begin{equation}
 \hat{N}_\vL
=
2
\left(N^\kappa_{\bm{L}}\right)^{2}
\int_\vl
F^\kappa_{\vl, \vL-\vl}
F^\kappa_{-\vl, -\vL+\vl}\
{\color{c5}{\underbrace{\left| T_\vl \right|^2
\left| T_{\vL - \vl} \right|^2}_{\rm (C)}}}.
\tag{\ref{eq:Nhat}}
\end{equation}
We can understand the term {\color{c5}$\sum_{i\ne j}x_i^2 x_j^2$} as all the contractions where we get information only about the disconnected Gaussian bias from the $(ii)(jj)$ contraction but never about the connected non-Gaussian kurtosis since $x_i$ and $x_j$ are independent for $i\ne j$.
This is analogous to $\color{c5} \int_\vl \left|T_\vl \right|^2 \left|T_{\vL- \vl}\right|^2$ since these terms are the terms where we get all our contribution to the disconnected Gaussian bias\footnote{You might notice a slight subtlety to this correspondence between our $\hat N$ and this toy model. 
The $|T_\vl|^2 |T_{\vL-\vl}|^2$ terms in our $\hat N$ also contain information about the connected non-Gaussian contribution since $\langle T_\vl T_\vl^* T_{\vL-\vl} T_{\vL-\vl}^*\rangle_c\ne 0$ due to mode couplings from lensing whereas $\langle x_ix_i x_j x_j\rangle_c = 0$ by assumption of the $x_i$ being iid in this toy model. 
However, the analogy still holds. 
This is because computing $\langle \hat\kappa\hat\kappa\rangle - \hat N$ removes only $N_{\rm modes}$ of the total $N_{\rm modes}^2$ contributions to the connected non-Gaussian contribution when we subtract $\hat N$. So, to roughly first order in $O(1 / N_{\rm modes})$, we can draw the connection between our $\hat N$ method and this $\hat{\mathcal K}_{\rm alt}$ despite the iid assumption in the toy model which is not present in the general case.}. Thus from this toy model and correspondence between $\hat{\mathcal K}_{\rm alt}$ and $\hat N$ we can intuit the fact that $\hat N$ is completely insensitive to misestimates of ``$\sigma^2$," as we established earlier, but \textit{is still a nearly optimal estimator for the connected trispectrum (``kurtosis").}

\section{Inhomogeneous noise/depth}
\label{sec:aniso-noise}
The complex scan strategy of CMB experiments produce maps with inhomogeneous noise levels.
These are difficult to model, resulting in inaccurate simulations, which can bias the standard $N^{(0)}$ subtraction techniques.
While the ${\rm RDN}^{(0)}$ is somewhat robust to this inaccurate modeling,
a very small bias of only $\sim 0.1\%$ on $N^{(0)}$
can result in a large bias of $\sim 10\%$ on the small-scale lensing power spectrum (\Fig{Nhat_unmasked}).

The $\hat N$ estimator is insensitive to any mismodeling of the observed power spectrum, since it does not rely on simulations.
However, our estimator implicitly assumes statistical isotropy which only approximately holds once noise inhomogeneities arise. 
This can be put in contrasts to the standard ${\rm RDN}^{(0)}$ method which is sensitive to mismodeling but does not assume statistical isotropy and instead utilizes the full two-point function when estimating $N^{(0)}$. 
So it is important to understand the tradeoff between (1) approximating maps with noise inhomogeneities as statistically isotropic and (2) the sensitivity of ${\rm RDN}^{(0)}$ to mismodeling of noise inhomogeneities.
This tradeoff will determine the relative performance between our method and the standard ${\rm RDN}^{(0)}$ in the presence of inhomogeneous instrument noise.
\begin{figure}
    \centering
    \includegraphics[width=\linewidth]{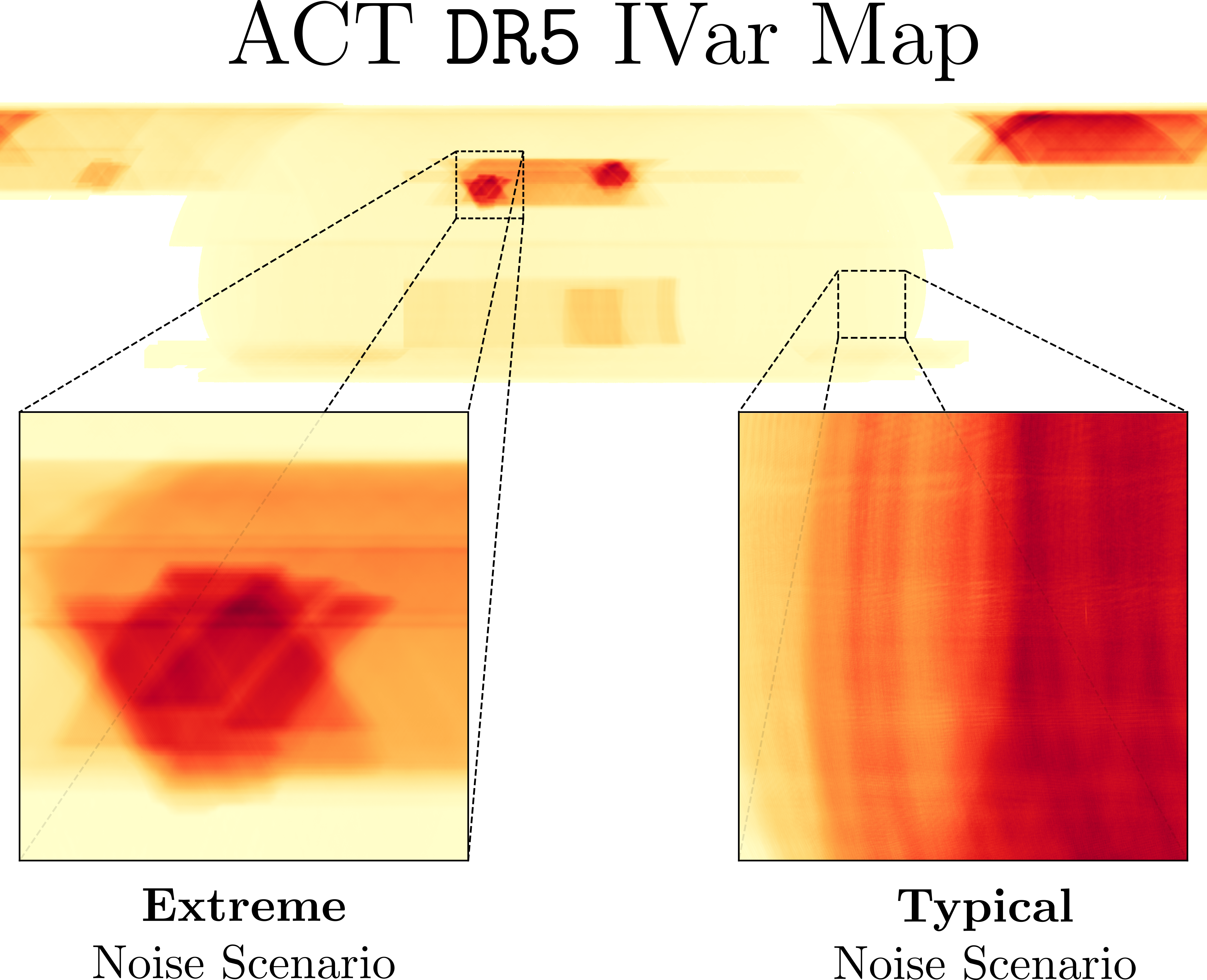}
    \caption{To test our proposed $\hat N$ subtraction method in the presence realistic anisotropic detector noise, we take two cutouts of the ACT \texttt{DR5} IVar Map: (1) a typical anisotropic noise scenario expected for SO and (2) a extreme anisotropic noise scenario. Both of these anisotropic noise maps are applied multiplicatively to a GRF with the expected noise power spectrum for SO such that $N^{\rm anisotropic}(\vx) \geq N^{\rm isotropic}(\vx)$. See \App{sims} for more details. {Note that in the \textbf{typical} noise scenario plot, we have enhanced the colors in order to emphasize that inhomogenous instrument noise can arise due to scan pattern.}}
    \label{fig:ACTDR5}
\end{figure}
In this section we study the robustness of our method to (1) typical or (2) extreme anisotropic detector noise.
For this, we use the actual depth maps from ACT \texttt{DR5} (see \Fig{ACTDR5}).
We perform a simulated analysis of the performance of our method and ${\rm RDN}^{(0)}$.
Specifically, we generate {anisotropic Gaussian} noise realizations by multiplying white noise maps with the {typical or extreme inhomogeneous depth maps shown in \Fig{ACTDR5}}.
We then add an independent lensed CMB realization to each of these, and a Gaussian random field representing the level of foregrounds at 150~GHz (neglecting their non-Gaussianity) resulting in 500 simulated maps.
\begin{figure}
    \centering
    \includegraphics[width=\linewidth]{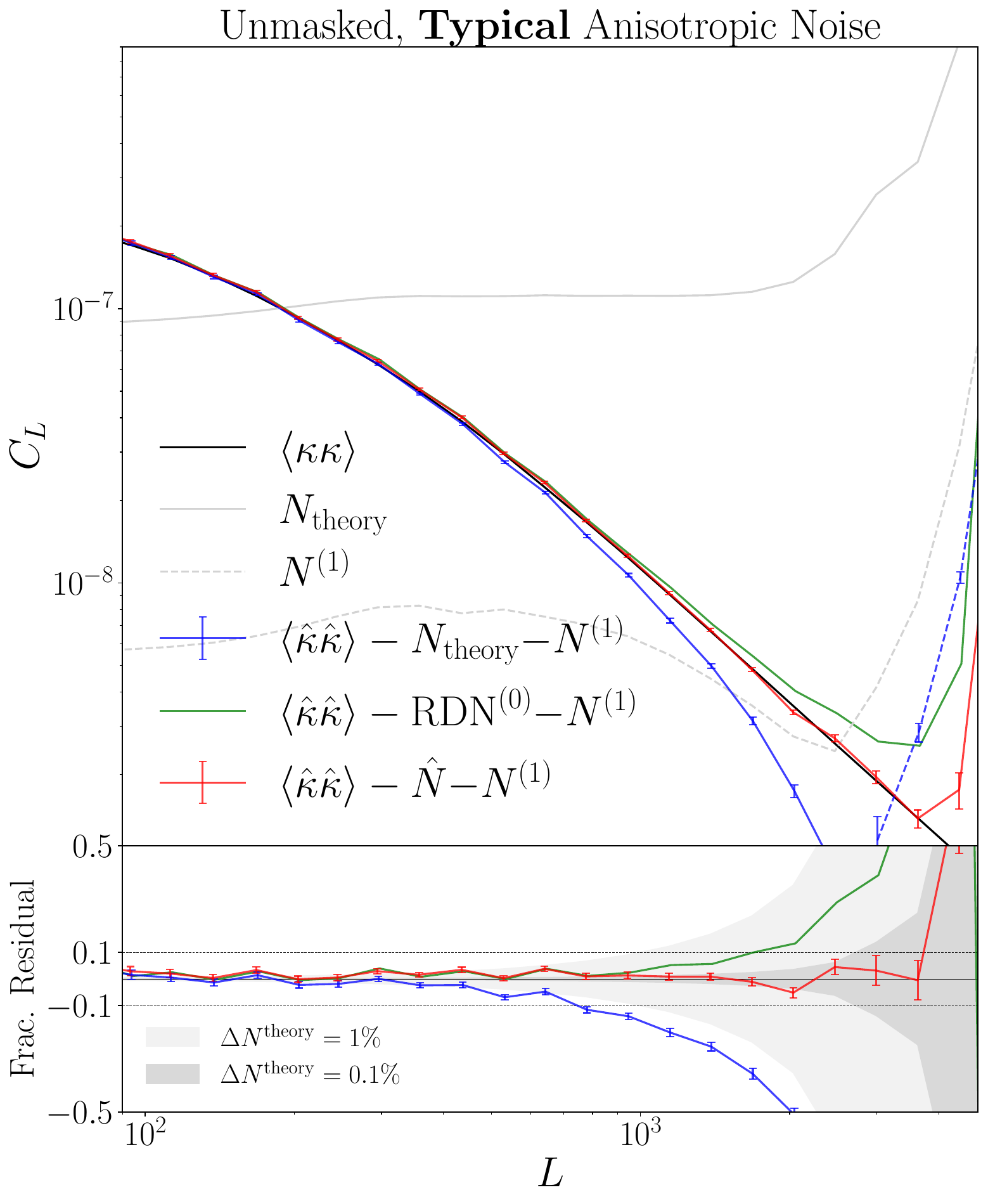}
    \caption{When CMB temperature anisotropy maps exhibit a \textbf{typical} level of anisotropic noise such as what's displayed in \Fig{ACTDR5}, there are additional mode couplings which one must address when estimating the lensing power spectrum $\langle \kappa\kappa\rangle$ (black) as a function of angular multipole $L$. 
    The standard method RDN0 subtraction ({\color{green}green}) and our new bias avoidance method ({\color{red}red}) both perform equally well in the presence of typical anisotropic noise levels whereas the naive $N^{\rm theory}$ subtraction ({\color{blue}blue}) fails at small scales.
    Dashed lines correspond to the same colored curve when that curve becomes negative. For similar reasons described in \Fig{Nhat_unmasked} we do no include error bars for the {\color{green}${\rm RDN}^{(0)}$} subtraction. The biased estimate for $\langle\kappa\kappa\rangle$ from ${\rm RDN}^{(0)}$ subtraction at small scales may be chalked up to both convergence of the MC correction as well as small errors in the assumed total temperature power spectrum including anisotropic noise needed for MC simulations. Even minor mismodeling can be magnified to percent level biases due to the fact that $N^{(0)}$ is orders of magnitude larger than the lensing signal $\langle\kappa\kappa\rangle$ at small scales. }
    \label{fig:anisonoise_unmasked}
\end{figure}
\Fig{anisonoise_unmasked} shows that our noise bias avoidance method ({\color{red}red}) is at least as robust to typical noise inhomogeneities as the ${\rm RDN}^{(0)}$ ({\color{green}green}) and $N_\text{theory}$ ({\color{blue}blue}) subtractions\footnote{{Note that in a real analysis one may use optimal anisotropic Wiener filtering \cite{Mirmelstein:2019sxi} as opposed to the naive quadratic estimator with isotropic weights as we do here. However, we do not believe the use optimal anisotropic Wiener filtering would change the qualitative picture shown in this section.}}.
While the ${\rm RDN}^{(0)}$ appears to fail at a lower multipole $L$, this might be improved with more simulations. 
However this comes at an increased computational cost which was prohibitive for us.
This highlights the benefit of the much lower computing cost of $\hat N$.

\begin{figure}
    \centering
    \includegraphics[width=\linewidth]{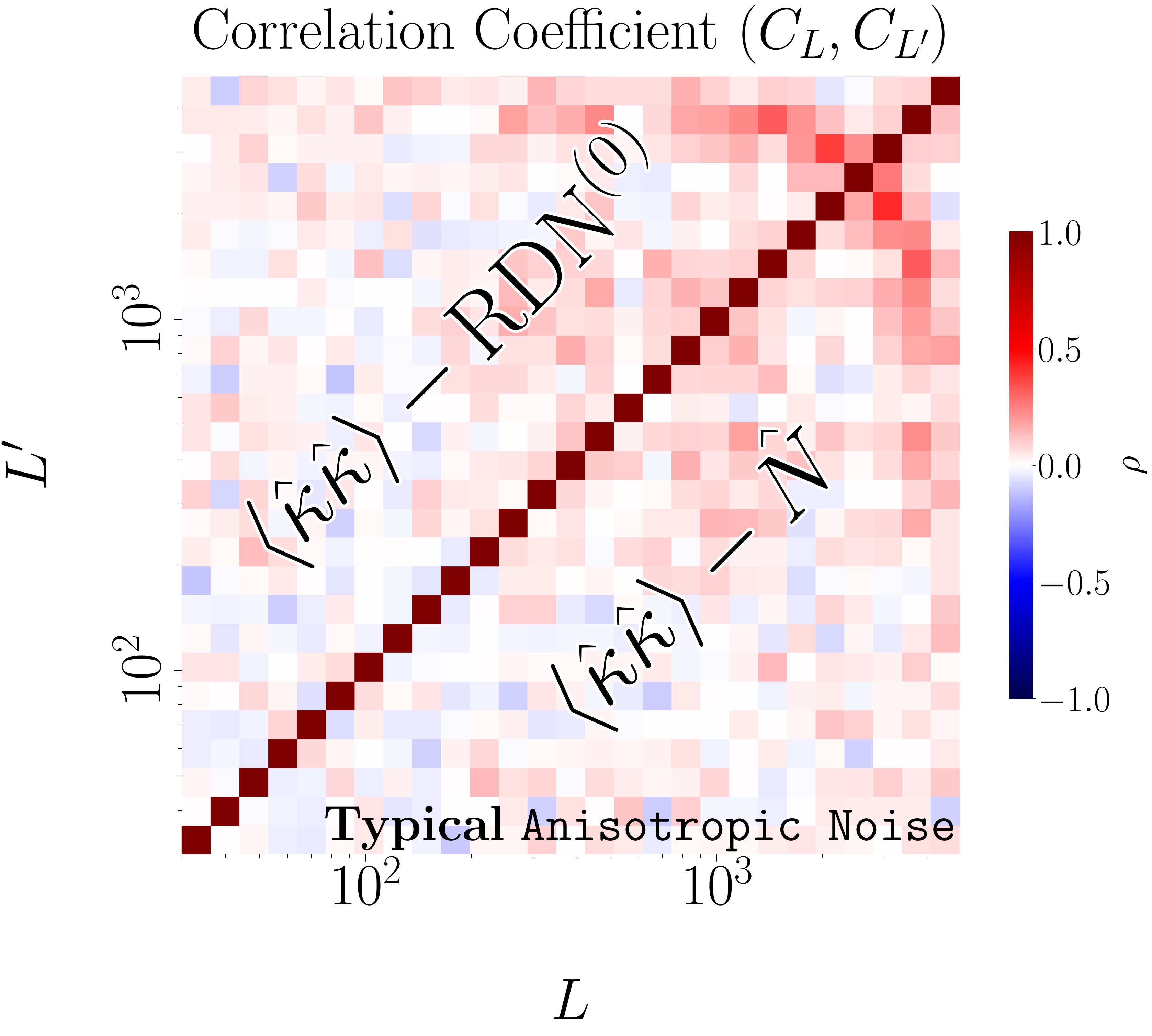}
    \caption{In the presence of typical anisotropic detector noise, the non-trivial covariance structure between different angular multipoles of $\langle\hat\kappa\hat\kappa\rangle \sim {\rm CMB\ lensing\ spectrum} + N^{(0)} + \dots$ shown in \Fig{corr_QEQEvGRF} may be modified. However this does not spoil the removal of the dominant contributor to these covariance by the standard ${\rm RDN}^{(0)}$ (upper left) subtraction and our proposed method of $N^{(0)}$ bias avoidance (lower right). 
    }
    \label{fig:corr_mNhatvmRDN0_aniso}
\end{figure}
In this typical noise inhomogeneity scenario, ${\rm RDN}^{(0)}$ and $\hat N$ both continue to suppress the off-diagonal covariances on the lensing power spectrum (\Fig{corr_mNhatvmRDN0_aniso}).
Indeed, the residual correlation structure is very similar to \Fig{corr_mNhatvmRDN0}.

\begin{figure}
    \centering
    \includegraphics[width=\linewidth]{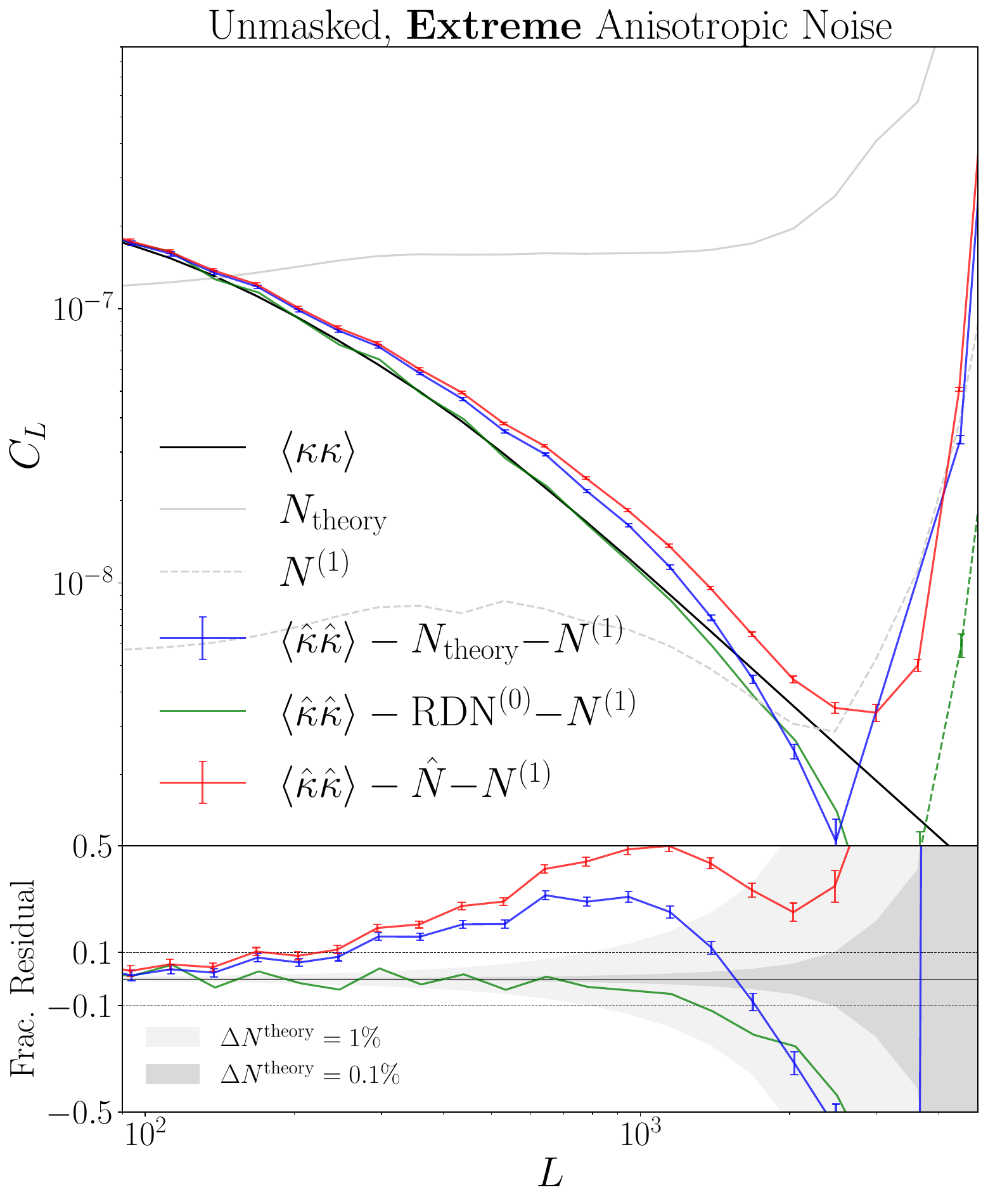}
     \caption{When CMB temperature anisotropy maps exhibit a \textbf{extreme} level of anisotropic noise such as what's displayed in \Fig{ACTDR5}, there are additional strong mode couplings which makes it difficult to estimate the lensing power spectrum $\langle \kappa\kappa\rangle$ (black) as a function of angular multipole $L$. 
    {At this level of extreme noise anisotropy, the naive $N_{\rm theory}$ subtraction ({\color{blue}blue}), and our new bias avoidance method ({\color{red}red}) fail while the standard ${\rm RDN}^{(0)}$ subtraction ({\color{green}green}), continues to work.}
    Dashed lines correspond to the same colored curve when that curve becomes negative. For similar reasons described in \Fig{Nhat_unmasked} we do no include error bars for the {\color{green}${\rm RDN}^{(0)}$} subtraction. It should be noted that this level of noise anisotropy is exactly what split-based methods described in \Sec{splits} and \Reff{Madhavacheril:2020ido} are designed to handle. We also showed in that section how one might combine split-based methods with our proposed noise bias avoidance method. Thus if one encounters such extreme anisotropic noise in analysis, we prescribe using split-based methods in combination with our proposed noise bias avoidance for split-based lensing power spectrum.
    {In the presence of such extreme anisotropic noise there will also be a mean-field which we estimate and subtract off. The origin of this mean-field is discussed more in \Sec{mean-field} for the comparable case of masking.}
    }
    \label{fig:anisonoise_unmasked_worst_case}
\end{figure}
In the extreme noise inhomogeneity scenario, \Fig{anisonoise_unmasked_worst_case} shows 
{that ${\rm RDN}^{(0)}$ works best if accurate noise simulations can be obtained. However, we shall see in the next section, that our $\hat{N}$ estimator can be combined with a split-based method \cite{Madhavacheril:2020ido} that is insensitive to the noise properties and does not rely on detailed understanding of the noise model}.
While this level of anisotropy was present in some parts of the ACT data, due to the coaddition of maps from multiple seasons with different footprints, we do not expect this level of inhomogeneity in current and upcoming {wide-field CMB surveys}.

\section{Comparison with the map split-based $N^{(0)}$ subtraction}
\label{sec:splits}

As we saw in \Fig{anisonoise_unmasked_worst_case}, once detector and atmospheric noise inhomogeneities become large, our method fails to remove the $N^{(0)}$ bias.
However one can combine our proposed method with split-based methods, proposed in \Reff{Madhavacheril:2020ido}, to handle these extreme noise anisotropies. \Reff{Madhavacheril:2020ido} proposed an estimator for the CMB lensing power spectrum with no detector noise bias $\hat C_\vL^{\kappa\kappa,\times}$. This estimator makes use of $m$ splits of the CMB map $T^{(i)}_\vl$ with independent instrument noise. One can then cross correlate these splits to make an estimator for the lensing potential that is insensitive to modelling of detector noise:
\begin{align}
\nonumber    \hat\kappa_\vL^{(ij)} 
&\equiv \frac 1 2 N^\kappa_\vL\int \frac{d^2 \vl}{(2\pi)^2}F^\kappa_{\vl, \vL-\vl}\\
&\times\left[T^{(i)}_\vl T^{(j)}_{\vL-\vl} + T^{(j)}_\vl T^{(i)}_{\vL-\vl} \right],\label{eq:kappaCross}\\
\nonumber   
\hat C_\vL^{\kappa\kappa, \times}
&\equiv \frac 1 {m(m-1)(m-2)(m-3)} \\&\times\sum_{ijkl}\gamma_{ijkl}C_\vL(\hat\kappa_\vL^{(ij)},(\hat\kappa_{\vL}^{(kl)})).
\end{align}
For the case of temperature, since $F^\kappa_{\vl, \vL-\vl}$ is symmetric, this symmetrization in \Eq{kappaCross} is trivial, but we use this notation since in general, e.g. when estimating $\kappa$ from TE or TB, $F$ is not symmetric. 
Following \Reff{Madhavacheril:2020ido}, we have defined $\gamma_{ijkl}$ as
\begin{equation}
    \gamma_{ijkl} =\begin{cases}
        1 & \textrm{if $(i,j,k,l)$ all distinct}\\
        0 & {\rm otherwise}
    \end{cases}.
\end{equation}
By ignoring contributions to the four-point function which repeat a split,
this method avoids the noise bias from atmospheric and detector noise, even if they are spatially inhomogeneous (e.g., due to the scan strategy).
This is otherwise a limiting factor for current analyses, even when temperature maps are dominated by the lensed CMB and foregrounds on the relevant scales.
This makes the split-based method very attractive.

It is also possible to define a version of our $\hat N$ subtraction method for this split-based lensing power spectrum. We can define:
\begin{align}
\nonumber\hat N_\vL^{ij,kl} &=  2(N^\kappa_\vL)^2 \int\frac{d^2\vl}{(2\pi)^2} F^\kappa_{\vl,\vL-\vl} F^\kappa_{-\vl,-\vL+\vl}\\
&\times\left|T^{(i)}_\vl T^{(j)}_{-\vl}\right|\left|T^{(k)}_{\vL-\vl} T^{(l)}_{-\vL+\vl}\right| .\label{eq:Nijkl}
\end{align}
Our estimate for the $N^{(0)}$ bias can be written as
\begin{equation}
    \hat N_\vL^\times = \frac 1 {m(m-1)(m-2)(m-3)}\times \sum_{ijkl} \gamma_{ijkl} \hat N^{ij,kl}_{\vL}.\label{eq:Nhatcross}
\end{equation}
{Thus} we can remove our estimated $N^{(0)}$ bias
\begin{equation}
    \hat C_\vL^{\kappa\kappa,\text{(no bias)}, \times}=\hat C_\vL^{\kappa\kappa,\times}- \hat N_\vL^\times.\label{eq:Cnbcross}
\end{equation}

Naively this is an $O(m^4)$ computation. However, analogous to the fast algorithm proposed by \Reff{Madhavacheril:2020ido}, we can construct a fast $O(m^2)$ algorithm to compute this $\hat N_\vL^\times$ which is described in \App{fastNhatX}.

\section{Impact of the mask}
\label{sec:masking}

The presence of a mask leads to additional complications in the analysis of the CMB lensing power spectrum by inducing additional difficult-to-model mode couplings and a mean field which must be removed.

\subsection{Biases from the mask mode coupling}
\label{sec:mean-field}

To illustrate these additional mode couplings consider a temperature field with a masking function applied to it:
\begin{equation}
    T^m(\vx)=M(\vx) T(\vx).
\end{equation}
Thus in Fourier space we have a convolution
\begin{equation}
    T^m_\vl = \int \frac{d^2\vl'}{(2\pi)^2} T_{\vl'}M_{\vl-\vl'},
\end{equation}
where we have introduced the Fourier transform of the mask. 
Now consider the mask
\begin{align}
    M(\vx) &= 1 - m(\vx)\\
    \Rightarrow M_\vl &= \delta^{(D)}(\vl) - m_\vl.
\end{align}
This tells us that
\begin{equation}
T^m_\vl = T_\vl - \int \frac{d^2 \vl'}{(2\pi)^2} T_{\vl'} m_{\vl-\vl'}.
\end{equation}
To leading order, this modifies the mode couplings from \Eq{TTspecial}:
\begin{align}
    \nonumber\langle T_\vl^m T_{\vL-\vl}^m\rangle &\approx \langle T_\vl T_{\vL-\vl}\rangle - m_\vL ( C_\vl^{TT} + C_{|\vL-\vl|}^{TT})\\
    &\equiv \langle T_\vl T_{\vL-\vl}\rangle + m_\vL f^m_{\vl,\vL}
\end{align}
Where in the lasts line we implicitly defined
\begin{equation}
    f^m_{\vl, \vL} = - C_\vl^{TT} - C_{|\vL-\vl|}^{TT}.
\end{equation}
Namely we get additional mode couplings due to masking. 
There are several methods to account for these mode coupling such as the pseudo-$C_\ell$ algorithm described in \Refs{Alonso:2018jzx, Hivon:2001jp} with modifications such as those done in \Reff{ACT:2023dou} or alternatively, the mask can be propagated through the quadratic estimator as shown in \Reff{noah}. 
We refer the reader to \Reff{ACT:2023dou} for details on how the latest measurement of the CMB lensing power spectrum by ACT accounts for mode couplings due to masking. Masking also leads to to a {\color{red}mean-field bias} in our quadratic estimator for $\kappa$, \Eq{kappa}:
\begin{equation}
    \langle\hat\kappa_\vL^m\rangle \approx \langle\hat\kappa_\vL \rangle + {\color{red}\underbrace{N^\kappa_\vL \left(\int \frac{d^2 \vl}{(2\pi)^2} F^\kappa_{\vl,\vL-\vl} f^m_{\vl,\vL}\right) m_\vL}_{\rm mean-field\ bias}}.
\end{equation}
Typically this mean field {from a fixed mask is estimated from many simulations with independent CMB and lensing realizations and subtracted off: 
\begin{equation}
    \textrm{Mean-field subtraction}:\hat\kappa \rightarrow\hat\kappa  - \langle\hat\kappa\rangle.
\end{equation}}
In this section we make use of this mean-field subtraction.
{Since our $\hat N$ (\Eq{Nhat}) is also derived from data, we must apply a mean-field correction $\Delta \hat N_{\rm mf}$ to it as well:
\begin{equation}
    \textrm{Mean-field subtraction}: \hat N \rightarrow \hat N/f_{\rm sky}^2 - \Delta \hat N_{\rm mf}.
\end{equation}
This correction can be computed from the same simulations used to compute the mean field bias $\langle\hat\kappa\rangle$ by averaging the difference between $\hat N$ run on masked and unmasked maps:
\begin{equation}
   \Delta \hat N_{\rm mf} = \langle\hat N(\textrm{Masked $T$'s})/f_{\rm sky}^2 - \hat N(\textrm{Unmasked $T$'s}) \rangle,
\end{equation}
where we discuss the origin of the $f_{\rm sky}^2$ factor in \Sec{fsky}. In this section we also make use of this mean-field correction to $\hat N$.
}

\begin{figure}
    \centering
    \includegraphics[width=\linewidth]{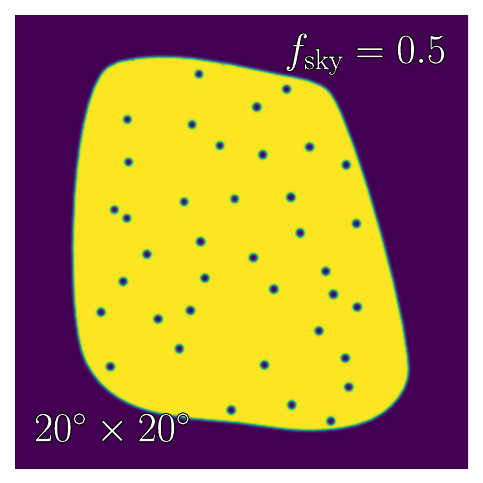}
    \caption{In our numerical studies in \Sec{masking} we use the above apodized $20^\circ\times 20^\circ$ mask which includes point sources of radius $\sim 10$ arcmin to roughly match SPT \cite{Millea:2020iuw}. Apodization is done using a Gaussian filter where the standard deviation for the Gaussian kernel is $3$ pixels. The unapodized mask and point sources were lovingly hand-drawn.}
    \label{fig:mask}
\end{figure}

\begin{figure}
    \centering
    \includegraphics[width=\linewidth]{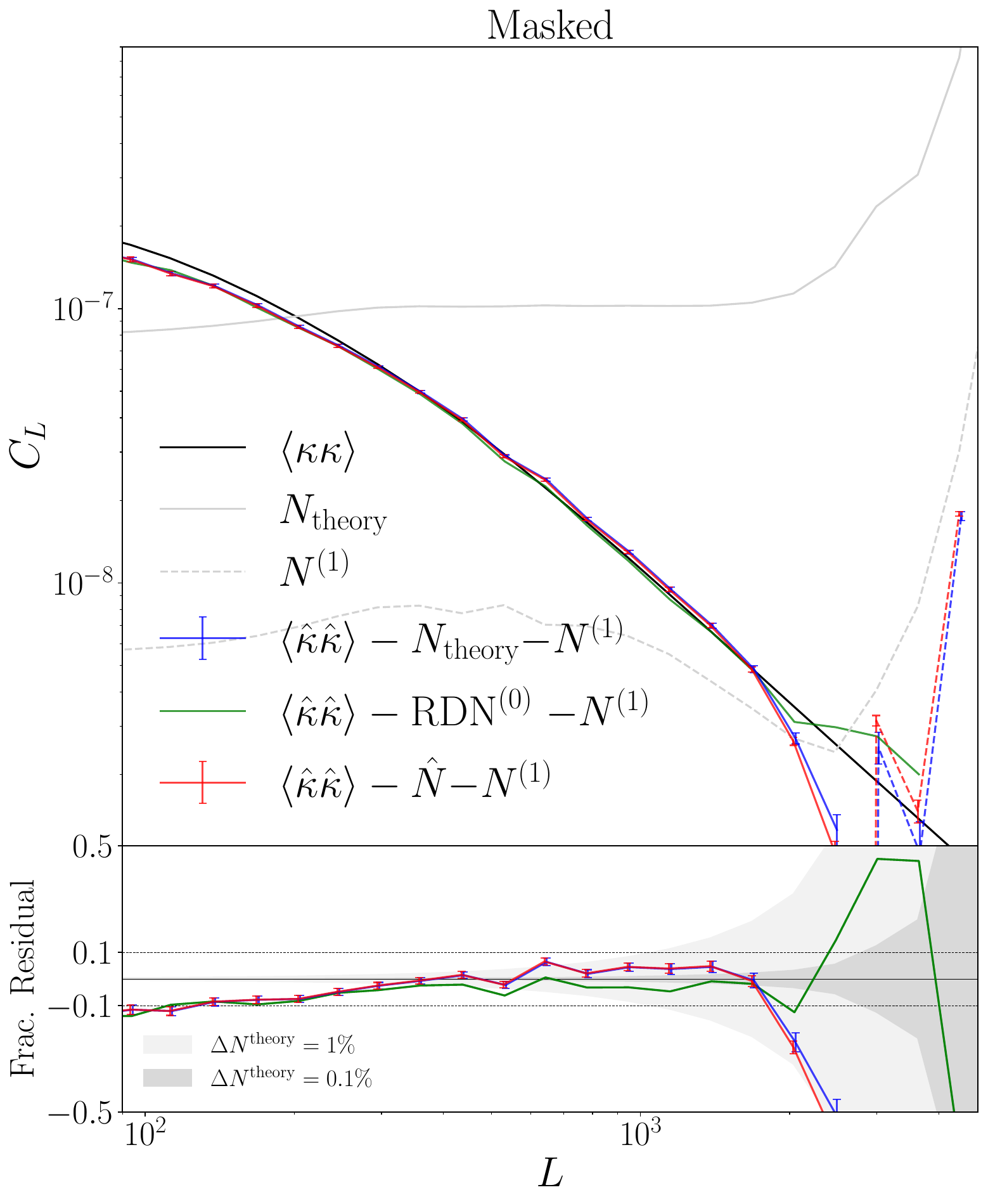}
    \caption{When masks such as the one shown in \Fig{mask} are applied CMB temperature anisotropy maps, there are additional mode couplings which one must address when estimating the lensing power spectrum $\langle \kappa\kappa\rangle$ (black) as a function of angular multipole $L$. We spell out roughly how these appear in \Sec{masking}.
    The standard method RDN0 subtraction ({\color{green}green}), the naive $N^{\rm theory}$ subtraction ({\color{blue}blue}), and our new bias avoidance method ({\color{red}red}) all perform comparably in the presence of masking. 
    Dashed lines correspond to the same colored curve when that curve becomes negative. For similar reasons described in \Fig{Nhat_unmasked} we do no include error bars for the {\color{green}${\rm RDN}^{(0)}$} subtraction. Since handling additional mask induced mode couplings are beyond the scope of this paper, we do not apply any corrections to account for these mask induced mode couplings. Instead the key result here is that {\color{red}our proposed method} has comparable performance to the standard {\color{green}${\rm RDN}^{(0)}$ subtraction} in the presence of masking.}
    \label{fig:masked}
\end{figure}
In \Fig{masked}, we show how our exact noise bias avoidance performs on a lensed CMB temperature map with the mask shown in \Fig{mask}, in comparison to the standard method. 
We see from this plot that our proposed method is able to match the performance the standard ${\rm RDN}^{(0)}$ method.

{In practice, the mask induces additional coupling between the observed Fourier modes of the temperature maps.
The full treatment of these mode coupling effects is beyond the scope of this paper (see \Reff{ACT:2023dou} for a thorough treatment of the effect of masking on CMB lensing reconstruction).}

\begin{figure}
    \centering
    \includegraphics[width=\linewidth]{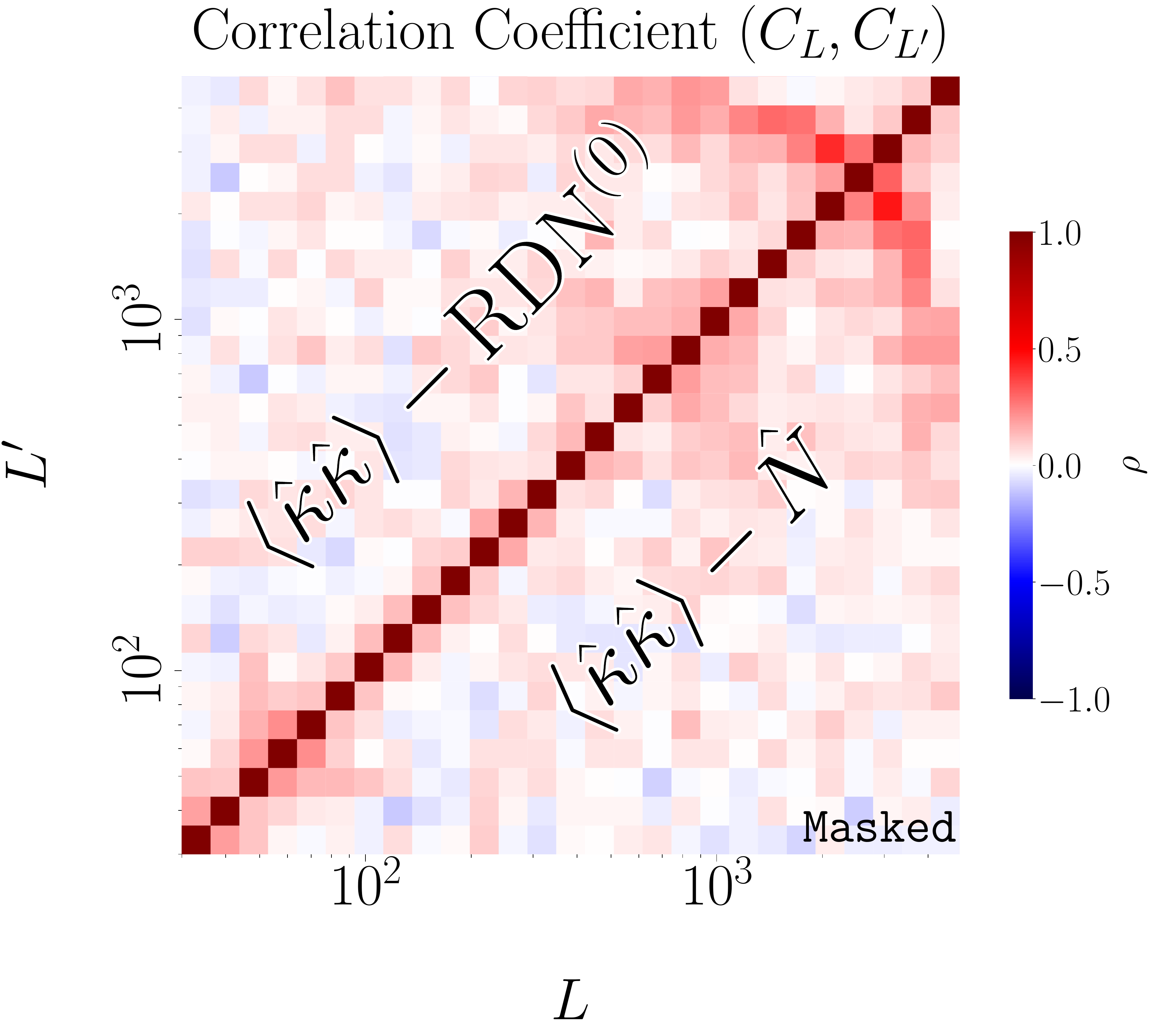}
    \caption{In the presence of masking, the non-trivial covariance structure between different angular multipoles of $\langle\hat\kappa\hat\kappa\rangle \sim {\rm CMB\ lensing\ spectrum} + N^{(0)} + \dots$ shown in \Fig{corr_QEQEvGRF} may be modified. However this does not completely spoil the removal of the dominant contributor to these covariance by the standard ${\rm RDN}^{(0)}$ (upper left) subtraction and our proposed method of $N^{(0)}$ bias avoidance (lower right). }
    \label{fig:masked_corr}
\end{figure}

In \Fig{masked_corr} we plot the correlation structure if we apply our $\hat N$ subtraction in comparison to the standard ${\rm RDN}^{(0)}$ subtraction in the presence of a mask. This is analogous to \Fig{corr_mNhatvmRDN0} and \Fig{corr_mNhatvmRDN0_aniso} where we plotted the same thing for isotropic  and typical anisotropic noise respectively. We can see that in this case, we are still able to remove covariances as before. 

\subsection{$f_{\rm sky}$ corrections for masking}
\label{sec:fsky}
In the presence of a mask, the lensing power spectrum from $\langle\hat\kappa\hat\kappa\rangle$ will be underestimated by approximately $f_{\rm sky}$ and our $\hat N$ will underestimate the $N^{(0)}$ bias by approximately $f_{\rm sky}^2$. In this paper we adopt the simple approximate correction in the presence of masking by modifying \Eq{ClmNhat}:
\begin{equation}
\label{eq:ClmNhatMasked}
  \hat C_\vL^{\kappa\kappa,\textrm{(no bias)}}  = \frac{\hat\kappa_\vL\hat\kappa_\vL^*}{f_{\rm sky}}  - \frac{\hat N_\vL}{f_{\rm sky}^2}.
\end{equation}
We note that a Monte-Carlo normalization or the use of pseudo-$C_{\ell}$ will automatically include this correction which won't need to be applied separately. 
We now illustrate why this is with the same toy model we considered in \Sec{toy}.
As in \Sec{toy}, consider $N$ independent measurements of a random variable $X$: $\{x_1,\dots,x_N\}$. 
This $X$ is again the 1D analogue of the {$T_\vl$ maps}.
The 1D analogue of the lensing power spectrum and our $\hat N$ estimator may be written as we did before in \Eq{Kalt}:
\begin{align}
\label{eq:1dKK}
   \left<\frac 1 N \sum_i x_i^4 \right>&= \langle X^4\rangle\\
   \label{eq:1dNhat} \left<\frac3 {N(N-1)}\sum_{i\ne j} x_i^2 x_j^2 \right> &= 3 \langle X^2\rangle^2.
\end{align}
Now suppose that we apply a mask that selects a fraction $f$ of our measurements $x_i$. 
Namely our masked dataset will have $x_i=0$ for $i> N f$. This $f$ is analogous to $f_{\rm sky}$\footnote{A subtlety in this claim is that the number of measurements in the toy model, $N$, is analogous to $N_{\rm modes}$. Since $N_{\rm modes} \propto {\rm Area}$ the analogy between $f$ and $f_{\rm sky}$ holds.}. Then \Eq{1dKK} would be modified as
\begin{equation}
    \left<\frac 1 N \sum_i x_i^4 \right> =    f\langle X^4 \rangle.
\end{equation}
{Similarly} if we approximate $N(N-1) \approx N^2$ then \Eq{1dNhat} would be modified as 
\begin{equation}
    \left< \frac 3 {N(N-1)}\sum_{i\ne j}x_i^2 x_j^2\right> \approx f^2 \times 3 \langle X^2\rangle^2 .
\end{equation}

{Thus,} to properly normalize our estimator we must include appropriate factors of $f^2$. For the general case of $\langle \hat \kappa \hat\kappa\rangle$ this of course is only an approximation but we believe error due to this approximation are subdominant to error from unaccounted for mask induced mode couplings in our numerical studies and so made use of \Eq{ClmNhatMasked} in this section.

\section{Polarization-based lensing}
\label{sec:polarization}

{One can estimate lensing statistics from other CMB random fields, such as polarization.} Indeed, for $\{X,Y\}\in \{T,E,B\}$ one can write a generalization  of \Eq{TTspecial}:
\begin{equation}
\langle X_\vl Y_{\vL-\vl}\rangle= f_{\vl,\vL-\vl}^{XY}\kappa_\vL + O(\kappa_\vL^2).
\end{equation}
We refer readers to \Refs{Hu:2001kj, Sailer:2022jwt} for the explicit form of these $f^{XY}$ and subsequent $F^{XY}$ which are generalizations of \Eq{weights}. Due to lensing's imprint on these correlations one may define a quadratic estimator using any pair of $\{T,E,B\}$. Thus, one may write a estimator for the lensing power spectrum using $X,Y,W,Z \in \{T,E,B\}$ which we will call $C_L^{XY,WZ}$. 

The lensing power spectrum estimated from $X,Y,W,Z\in\{T,E,B\}$ will also have it's own $N^{(0)}$ bias that our proposed $\hat N$ subtraction can be generalized to remove. Namely we can write as a generalization of \Eq{Nhat}
\begin{align}
\nonumber    \hat N_\vL^{XY,WZ} &=  (N^\kappa_\vL)^2 \int\frac{d^2 \vl}{(2\pi)^2} F^{XY}_{\vl,\vL-\vl} F^{WZ}_{-\vl,-\vL+\vl} \\
\nonumber    \times\Big( &\left|X_\vl W_{-\vl}\right|\left|Y_{\vL-\vl} Z_{-\vL+\vl}\right| \\
    + &\left| X_\vl Z_{-\vl} \right| \left| Y_{\vL-\vl} W_{-\vL+\vl}\right|\Big).
\end{align}
{Correspondingly} \Eq{ClmNhat} becomes
\begin{equation}
    \hat C_\vL^{XY,WZ\textrm{(no bias)}} \equiv  \hat C_\vL^{XY,WZ} - \hat N^{XY,WZ}_\vL
\end{equation}
We leave the numerical exploration of the performance of $\hat N^{XY,WZ}$ subtraction to future work.

\section{Conclusions}
\label{sec:conclusion}

In this paper we have presented an estimator of the CMB lensing power spectrum that avoids the Gaussian $N^{(0)}$ noise bias. This is done by isolating which terms contribute to the $N^{(0)}$ bias when estimating the lensing power spectrum $\langle\kappa\kappa\rangle$ and avoiding these terms. We showed that avoiding these terms is computationally efficient and negligibly reduces the signal-to-noise. This estimator is run only on data thus avoiding the need for bias subtraction using simulations. {Since} our estimator avoids simulations{, we avoid} (1) sensitivity to misestimates in simulated CMB and noise models and (2) the large computational cost that current simulation-based methods like ${\rm RDN}^{(0)}$ require.

Our estimator is as robust as the standard ${\rm RDN}^{(0)}$ to (1) {{masking of the CMB map}} and (2) {the level of }{noise inhomogeneity}{ expected for current and upcoming wide-field CMB surveys}. 
{To handle extreme levels of noise inhomogeneity,} we also show how our estimator may be combined with split-based methods (\Reff{Madhavacheril:2020ido}) which make it insensitive to any assumptions made in modeling or simulating the instrument noise. 
In addition, even though we focused mostly on estimating lensing with CMB temperature, we show how our estimator can be generalized to include polarization information as well.

We also discuss the connection between our estimator, ${\rm RDN}^{(0)}$, and optimal trispectrum/four-point function estimation in the context of an illuminating toy model presented in \Reff{Smith:2015uia}. This toy model allowed us to argue that our proposed estimator has the same variance as the optimal minimum variance unbiased estimator of the connected trispectrum/four-point function to first order in $1/N_{\rm modes}$ but with no parametric dependence on an assumed power spectrum/two-point function. Thus we believe our estimator is applicable to analogous problems in other fields where estimation of connected trispectra/four-point functions are of interest such as in large-scale structure.

There are several key directions for further explorations of our proposed method. 
First, for the usage of our method in a full-sky analysis of CMB lensing power spectra, a generalization of this method to curved sky must be worked out. 
Additionally, though we presented the theoretical framework for how our method may be naturally combined with split-based methods such as those presented in \Reff{Madhavacheril:2020ido}, numerical studies of combining our proposed estimator and split-based methods are needed to verify their compatibility. 
{The use of bias hardening techniques \cite{Namikawa:2012pe, Sailer:2020lal}, optimal anisotropic Wiener filtering \cite{Mirmelstein:2019sxi}, and global-minimum-variance lensing quadratic estimators \cite{Maniyar:2021msb} are also often employed to enhance CMB weak lensing analyses.
We do not explore how our proposed method may be combined with such techniques and leave this to future work.}
Finally, there is the interesting work of considering how our proposed estimator which in general may be thought of as an estimator for the connected trispectrum/four-point function could feed into analysis of large-scale structure data where higher order statistics have been an area of recent interest.

\section*{Acknowledgments}

We thank Federico Bianchini, Anthony Challinor, Mathew Madhavacheril, Abhishek Maniyar, Antony Lewis, Frank Qu, Noah Sailer, Blake Sherwin, and Kimmy Wu for useful discussions.
This work received support from the U.S. Department of Energy under contract number DE-AC02-76SF00515 to SLAC National Accelerator Laboratory.
D.S. is supported by the National Science Foundation Graduate Research Fellowship under Grant No. DGE-2146755.
S.F. is supported by Lawrence Berkeley National Laboratory and the Director, Office of Science, Office of High Energy Physics of the U.S. Department of Energy under Contract No.\ DE-AC02-05CH11231.

\appendix
\section{Convention}
Throughout this paper we use the convention
\begin{align}
f(\vx) &= \int \frac{d^2 \vl}{(2\pi)^2}  e^{i \vl\cdot \vx} f_\vl\\
f_\vl &= \int d^2 \vx\ e^{-i\vl\cdot \vx} f(\vx)\\
\Rightarrow \delta^D(\vl) &= \int \frac{d^2 \vx}{(2\pi)^2}\ e^{i\vl\cdot \vx} 
\end{align}
Our square CMB maps have finite area $A$ and resolution. When Fourier transforming these maps, there is a discrete set of modes we can resolve. These modes are integer multiples of the smallest {fundamental mode} $k_F=2\pi/\sqrt{A}$ which we can measure. We may rewrite the Dirac delta for these discrete set of modes. For two modes $\vk_i = k_F\mathbf i $ and $\vk_j = k_F \mathbf j$, where $\mathbf i$ and  $\mathbf j$ are a vector of integers,
\begin{equation}
    \delta^{(D)}(\vk_i - \vk_j)=\delta^{(D)}(k_F(\mathbf i - \mathbf j))=  \frac A{(2\pi)^2}\delta^{(K)}_{\mathbf i, \mathbf j},
    \label{eq:deltaFinite}
\end{equation}
where $\delta^{(K)}$ is the Kronecker delta.

\section{$N^{(i)}$ Biases}
\label{app:Ni}
In this section we will spell out explicitly the origin of the $N^{(i)}$ biases. 
These noise biases fall out when using \Eq{T} to expand the four-point function within the integral of \Eq{kk} and applying Wick's theorem to the Gaussian fields that fall out. 
First lets expand \Eq{Tx} out to another order:
\begin{align}
\nonumber       T(\vx) &= T^0(\vx + \vd(\vx)) = T^0(\vx) + (\partial_i \psi(\vx)) (\partial^i T^0(\vx)) \\
       &+ \frac 1 2 (\partial^i\psi(\vx))(\partial^j \psi(\vx)) (\partial_i\partial_j T^0(\vx)) + O(\vd^3). 
\end{align}
From here we can rewrite \Eq{T} as 
\begin{equation}
    T_\vl = T^{(0)}_\vl + T^{(1)}_\vl+ T^{(2)}_\vl + O(\kappa^3)
\end{equation}
where we have defined 
\begin{align}
T^{(0)}_\vl &\equiv T^0_\vl\\
   T^{(1)}_\vl &\equiv - \int \frac{d^2 \vl'}{(2\pi)^2}  \vl'\cdot(\vl - \vl') \frac{2 \kappa_{\vl - \vl'}}{(\vl - \vl')^2} T^0_{\vl'}  \label{eq:T1}
\end{align}
which is the same as in \Eq{T} and
\begin{align}
\nonumber T^{(2)}_\vl \equiv - \frac 1 2 \int &\frac{d^2 \vl_1}{(2\pi)^2}\frac{d^2\vl_2}{(2\pi)^2} T^0_{\vl_1}\frac{2\kappa_{\vl_2}}{\vl_2^2}\frac{2\kappa^*_{\vl_1+\vl_2-\vl}}{(\vl_1+\vl_2-\vl)^2}\\
   &\times[\vl_1\cdot \vl_2][\vl_1\cdot(\vl_1+\vl_2-\vl)]\label{eq:T2}
   .
\end{align}
Now, instead of keeping $\kappa_\vL$ fixed as we did in \Sec{standard}, lets promote $\kappa_\vL$ to a Gaussian random field with power spectrum $ C^{\kappa\kappa}$:
\begin{equation}
\kappa_\vL \sim \mathcal N(0, C_\vl^{\kappa\kappa}).
\end{equation}
We will be computing $\langle TTTT\rangle$ which at each order has terms like:
\begin{align}
\nonumber    \langle T_\vlo T_\vlt T_\vlth T_\vlf \rangle &\approx \underbrace{\langle \Tz_\vlo \Tz_\vlt \Tz_\vlth \Tz_\vlf \rangle}_{O(\kappa^0)}\\
\nonumber    &+\underbrace{\langle \To_\vlo \Tz_\vlt \Tz_\vlth \Tz_\vlf \rangle+3\,{\rm perms.}}_{O(\kappa^1)}\\
\nonumber    &+{\langle \Tt_\vlo \Tz_\vlt \Tz_\vlth \Tz_\vlf \rangle+3\,{\rm perms.}}\\
\nonumber    &+{\langle \To_\vlo \To_\vlt \Tz_\vlth \Tz_\vlf \rangle}+{\langle \Tz_\vlo \Tz_\vlt \To_\vlth \To_\vlf \rangle}\\
\nonumber    &+{\langle \To_\vlo \Tz_\vlt \To_\vlth \Tz_\vlf \rangle}+{\langle \Tz_\vlo \To_\vlt \Tz_\vlth \To_\vlf \rangle}\\
\nonumber    &+\underbrace{\langle \Tz_\vlo \To_\vlt \To_\vlth \Tz_\vlf \rangle+\langle \To_\vlo \Tz_\vlt \Tz_\vlth \To_\vlf \rangle}_{O(\kappa^2)}\\
\label{eq:TTTT}
\end{align}
We could use Wick contractions to compute each of theses terms. For example for the $O(\kappa^0)$ term.
\begin{align}
\nonumber        &\langle T^{(0)}_{\vl_1}T^{(0)}_{\vl_2}T^{(0)}_{\vl_3}T^{(0)}_{\vl_4}\rangle = (2\pi)^4\\
\nonumber        \times \Big[&\delta^{(D)}(\vl_1+\vl_2) \delta^{(D)}(\vl_3+\vl_4)C_{\vl_1}^{TT}C_{\vl_3}^{TT} \\
\nonumber        + &\delta^{(D)}(\vl_1+\vl_3)\delta^{(D)}(\vl_2+\vl_4) C_{\vl_1}^{TT}C_{\vl_2}^{TT} \\
        + &\delta^{(D)}(\vl_1+\vl_4)\delta^{(D)}(\vl_2+\vl_3) C_{\vl_1}^{TT}C_{\vl_2}^{TT} \Big].\label{eq:T04}
\end{align}
These terms are what will lead to the $N^{(0)}$ bias once pluggged into \Eq{kk} as we shall see shortly.
Since $\langle\kappa\rangle = 0$ and $\langle \To \Tz\Tz\Tz\rangle$ will generically be of the form $\int \langle\kappa\rangle \langle \Tz\Tz\Tz\Tz\rangle$ we have
\begin{equation}
   \langle \To_\vlo \Tz_\vlt \Tz_\vlth \Tz_\vlf \rangle = 0 .
\end{equation}
At $O(\kappa^2)$ the Wick contraction machinery alone becomes unwieldy. There is a much more conceptually transparent method to understand higher order terms and consequently $N^{(i)}$ biases through the use of Feynman diagrams. We will connect our Wick contraction machinery stated so far to this diagrammatic language. This diagrammatic approach was derived in \Refs{Jenkins:2014hza, Jenkins:2014oja} and used to understand the CMB lensing bispectrum in \Reff{Kalaja:2022xhi}. To start, we will restate the Feynman rules derived in \Refs{Jenkins:2014hza, Jenkins:2014oja} here:
\begin{itemize}
    \item Each lensed CMB temperature field $T_\vl$ corresponds to a vertex with a momentum $\vl$ flowing into that vertex. 
    \item Two vertices connected by a solid or dashed with momentum $\vl$ come with a factor of $C_\vl^{TT}$ or $C_\vl^{\kappa\kappa}$ respectively. These are our propagators. 
    \begin{align}
        {\hbox{\includegraphics{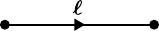}}} &= (2\pi)^2C_\vl^{TT}\\
        {\hbox{\includegraphics{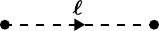}}} &= (2\pi)^2C_\vl^{\kappa\kappa}
    \end{align}  
    \item Uncontracted external legs come with a factor of their corresponding field
     \begin{align}
        {\hbox{\includegraphics{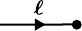}}} &= T^0_\vl\\
        {\hbox{\includegraphics{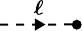}}} &= \kappa_\vl
    \end{align}     
    \item Coming out of each vertex $T_\vl$ we can draw one solid line corresponding to the unlensed CMB temperature field $T_{\vl'}$ and an arbitrary number $n$ of dashed lines with momentum $\vk_i$ where $i=1,\dots,n$. These dashed lines correspond to the order $\kappa^n$ we are expanding to. For such a vertex we must assert momentum conservation $\sum_k \vk_i + \vl' = \vl$. Each vertex also comes with a factor $\prod_{i=1}^n -{2\vk_i\cdot \vl'}/{\vk_i^2}$. Such a factor can be seen for the $n=1$ case in \Eq{T1} and $n=2$ in \Eq{T2}.
    \begin{equation}
        \vcenter{\hbox{\includegraphics{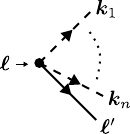}}} = \delta^{(D)}\left(\sum_i \vk_i + \vl'-\vl\right)\prod_{i=1}^n - \frac{2\vk_i\cdot\vl'}{\vk_i^2}
    \end{equation}

    \item Unconstrained momenta $\vk$ are integrated over with $\int d^2\vk / ( 2\pi)^2$
\end{itemize}
We can now understand the terms in \Eq{TTTT} in a more transparent way:
\begin{widetext}
To start we can consider the contraction of purely Gaussian random field \Eq{T04}:
\begin{align}
     \langle T^{(0)}_{\vl_1}T^{(0)}_{\vl_2}T^{(0)}_{\vl_3}T^{(0)}_{\vl_4}\rangle &= 
        \vcenter{\hbox{\includegraphics{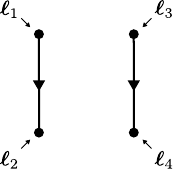}}}  + \vcenter{\hbox{\includegraphics{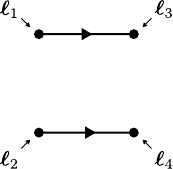}}}  + 
        \vcenter{\hbox{\includegraphics{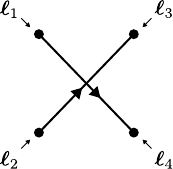}}}.\label{eq:TTfd}
\end{align}
Which recovers exactly what's expected. Similarly the $\langle \To \Tz\Tz\Tz\rangle$ term which we know evaluates to $0$ after taking average $\langle\kappa\rangle$:
\begin{align}
    \langle T^{(1)}_{\vl_1}T^{(0)}_{\vl_2}T^{(0)}_{\vl_3}T^{(0)}_{\vl_4}\rangle &=         
    \vcenter{\hbox{\includegraphics{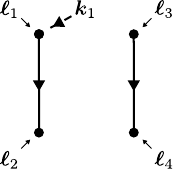}}}  +
    \vcenter{\hbox{\includegraphics{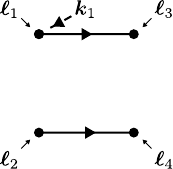}}}  + 
    \vcenter{\hbox{\includegraphics{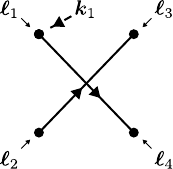}}} .
\end{align}
{Here we} get to the real meat of the usefulness of this diagrammatic approach. First the $\langle \Tt \Tz\Tz\Tz\rangle$ terms which include a single loop:
\begin{align}
    \langle T^{(2)}_{\vl_1}T^{(0)}_{\vl_2}T^{(0)}_{\vl_3}T^{(0)}_{\vl_4}\rangle &=         
    \vcenter{\hbox{\includegraphics{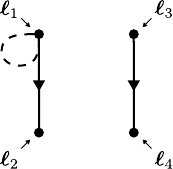}}}  +
    \vcenter{\hbox{\includegraphics{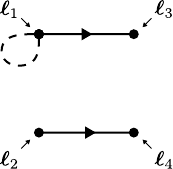}}}  + 
    \vcenter{\hbox{\includegraphics{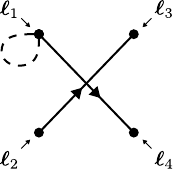}}} .\label{eq:T203}
\end{align}
Through working out the Wick contraction or from looking at these diagrams one can see that these terms go like $\langle \Tt \Tz\Tz\Tz\rangle\sim \langle \Tz\Tz\Tz\Tz\rangle \times \int C^{\kappa\kappa}$. 
As stated in \Sec{Ni} the $N^{(1)}$ bias arises from integrals of the lensing power spectrum. 
So naively, one might think this term and its permutations would contribute to $N^{(1)}$.
However, it turns out that if one replaces the unlensed CMB temperature power spectrum $C_\vl^{TT}$ with the lensed spectrum $\tilde C_\vl^{TT}$ in \Eqs{TT}{TTfd}, terms like $\langle \Tt\Tz\Tz\Tz\rangle$ are automatically taken care of and do not contribute to $N^{(1)}$. 
Roughly this is because the lensed spectra $\tilde C_\vl^{TT}$, which we will denote with a double line, expanded to $O(\kappa^2)$ looks like
\begin{align}
\nonumber    \tilde C_\vl^{TT} = 
    {\hbox{\includegraphics{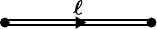}}} &= 
    {\hbox{\includegraphics{figures/lensing_frule3.pdf}}} +
    {\hbox{\includegraphics{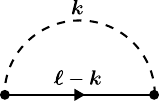}}}  \\
    &+
    {\hbox{\includegraphics{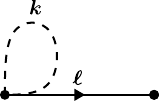}}}  +
    {\hbox{\includegraphics{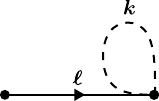}}}  .\label{eq:Cltfd}
\end{align}
{Terms} like \Eq{T203} are accounted for by contributions coming from the second line of \Eq{Cltfd} if we reorganize our expansion with $\tilde C_\vl^{TT}$. We again refer the reader to \Refs{Jenkins:2014hza, Jenkins:2014oja} for more details on this point. Given this point, we will neglect these $\langle \Tt\Tz\Tz\Tz\rangle$ terms going forward.

The diagrams contributing to the final terms at $O(\kappa^2)$ in \Eq{TTTT} are
\footnote{
Here, we neglect some diagrams here which contain a loop like the $\vcenter{\hbox{\includegraphics{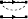}}}$ diagram contributing to $\langle \To_{\vl_1} \Tz_{\vl_2} \To_{\vl_3} \Tz_{\vl_4}\rangle$. We neglect these diagrams for the same reason we neglect \Eq{T203}.
The neglected diagrams are handled by the second term in the first line of \Eq{Cltfd} when replacing the unlensed with the lensed temperature power spectrum.\label{footnote:negdiag}}
\begin{align}
    \begin{array}{r}
    \langle T^{(1)}_{\vl_1}T^{(1)}_{\vl_2}T^{(0)}_{\vl_3}T^{(0)}_{\vl_4}\rangle \\[6pt] +\ \langle T^{(0)}_{\vl_1}T^{(0)}_{\vl_2}T^{(1)}_{\vl_3}T^{(1)}_{\vl_4}\rangle \end{array}&=
    \vcenter{\hbox{\includegraphics{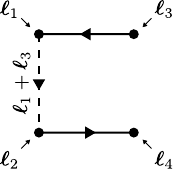}}}  +
    \vcenter{\hbox{\includegraphics{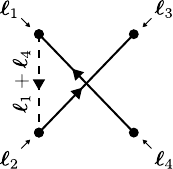}}}  +
    \vcenter{\hbox{\includegraphics{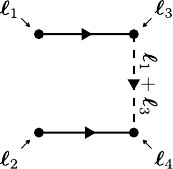}}}  +
    \vcenter{\hbox{\includegraphics{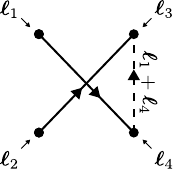}}} \label{eq:k2e1} \\\nonumber \\\nonumber \\
     \begin{array}{r}
    \langle T^{(1)}_{\vl_1}T^{(0)}_{\vl_2}T^{(1)}_{\vl_3}T^{(0)}_{\vl_4}\rangle \\[6pt] +\ \langle T^{(0)}_{\vl_1}T^{(1)}_{\vl_2}T^{(0)}_{\vl_3}T^{(1)}_{\vl_4}\rangle \end{array}&=
    \vcenter{\hbox{\includegraphics{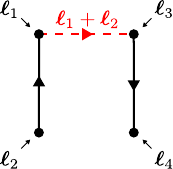}}}  +
    \vcenter{\hbox{\includegraphics{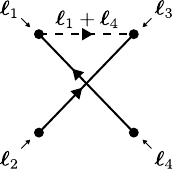}}}  +
    \vcenter{\hbox{\includegraphics{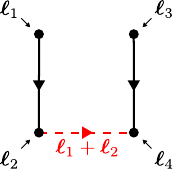}}}  +
    \vcenter{\hbox{\includegraphics{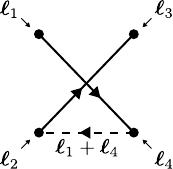}}} \label{eq:k2e2} \\\nonumber \\\nonumber \\   
     \begin{array}{r}
    \langle T^{(0)}_{\vl_1}T^{(1)}_{\vl_2}T^{(1)}_{\vl_3}T^{(0)}_{\vl_4}\rangle \\[6pt] +\ \langle T^{(1)}_{\vl_1}T^{(0)}_{\vl_2}T^{(0)}_{\vl_3}T^{(1)}_{\vl_4}\rangle \end{array}&=
    \vcenter{\hbox{\includegraphics{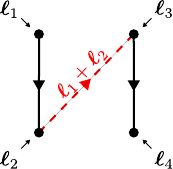}}}  +
    \vcenter{\hbox{\includegraphics{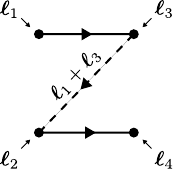}}}  +
    \vcenter{\hbox{\includegraphics{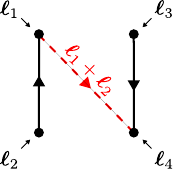}}}  +
    \vcenter{\hbox{\includegraphics{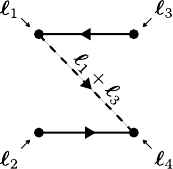}}} \label{eq:k2e3}.
\end{align}
In {\color{red}red} we have called out the diagrams that will contribute to $C_\vL^{\kappa\kappa}$ once we plug \Eq{TTTT} in \Eq{kk}. Adding all these diagrams together allows us to make a very nifty simplification. We follow \Refs{Jenkins:2014hza, Jenkins:2014oja} in defining a composite vertex which restate \Eq{fK} diagrammatically:
\begin{equation}
    f^\kappa_{\vl_1,\vl_2} = \vcenter{\hbox{\includegraphics{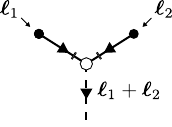}}}   =  \vcenter{\hbox{\includegraphics{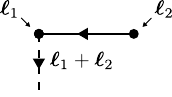}}}  +  \vcenter{\hbox{\includegraphics{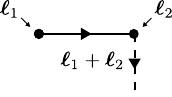}}}.
\end{equation}
The slashes on the line denote the fact that these lines do not produce a temperature power spectrum when translated into equations. 
Now note that all of the diagrams in Eqs.~\ref{eq:k2e1}, \ref{eq:k2e2}, and \ref{eq:k2e3} can better organized in three diagrams which utilize this composite vertex:
\begin{align}
    \vcenter{\hbox{\includegraphics{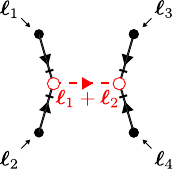}}}  &= \underbrace{\vcenter{\hbox{\includegraphics{figures/1010_d1.pdf}}}}_{\subset \langle\To\Tz\To\Tz\rangle}  
    + \underbrace{\vcenter{\hbox{\includegraphics{figures/0101_d1.pdf}}}  }_{\subset \langle\Tz\To\Tz\To\rangle}
    + \underbrace{\vcenter{\hbox{\includegraphics{figures/0110_d1.pdf}}}  }_{\subset \langle\Tz\To\To\Tz\rangle}
    + \underbrace{\vcenter{\hbox{\includegraphics{figures/1001_d1.pdf}}}  }_{\subset \langle\To\Tz\Tz\To\rangle}\\\nonumber\\\nonumber\\
     \vcenter{\hbox{\includegraphics{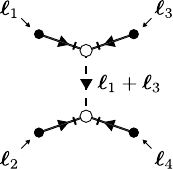}}}  &= \underbrace{\vcenter{\hbox{\includegraphics{figures/1100_d1.pdf}}}}_{\subset \langle\To\To\Tz\Tz\rangle}  
    + \underbrace{\vcenter{\hbox{\includegraphics{figures/0011_d1.pdf}}}  }_{\subset \langle\Tz\Tz\To\To\rangle}
    + \underbrace{\vcenter{\hbox{\includegraphics{figures/0110_d2.pdf}}}  }_{\subset \langle\Tz\To\To\Tz\rangle}
    + \underbrace{\vcenter{\hbox{\includegraphics{figures/1001_d2.pdf}}}  }_{\subset \langle\To\Tz\Tz\To\rangle}\\\nonumber\\\nonumber\\
      \vcenter{\hbox{\includegraphics{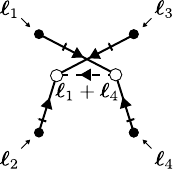}}}  &= \underbrace{\vcenter{\hbox{\includegraphics{figures/1100_d2.pdf}}}}_{\subset \langle\To\To\Tz\Tz\rangle}  
    + \underbrace{\vcenter{\hbox{\includegraphics{figures/0011_d2.pdf}}}  }_{\subset \langle\Tz\Tz\To\To\rangle}
    + \underbrace{\vcenter{\hbox{\includegraphics{figures/1010_d2.pdf}}}  }_{\subset \langle\To\Tz\To\Tz\rangle}
    + \underbrace{\vcenter{\hbox{\includegraphics{figures/0101_d2.pdf}}}  }_{\subset \langle\Tz\To\Tz\To\rangle}
\end{align}
{The} sum of Eqs.~\ref{eq:k2e1}, \ref{eq:k2e2}, and \ref{eq:k2e3} can be written as
\begin{align}
\begin{array}{r}
{\langle \To_\vlo \To_\vlt \Tz_\vlth \Tz_\vlf \rangle}+{\langle \Tz_\vlo \Tz_\vlt \To_\vlth \To_\vlf \rangle}\\[6pt]+
{\langle \To_\vlo \Tz_\vlt \To_\vlth \Tz_\vlf \rangle}+{\langle \Tz_\vlo \To_\vlt \Tz_\vlth \To_\vlf \rangle}\\[6pt]+
{\langle \Tz_\vlo \To_\vlt \To_\vlth \Tz_\vlf \rangle+\langle \To_\vlo \Tz_\vlt \Tz_\vlth \To_\vlf \rangle}
\end{array}&=
    \vcenter{\hbox{\includegraphics{figures/c1.pdf}}}  +  \vcenter{\hbox{\includegraphics{figures/c2.pdf}}} +  \vcenter{\hbox{\includegraphics{figures/c3.pdf}}} \\\nonumber \\
    &= (2\pi)^2\delta^{(D)}(\vl_1+\vl_2+\vl_3+\vl_4)\left( 
    \begin{array}{r}
    {\color{red}f^\kappa_{\vl_1,\vl_2}C_{\vl_1+\vl_2}^{\kappa\kappa} f^{\kappa}_{\vl_3,\vl_4}}\\
    + f^\kappa_{\vl_1,\vl_3} C^{\kappa\kappa}_{\vl_1+\vl_3} f^{\kappa}_{\vl_2,\vl_4}\\
    + f^\kappa_{\vl_1,\vl_4} C^{\kappa\kappa}_{\vl_1+\vl_4} f^{\kappa}_{\vl_2,\vl_3}\end{array}\right)\label{eq:ok2}
\end{align}

With all of this in hand, we can now compute the $N^{(0)}$ and $N^{(1)}$ bias. As stated earlier the $N^{(i)}$ come out when we plug \Eq{TTTT} into \Eq{kk}. At $O(\kappa^0)$ when we plug \Eq{T04} (equivalently \Eq{TTfd}) we will get the $N^{(0)}$ bias:
\begin{align}
\nonumber N^{(0)}_\vL &=(N^\kappa_\vL)^2 \int \frac{d^2\vl_1}{(2\pi)^2}\int\frac{d^2\vl_2}{(2\pi)^2} F^\kappa_{\vl_1,\vL-\vl_1}F^\kappa_{-\vl_2,-\vL+\vl_2}\times \langle \Tz_{\vl_1}\Tz_{\vL-\vl_1} \Tz_{-\vl_2}\Tz_{-\vL+\vl_2}\rangle \\
\nonumber&= (2\pi)^4(N^\kappa_\vL)^2 \int \frac{d^2\vl_1}{(2\pi)^2}\int \frac{d^2\vl_2}{(2\pi)^2}F^{\kappa}_{\vl_1,\vL-\vl_1}F^\kappa_{-\vl_2,-\vL+\vl_2}\left[\underbrace{\delta^{(D)}(\vL)^2C_{\vl_1}C_{\-\vl_2}}_{=0\textrm{ since $\vL\ne 0$}} + \underbrace{2\delta^{(D)}(\vl_1-\vl_2)\delta^{(D)}(\vl_2-\vl_1)C_{\vl_1}^{TT}C_{\vL-\vl_1}^{TT}}_{\vl_1=\vl_2\ \textrm{(parallelograms)}}\right]\\
&= (2\pi)^2 \delta^{(D)}(0)\times 2(N^\kappa_\vL)^2 \int \frac{d^2\vl_1}{(2\pi)^2} (F^\kappa_{\vl_1,\vL-\vl_1})^2 C_{\vl_1}^{TT} C_{\vL-\vl_1}^{TT}
\label{eq:Nhat_explicit}
\end{align}
where in the second line we used the $\vl_2\rightarrow \vL-\vl_2$ symmetry of the integral. A key point of this calculation is that of all the $\{\vl_1, \vl_2\}$ used to compute $\langle\hat\kappa\hat\kappa\rangle$, only the $\vl_1=\vl_2$ contribute to the $N^{(0)}$ noise bias. This explicitly shows the motivation for our $N^{(0)}$ noise bias avoidance principle outlined in \Fig{summary} and \Sec{Nhat}. 

Since we have only vanishing terms at $O(\kappa^1)$ we go straight to $O(\kappa^2)$ which gives us both $C_\vL^{\kappa\kappa}$ and the $N^{(1)}$ bias. Plugging in \Eq{ok2} into \Eq{kk} yields
\begin{align}
\nonumber &{(2\pi)^2 \delta^{(D)}(0)}\times(N^\kappa_\vL)^2 
\Big({\color{red}C_\vL^{\kappa\kappa}}\int \frac{d^2\vl_1}{(2\pi)^2}F^\kappa_{\vl_1,\vL-\vl_1} {\color{red}f^\kappa_{\vl_1,\vL-\vl_1}}\times \int\frac{d^2\vl_2}{(2\pi)^2} F^\kappa_{-\vl_2,-\vL+\vl_2}{\color{red}f^\kappa_{-\vl_2, -\vL+\vl_2} }\\
+ &\int \frac{d^2\vl_1}{(2\pi)^2}\int\frac{d^2\vl_2}{(2\pi)^2} F^\kappa_{\vl_1,\vL-\vl_1}F^\kappa_{-\vl_2,-\vL+\vl_2}\times \left[f^\kappa_{\vl_1,-\vl_2}C_{\vl_1-\vl_2}^{\kappa\kappa}f^\kappa_{\vL-\vl_1, -\vL+\vl_2}+  f^\kappa_{\vl_1,\vl_2-\vL} C_{\vl_1+\vl_2-\vL}^{\kappa\kappa}f^\kappa_{\vL-\vl_1, -\vl_2}\right] \Big)\label{eq:Ok2}.
\end{align}
From here we see two things. The first line when combined with \Eq{NkappaL} gives us $(2\pi)^2\delta^{(D)}(0) C_\vL^{\kappa\kappa}$, the CMB lensing power spectrum, as promised. The second line gives us the $N^{(1)}$ bias:
\begin{align}
\nonumber   N^{(1)}_\vL =  {(2\pi)^2 \delta^{(D)}(0)}\times(N^\kappa_\vL)^2&\int \frac{d^2\vl_1}{(2\pi)^2}\int\frac{d^2\vl_2}{(2\pi)^2} F^\kappa_{\vl_1,\vL-\vl_1}F^\kappa_{-\vl_2,-\vL+\vl_2}\\
\nonumber   &\times \left[f^\kappa_{\vl_1,-\vl_2}C_{\vl_1-\vl_2}^{\kappa\kappa}f^\kappa_{\vL-\vl_1, -\vL+\vl_2}+  f^\kappa_{\vl_1,\vl_2-\vL} C_{\vl_1+\vl_2-\vL}^{\kappa\kappa}f^\kappa_{\vL-\vl_1, -\vl_2}\right] \\
\nonumber=  (2\pi)^2 \delta^{(D)}(0) \times (N^\kappa_\vL)^2 &\int \frac{d^2\vl_1}{(2\pi)^2}\int \frac{d^2\vl_2}{(2\pi)^2} F^\kappa_{\vl_1, \vL-\vl_1} F^\kappa_{-\vl_2, -\vL+\vl_2}\\&\times 2 [f^\kappa_{\vl_1,-\vl_2} C^{\kappa\kappa}_{\vl_1,-\vl_2} f^\kappa_{\vL-\vl_1, -\vL+\vl_2} ],
   \label{eq:n1}
\end{align}
where in the final line we used the $\vl_2 \rightarrow \vL-\vl_2$ symmetry of the integral to combine the two terms in the brackets. This result is known in the literature and first computed by \Reff{Kesden:2003cc} and also reproduced via the use of Feynman diagrams in \Refs{Jenkins:2014hza, Jenkins:2014oja}. From this equation, we can explicitly see the fact that $N^{(1)}\sim \iint \langle\kappa\kappa\rangle$  as stated in \Sec{Ni}. Higher order biases such as $N^{(2)}$ are also computed in \Refs{Jenkins:2014hza, Jenkins:2014oja} and we refer the interested reader to these sources.

\end{widetext}
\subsection{Why $\hat N$ subtraction has no effect on $N^{(1)}$}
\label{app:Nigeq1}
Our $\hat N$'s effect on $N_\vL^{(1)}$, $\Delta N^{(1)}_{\vL}$, can be computed by inserting a $\delta^{(D)}(\vl_1 - \vl_2)$ into the integrand:
\begin{align}
    \nonumber \Delta N^{(1)}_\vL &= (2\pi)^2 \delta^{(D)}(0)\times 2 (N^\kappa_\vL)^2 \\
\nonumber    &\times\int \frac{d^2 \vl_1}{(2\pi)^2} (F^\kappa_{\vl_1,\vL-\vl_1})^2  \left[f^\kappa_{\vl_1,-\vl_1} C_{0}^{\kappa\kappa} f^\kappa_{\vL-\vl_1,-\vL+\vl_1} \right]\\
    &= 0,
\end{align}
where in the second line we used \Eq{fK} to see that $f^{\kappa}_{\vl,-\vl}=0$. 
This means that $\hat N$ does not affect the $N^{(1)}$ biases. 
In general when our proposed noise bias avoidance method removes a small subset of terms in any integration that results in a $N^{(i\geq 1)}$ bias. 
So, we believe that our proposed bias avoidance method will also have a negligible effect on $N^{(i> 1)}$ biases similar to its effect on the $N^{(1)}$. 
Thus, standard methods of avoiding or estimating higher order noise bias such as the use of $T\nabla T$ weights 
and Monte-Carlo computations of $N^{(1)}$
can be naturally combined with our proposed method.

\section{Details on Map Simulations}
\label{app:sims}
In this section we describe how we generate our simulated maps used for our numerical studies in this paper. 
{All of our maps span $20^\circ \times 20^\circ$ on the sky and have $1200\times 1200$ pixels.}
We start by computing the unlensed, lensed, and lensed gradient CMB Temperature power spectrum $C_\vl^{TT}$, $C_\vl^{TT,L}$, and $C_\vl^{T\nabla T}$ as well as the lensing potential spectrum $C_\vl^{\kappa\kappa}$ with \texttt{CAMB} (\Refs{Lewis:1999bs, 2011JCAP...03..018L, 2011ascl.soft02026L, 2012JCAP...04..027H}). 
Then using \href{https://github.com/DelonShen/LensQuEst}{\texttt{LensQuEst}} (\Reff{Schaan:2018tup}) we compute the expected foreground $C_\vl^{F}$ and noise spectrum $C_\vl^{N}$ for a SO-like survey\footnote{Gaussian beam with full-width-at-half-maximum of $1.4$ arcminutes and white noise levels of $7\ \mu$K-arcmin}.

With these spectra in hand, our recipe to generate a lensed temperature map is:
\begin{enumerate}
    \item Generate a Gaussian random field with the unlensed temperature field power spectra $C_\vl^{TT}$:
    \begin{equation}
        T_\vl^0 \sim \mathcal N(0, C_\vl^{TT})
    \end{equation}
    \item Generate a Gaussian random field with the lensing potential power spectra $C_\vl^{\kappa\kappa}$
    \begin{equation}
        \kappa_\vl \sim \mathcal N(0, C_\vl^{\kappa\kappa})
    \end{equation}
    \item Generate the lensed temperature map $T_\vl$ by first computing the deflection field $\vd(\vx)$ due to the lensing potential $\kappa_\vl$ with
    \begin{equation}
        \vd_\vl = -\frac{2i \vl}{\ell^2} \kappa_\vl.
    \end{equation}
    Then generate the lensed temperature $T_\vl^L$ by computing $T^L(\vx)$ as in \Eq{Tx}:
    \begin{equation}
        T^L(\vx) = T^0(\vx+\vd(\vx))   \tag{\ref{eq:Tx}}.
    \end{equation}
    \item Generate two Gaussian random fields corresponding to foregrounds and detector noise with the corresponding spectra $C_\vl^F$ and $C_\vl^N$. 
    \begin{align}
        F_\vl &\sim \mathcal N(0, C_\vl^{F})\\
        N_\vl &\sim \mathcal N(0, C_\vl^N)\label{eq:noise_map}
    \end{align}
    \item Generate the total lensed CMB Temperature map $T_\vl$ by adding all these maps together:
    \begin{equation}
        T_\vl = T^L_\vl + F_\vl + N_\vl\label{eq:total_map}
    \end{equation}
\end{enumerate}
This is the recipe we used to generate 500 lensed CMB temperature maps used in the numerical studies for Figs.~\ref{fig:Nhat_unmasked}, \ref{fig:corr_QEQEvGRF}, and \ref{fig:corr_mNhatvmRDN0}. Note that the above recipe implies that the total CMB temperature power spectrum is
\begin{equation}
    \tilde C_\vl^{TT} = C_\vl^{TT,L} + C_\vl^F + C_\vl^N
    \label{eq:Cltot}
\end{equation}
\begin{figure}
    \includegraphics[width=\linewidth]{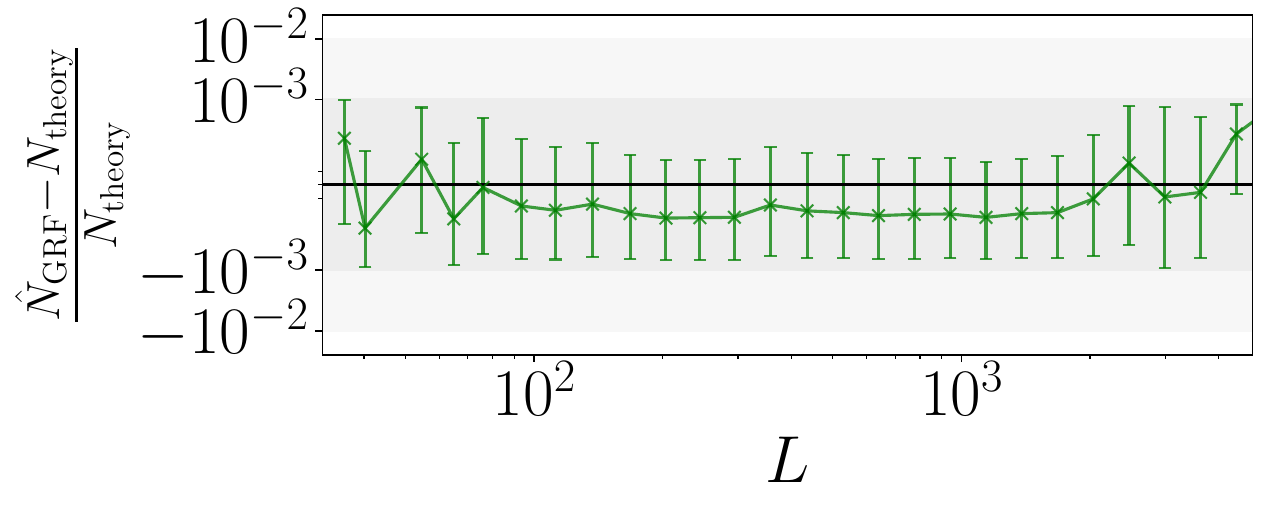}
    \caption{In \Eq{Clnobias} we claimed that our estimator $\hat N$ when run on a Gaussian random field $\hat N^{\rm GRF}$ would equal $\langle\hat\kappa\hat\kappa^*\rangle_{\rm GRF} \sim N_{\rm theory}$ (see \Eq{kLkLGRF}). In this plot we numerically check this claim. We also point out that $\hat N$ is highly correlated between different $L, L'$. This is explicitly shown in \Fig{nhat_grf}. This off-diagonal correlation makes $\chi^2$ by eye hard to do here. We explain this correlation structure in \App{off_diagonal}.}
    \label{fig:NhatGRF}
\end{figure}

In \Eq{Clnobias} we claimed that $\hat N_{\rm GRF} = \langle\hat\kappa\hat\kappa^*\rangle_{\rm GRF}\sim N_{\rm theory}$ (see \Eq{kLkLGRF}). With \Eq{Cltot} we are able to confirm this numerically by generating 500 Gaussian random fields $T_\vl^{\rm GRF}\sim \mathcal N(0, \tilde C_\vl^{TT})$ and computing $\hat N$ on these fields. We plot the results in \Fig{NhatGRF} and find that indeed \Eq{Clnobias} holds. Something to note is that, as we will argue in \App{off_diagonal}, $\hat N$ is highly correlated between different $L,L'$ which makes doing $\chi^2$ by eye tests deceptive for this plot.

In \Sec{aniso-noise} we study the effect of anisotropic instrument noise on our proposed CMB lensing spectrum estimator.
To generate CMB maps with realistic anisotropic detector noise we use cutouts from the ACT DR5 IVar map as shown in \Fig{ACTDR5}. Let the anisotropic noise pattern we extract from the ACT DR5 maps be called $A(\vx)$. Using this we modify our noise map computed in \Eq{noise_map} as:
\begin{equation}
    N^{\rm aniso}(\vx) = N(\vx)\times \frac{A(\vx)}{{\rm min}[A(\vx)]}.
\end{equation}
In this way $N_{\rm aniso}(\vx) \geq N(\vx)$. From here \Eq{total_map} is then modified as 
\begin{equation}
    T_\vl^{\rm aniso} = T^L_\vl + F_\vl + N_\vl^{\rm aniso}\label{eq:total_map_aniso}.
\end{equation}
Note that this means \Eq{Cltot} is no longer true. Instead we estimate $\tilde C_\vl^{TT,{\rm aniso}}$ by generating 500 maps using \Eq{total_map_aniso} and averaging the power spectra of these maps.

In \Sec{masking} we study the effect of masking on our proposed CMB lensing spectrum estimator. To generate CMB maps with masking we use the mask shown in \Fig{mask}. Let this mask be called $M(\vx)$. After \Eq{total_map} we apply this mask:
\begin{equation}
   T^{\rm masked}(\vx) =   M(\vx) \times T(\vx).
\end{equation}
We perform our numerical studies in \Sec{masking} using these maps.


\section{Neglected additional contractions in $\hat{N}$ are negligible}
\label{app:lL2}
In footnote \ref{footnote:l2} we commented that \Eq{Clnobias} is not exact since we do not include a $\vl = \vL/ 2$ term but that this term is suppressed like $1 / N_{\rm modes}$. In this appendix we shall show this suppression thus concluding that the neglected additional contraction is negligible.

Recall from \Eq{kk}
\begin{align}
    \nonumber\langle \hat\kappa_\vL \hat\kappa_{\vL}^*\rangle = (N^\kappa_\vL)^2 \int &\frac{d^2\vl_1}{(2\pi)^2}\int\frac{d^2 \vl_2}{(2\pi)^2} F^\kappa_{\vl_1,\vL-\vl_1}F^\kappa_{-\vl_2,-\vL+\vl_2}\\
    &\times \langle T_{\vl_1} T_{\vL-\vl_1} T_{-\vl_2}T_{-\vL+\vl_2}\rangle\tag{\ref{eq:kk}}.
\end{align}

For a Gaussian random field with no lensing but with a power spectrum equal to the total CMB power spectrum $T_\vl^G \sim \mathcal N(0, \tilde C_\vl^{TT})$ we can evaluate the four-point function in the integral above:
\begin{align}
\nonumber\langle T^G_{\vl _ 1} T^G_{\vL - \vl_1} T^G_{-\vl_2} T^G_{-\vL + \vl_2}\rangle 
\nonumber&= \underbrace{\langle T^G_{\vl_1} T^G_{\vL - \vl_1}\rangle\langle T^G_{-\vl_2} T^G_{-\vL + \vl_2}\rangle}
_{=0\textrm{ since we assume $L\ne 0$}}\\
 \nonumber&+ \langle T^G_{\vl_1} T^G_{-\vl_2} \rangle \langle T^G_{\vL - \vl_1} T^G_{-\vL + \vl_2} \rangle \\ 
 \nonumber&+ \langle T^G_{\vl_1}T^G_{-\vL+\vl_2} \rangle\langle T^G_{\vL-\vl_1} T^G_{-\vl_2}\rangle\\
 &+ \underbrace{\langle T^G_{\vl _ 1} T^G_{\vL - \vl_1} T^G_{-\vl_2} T^G_{-\vL + \vl_2}\rangle_c}_{=0\textrm{ for a GRF}}
.
\end{align}
We know from \Eq{T0T0} that this then evaluates to 
\begin{align}
 \nonumber\langle T^G_{\vl _ 1}& T^G_{\vL - \vl_1}  T^G_{-\vl_2} T^G_{-\vL + \vl_2}\rangle 
= (2\pi)^4 \tilde C_{\vl_1}^{TT}\tilde C_{\vL-\vl_1}^{TT}\\
 \times\Big[&(\delta^{(D)}(\vl_1 - \vl_2) )^2+(\delta^{(D)}(\vl_1 -\vL+\vl_2) )^2\Big] 
\end{align}
Plugging this into \Eq{kk} to get
\begin{align}
\nonumber &\langle \hat\kappa_\vL \hat\kappa_{\vL}^*\rangle_{\rm GRF} = \\
\nonumber &=(N^\kappa_\vL)^2 \int\frac{d^2\vl_1}{(2\pi)^2}\int\frac{d^2 \vl_2}{(2\pi)^2} F^\kappa_{\vl_1,\vL-\vl_1}F^\kappa_{-\vl_2,-\vL+\vl_2}\\
\nonumber&\times (2\pi)^4  \tilde C_{\vl_1}^{TT}\tilde C_{\vL-\vl_1}^{TT}\times 2 (\delta^{(D)}(\vl_1 - \vl_2))^2\\
\nonumber &= (2\pi)^2\delta^{(D)}(0)\times (N^\kappa_\vL)^2 \int\frac{d^2 \vl_1}{(2\pi)^2} F^\kappa_{\vl_1, \vL-\vl_1} f^\kappa_{\vl_1, \vL-\vl_1}\\
\nonumber&= (2\pi)^2\delta^{(D)}(0)\times N^\kappa_\vL\quad\textrm{(Using \Eq{NkappaL})}\\
&\equiv (2\pi)^2 \delta^{(D)}(0)\times  N_{\rm theory}\label{eq:kLkLGRF}
\end{align}
Where in the second line we have a factor of two by exploiting the $\vl_2\rightarrow \vL-\vl_2$ symmetry of the integral allowing us to combine both Dirac delta's. Now, recall our definition of $\hat N$:
\begin{align}
 \hat{N}_\vL
=
\frac{2
\left({N^\kappa_{\bm{L}}}\right)^{2}}{{\color{blue}(2\pi)^2\delta^{(D)}(0)}}
\int \frac{d^2 \vl}{(2\pi)^2}
&F^\kappa_{\vl, \vL-\vl}
F^\kappa_{-\vl, -\vL+\vl}
\left| T_\vl \right|^2
\left| T_{\vL - \vl} \right|^2
.
\tag{\ref{eq:Nhat}}   
\end{align}

Where in {\color{blue}blue} we include the additional finite area correction we suppressed in \Eq{Nhat} as pointed out in footnote \ref{footnote:area}.  
Lets also evaluate $\langle \hat N_\vL\rangle$ on the same Gaussian random field $T_\vl^G \sim \mathcal N(0, \tilde C_\vl^{TT})$. To start, we can evaluate the four-point function in the integral:
\begin{align}
\nonumber    \langle T^G_\vl T^G_{-\vl} &T^G_{\vL-\vl} T^G_{-\vL+\vl}\rangle  = (2\pi)^4 \tilde C_\vl^{TT}\tilde C_{\vL-\vl}^{TT}\\
&\times \left[(\delta^{(D)}(0))^2 + (\delta^{(D)}(2\vl - \vL))^2 \right]
\end{align}
Plugging this into \Eq{Nhat} gives us 
\begin{align}
\nonumber    \langle\hat N_\vL \rangle_{\rm GRF} &= \frac{ \left({N^\kappa_\vL} \right)^2}{(2\pi)^2 \delta^{(D)}(0)}  \int \frac{d^2 \vl}{(2\pi)^2}F^\kappa_{\vl,\vL-\vl} f^\kappa_{-\vl,-\vL+\vl}\\
&\times (2\pi)^4  \left[(\delta^{(D)}(0))^2 + (\delta^{(D)}(2\vl - \vL))^2 \right]\\
\nonumber &=  (2\pi)^2\times \frac{\left({N^\kappa_\vL} \right)^2 }{\delta^{(D)}(0)} \Big[ (N^\kappa_\vL)^{-1} (\delta^{(D)}(0))^2 \\
&+ \frac 1 {4(2\pi)^2}\delta^{(D)}(0) F^\kappa_{\vL/2, \vL/2} f^\kappa_{-\vL/2, -\vL/2} \Big]\\
&\equiv (2\pi)^2 \delta^{(D)}(0)\times N_{\rm theory}(\vL) \left(1 + {\color{red}\mathcal C} \right)
\end{align}
So we see that there is an extra $\vL/2$ contraction leads to a {\color{red}extra term $\mathcal C$} which makes \Eq{Clnobias} not exact. However, note that $\mathcal C\sim N_{\rm theory}\sim [\int (C_\vl/\tilde C_\vl)^2]^{-1}$. {If we approximate $C_\vl \approx \tilde C_\vl$ then the integral evaluates to $N_{\rm modes}$ meaning $\mathcal C$ is suppressed by a factor of $1/N_{\rm modes}\sim 10^{-6}$:}
\begin{equation}
    \mathcal C \sim N_{\rm theory} \sim \frac 1 {N_{\rm modes}}\ll 1.
\end{equation}
Thus we neglect this extra contraction coming from $\delta^{(D)}(2\vl - \vL)$. 

\section{Origin of off-diagonal covariances of the lensing power spectrum}
\label{app:off_diagonal}

\begin{figure}
    \centering
    \includegraphics[width=\linewidth]{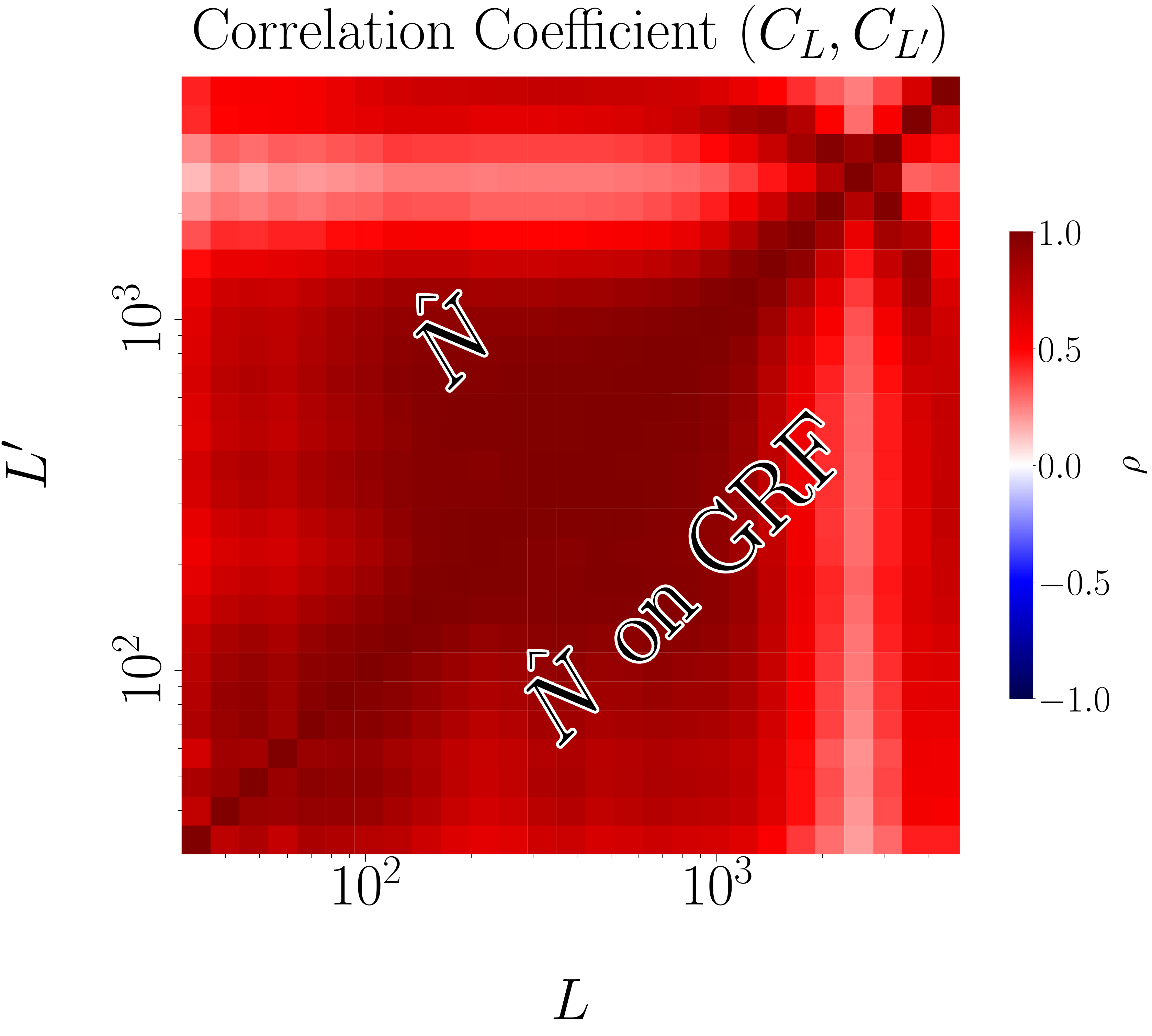}
    \caption{We saw in \Sec{cov} that there is strong correlation between $\hat N_L$ and $\hat N_{L'}$. Here we explicitly show this off-diagonal correlation for $\hat N$ run on lensed maps (upper left) and $\hat N$ run on Gaussian random fields with the same total power spectrum as a lensed map (lower right). We analytically explain the origin of this correlation structure in \App{off_diagonal}}
    \label{fig:nhat_grf}
\end{figure}

In \Sec{cov} we point out the fact that the $N^{(0)}$ bias and correspondingly our $\hat N$ has significant off-diagonal correlation ${\rm cov}[\hat N_L,\hat N_{L'}]$.  
We explicitly plot this strong off-diagonal correlation structure in \Fig{nhat_grf}.
In this Appendix we shall explain the reason for this strong off-diagonal correlation. 

Let's consider the covariance between $\hat N_{\vL}$ and $\hat N_{\vL'}$. 
Recall the language of summing over quadrilaterals to estimate the lensing potential spectrum that we introduced in \Eq{kappa} and \Fig{summary}. 
In this language, as we stated in \Fig{summary}, $\hat N$ can be thought of summing over only parallelograms: 
\begin{equation}
    \hat N_\vL \sim    \vcenter{\hbox{\includegraphics{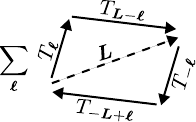}}} .
\end{equation}
In this language, we can write ${\rm cov}[\hat N_\vL, \hat N_{\vL'}]$ as
\begin{widetext}
\begin{align}
\nonumber    {\rm cov}[\hat N_\vL, \hat N_{\vL'}]&\sim    
    \sum_{{\color{red}\vl},{\color{blue}\vl'}}\ \ \vcenter{\hbox{\includegraphics{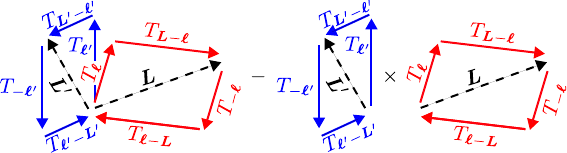}}}\\\nonumber \\
    &= \sum_{{\color{red}\vl},{\color{blue}\vl'}}{\rm cov}\left[T_{\color{red}\vl} T_{\vL-\color{red}\vl} T_{-\color{red}\vl}T_{{\color{red}\vl}-\vL} , T_{\color{blue} \vl'}T_{\vL'-\color{blue}\vl'} T_{-\color{blue}\vl'}T_{{\color{blue}\vl'} - \vL'} \right],\label{eq:covNhat}
\end{align}
\end{widetext}
where we implicitly take expectation values over each disjoint diagram.
Since we argued in \Fig{corr_QEQEvGRF} the dominant contribution to the covariance is $N^{(0)}$ we will approximate $T_\vl$ to be a Gaussian random field since this approximation still captures the $N^{(0)}$'s effect on the covariance structure. 
The only time the term we are summing over in \Eq{covNhat} is non-zero\footnote{Once again neglecting additional contributions that are similar to $\vl=\vL/2$ in the eight-point function since we believe they are negligible just like they were in the four-point function as we argued for in \App{lL2}.} 
is when both parallelogram share a leg: ${\color{red}\vl}={\color{blue}\vl'}$ (or symmetrically ${\color{red}\vl}=\vL'-{\color{blue}\vl'}$). 
All other $\{{\color{red}\vl}, {\color{blue}\vl'}\}$ force the parallelograms to be statistically independent and thus do not contribute to the covariance. So we have
\begin{widetext}
\begin{align}
\nonumber    {\rm cov}[\hat N_\vL, \hat N_{\vL'}]&\sim    
    \sum_{{\color{magenta}\vl}}\ \ \vcenter{\hbox{\includegraphics{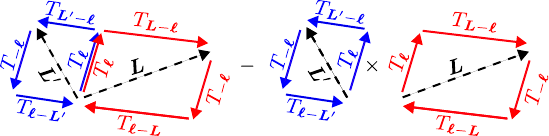}}}\\\nonumber \\
    &= \sum_{\color{magenta}\vl}{\rm var}\left[T_{\color{magenta}\vl} T_{\vL-\color{magenta}\vl} T_{-\color{magenta}\vl}T_{{\color{magenta}\vl}-\vL} , T_{\color{magenta} \vl}T_{\vL'-\color{magenta}\vl} T_{-\color{magenta}\vl}T_{{\color{magenta}\vl} - \vL'} \right],
\end{align}
\end{widetext}

Since for a fixed $\vL$, there is a unique parallelogram associated with having a leg of length $\vl$, we see that \textbf{each diagram contributing to $\hat N_\vL$ is correlated with exactly one diagram contributing to $\hat N_{\vL'}$ and vice versa}\footnote{This is true if we had access to all modes. Some modes we cannot see due to finite resolution/size measurements but for the bulk of the modes this is a negligible effect.}.  
This is the origin of the strong off-diagonal covariance due to $N^{(0)}$ and consequently removed by our $\hat N$ as well as ${\rm RDN}^{(0)}$.

\section{Deriving the optimal Kurtosis estimator}
\label{app:Kopt_deriv}
In \Sec{toy} we were considering how to estimate the small kurtosis of a weakly non-Gaussian random variable. This was used as a toy model to understand optimal trispectrum estimation. During this discussion we recalled a result from \Reff{Smith:2015uia} that stated the optimal unbiased and minimum variance estimator for the kurtosis in this toy model is:
\begin{equation*}
    {\hat {\mathcal K}_{\rm opt} = \frac 1 N \sum_i x_i^4- \frac {6\sigma^2} N \sum_i x_i^2 + 3 \sigma^4}\tag{\ref{eq:Kopt}}.
\end{equation*}
In this appendix we will spell out how to arrive at this result. Obtaining the form of ${\rm RDN}^{(0)}$ described in \Eq{RDN0} requires a similar argument and we will spell this out in \App{RDN0}. Our derivation in this section is directly inspired by a similar discussion for a skewness estimator in \Reff{1996MNRAS.283..983A}.

The optimal kurtosis estimator $\hat{\mathcal K}_{\rm opt}$ stated in \Eq{Kopt} is derived via Edgeworth expansion. Lets add a small kurtosis $\mathcal K$ to a normal random variable and call this perturbed random variable $X$. Then it's moment generating function is
\begin{align}
    &M_X(J) =  {\rm exp}\left\{ -\frac 1 2 J^2 \sigma^2 + \frac 1 {4!} J^4 \mathcal K\right\} \\
     &\approx {\rm exp}\left\{ -\frac 1 2 J^2 \sigma^2 \right\} \left( 1 + \frac 1 {4!}J^4 \mathcal K + \frac 1 {2\times  4! 4!} J^8 \mathcal K^2\right)
     \label{eq:MXJ}
\end{align}
This reconstructs the random variable we were considering in \Sec{toy}. 
\begin{align*}
  \langle X^2\rangle = \sigma^2\tag{\ref{eq:toyX2}}\\ 
  \langle X^4 \rangle = 3 \sigma^4 + \mathcal K.\tag{\ref{eq:toyX4}}
\end{align*}
 Before we continue, recall the definition of the Hermite polynomials:
\begin{equation}
    h_n(x) = (-1)^n p_{\mathcal N}^{-1}(x) \frac{\partial^n}{\partial x^n} p_{\mathcal N}(x).
\end{equation}
Here, $p_{\mathcal N}(x)$ is the probability density function of Gaussian random variable with zero mean and variance $\sigma^2$
\begin{equation}
    p_{\mathcal N}(x) = \frac{1}{\sqrt{2\pi\sigma^2}}  {\rm exp}\left\{- \frac 1 2 \frac{x^2}{\sigma^2} \right\}.
\end{equation}

Inverse transforming the moment generating function \Eq{MXJ} gives us the PDF
\begin{align}
   \frac{p_X(x)}{p_{\mathcal N} (x)} &= {1 + \frac{\mathcal K}{4!} h_4(x)  + \frac{\mathcal K^2}{2\times 4!4!}h_8(x)},\label{eq:hermite}
\end{align}
where we have
\begin{align}
    h_4(x)&=\frac{x^4-6 \sigma ^2 x^2+3 \sigma ^4}{\sigma ^8}\label{eq:h4}\\
    h_8(x)&=\frac{x^8-28 \sigma ^2 x^6+210 \sigma ^4 x^4-420 \sigma ^6
   x^2+105 \sigma ^8}{\sigma ^{16}}.
\end{align}
If we have $N$ independent realization of this random variable $\{x_i\}$ where each $x_i\sim X$ then the log likelihood $\mathcal L=\sum \log p_X(x_i)$ becomes
\begin{align}
    \nonumber\mathcal L-\mathcal L_{\mathcal N} &=  \sum_i \left[\log\left(1 + \frac {\mathcal K}{4!} h_4(x_i) + \frac{\mathcal K^2}{2\times 4!4!}h_8(x_i) \right)\right]\\
    &\approx \sum_i \left[ \frac{\mathcal K}{4!}h_4(x_i) + \frac{\mathcal K^2}{2\times 4!4!}(h_8(x_i)-h_4^2(x_i))\right].
\end{align}
Asserting $\partial \mathcal L / \partial \mathcal K = 0$ we find the ML estimate for $\mathcal K$ is 
\begin{equation}
    \hat {\mathcal K}_{\rm opt} = - \frac{4!\sum_i h_4(x_i)}{\sum_i [h_8(x_i) - h_4^2(x_i)]}\label{eq:kinterm}
\end{equation}
Now let us approximate 
\begin{equation}
 \sum_{i=1}^N x_i^n = N\sigma^n \times (n-1)!!\label{eq:appro1}
\end{equation}
for even $n$ to derive a relation between $h_4$ and $h_8$. Essentially we are saying we know the variance exactly and so can replace anything where we're estimating the variance with the true variance. Non-Gaussian corrections will be subleading for the purpose of deriving this relation between $h_4$ and $h_8$. With this approximation we get
\begin{align}
    \sum_{i=1}^N h_8(x_i) &=  0\\
    \sum_{i=1}^N h_4^2(x_i)&= \frac {4!\times N } {\sigma^8}\\
    \Rightarrow \sum_i [h_8(x_i) - h_4^2(x_i)] &= -\frac{4!\times N}{\sigma^8}\label{eq:h8h4}
\end{align}
Plugging this into \Eq{kinterm} gives us the optimal estimator asserted by \Reff{Smith:2015uia}:
\begin{equation*}
    \boxed{\hat {\mathcal K}_{\rm opt} = \frac 1 N \sum_i x_i^4- \frac {6\sigma^2} N \sum_i x_i^2 + 3 \sigma^4}\tag{\ref{eq:Kopt}}.
\end{equation*}
A slight curiosity is that this estimator looks very similar to $h_4(x)$ written in \Eq{h4}.

\section{Deriving ${\rm RDN}^{(0)}$}
\label{app:RDN0}
The derivation of ${\rm RDN}^{(0)}$ in \Eq{RDN0} follows from a multivariate Edgeworth expansion of the CMB likelihood (\Refs{Namikawa:2012pe, Regan:2010cn, Schmittfull:2013uea}). In this section we shall spell out this calculation. This derivation contains the multivariate generalization of the derivation we performed in \App{Kopt_deriv} for the optimal kurtosis estimator \Eq{Kopt}. Because of this, we shall parallel the structure of \App{Kopt_deriv} in this section for clarity.

For discrete data, we can write an arbitrary Fourier mode $\vl_i$ as a vector of integers times the fundamental $\ell_F$: 
\begin{equation}
    {\vl_i} = {\vi \ell_F}
\end{equation}
so in this section for brevity we shall use the notation $T_{\vl_i}\equiv T_{\vi}$.
Suppose our lensed temperature map $T_\vl$ is approximately described by a Gaussian random field $\mathcal N(0,\tilde C_{\vi\vj})$ but perturbed such that it has small connected four-point function $\mathcal T$: 
\begin{equation}
    \mathcal T_{\vi\vj\vk\vll}=\langle T_{\vl_i}T_{\vl_j}T_{\vl_k}T_{\vl_l}\rangle_c.
\end{equation}
In this case, the moment generating function $M$ of $T_\vl$ is
\begin{align}
\nonumber    &M(\vJ) = {\rm exp} \Big\{-\frac 1 2 J_{\vi}\tilde C_{\vi\vj} J_{\vj} + \frac 1 {4!} J_{\vi}J_{\vj}J_{\vk}J_{\vll} \mathcal T_{\vi\vj\vk\vll}  \Big\}\\
\nonumber    &\approx {\rm exp} \Big\{-\frac 1 2 J_{\vi}\tilde C_{\vi\vj}J_{\vj}\Big\} \Big(1 + \frac 1 {4!} J_{\vi}J_{\vj}J_{\vk}J_{\vll} \mathcal T_{\vi\vj\vk\vll} \\
    &+ \frac 1 {2\times 4!4!} J_{\vi}J_{\vj}J_{\vk}J_{\vll} \mathcal T_{\vi\vj\vk\vll}  \times J_{\va}J_{\vb}J_{\vc}J_{\vd} \mathcal T_{\va\vb\vc\vd}\Big),
    \label{eq:MvJ}
\end{align}
where repeated indices are summed over. 
Before we continue, we can define the multivariate generalization of the Hermite polynomial \Eq{hermite}, the Hermite tensor $h_{\vi\vj\dots}$ which has $r$ subscripts:
\begin{equation}
    h_{\vi\vj\dots} =  (-1)^r p_{\mathcal N}^{-1}(T_\vl) \left[\frac \partial {\partial T_{\vi} }   \frac{\partial}{\partial T_\vj} \dots\right]p_{\mathcal N}(T_\vl).
\end{equation}
Here $p_{\mathcal N}(T_\vl)$ is the probability density function of a Gaussian random field with zero mean and covariance matrix $\tilde C_{\vi\vj}$:
\begin{equation}
    p_{\mathcal N}(T_\vl) = \frac 1 {\sqrt{(2\pi)^N {\rm det}[\tilde C_{\vi\vj}]}} {\rm exp} \left \{-\frac 1 2 T_\vi \tilde D_{\vi\vj} T_\vj \right\},
\end{equation}
where $N$ is the number of $T_\vl$ we have and we have defined the inverse covariance matrix $\tilde D \equiv (\tilde C)^{-1}$.

Inverse transforming the moment generating function \Eq{MvJ} gives us the PDF
\begin{align}
    \frac{p(T_\vl)}{p_{\mathcal N}(T_\vl)} &= 1 + \frac {\mathcal T_{\vi\vj\vk\vl} } {4!} h_{\vi\vj\vk\vl} + \frac {\mathcal T_{\vi\vj\vk\vl}\mathcal T_{\va\vb\vc\vd}} {2\times 4! 4!}h_{\vi\vj\vk\vl\va\vb\vc\vd},
\end{align}
where we have
\begin{align}
h_\vi &= T_\va \tilde D_{\va\vi}\\
\nonumber h_{\vi\vj\vk\vl} &= h_\vi h_\vj h_\vk h_\vl-[h_\vi h_\vj\tilde D_{\vk\vl} + ({\rm 5\ perms.})]\\
&+[\tilde D_{\vi\vj} \tilde D_{\vk\vl} + ({\rm 2\ perms.})]\label{eq:hijkl}\\
\nonumber h_{\vi\vj\vk\vl\va\vb\vc\vd} &= h_\vi h_\vj h_\vk h_\vl h_\va h_\vb h_\vc h_\vd \\
\nonumber &-[h_\vi h_\vj h_\vk h_\vl h_\va h_\vb \tilde D_{\vc\vd} + ({\rm 27\ perms.})]\\
\nonumber &+[h_\vi h_\vj h_\vk h_\vl \tilde D_{\va\vb} \tilde D_{\vc\vd} + ({\rm 209\ perms.})] \\
\nonumber &-[h_\vi h_\vj  \tilde D_{\vk\vl}\tilde D_{\va\vb} \tilde D_{\vc\vd} + ({\rm 419\ perms.})] \\
&+[ \tilde D_{\vi\vj} \tilde D_{\vk\vl}\tilde D_{\va\vb} \tilde D_{\vc\vd} + ({\rm 104\ perms.})] .
\end{align}
This then yields a log likelihood $\mathcal L=\log p(T_\vl)$ of
\begin{align}
\nonumber    \mathcal L - \mathcal L_{\mathcal N} &= \log \Big(1 + \frac 1 {4!} \mathcal T_{\vi\vj\vk\vl} h_{\vi\vj\vk\vl} + \frac {\mathcal T_{\vi\vj\vk\vl} \mathcal T_{\va\vb\vc\vd}} {2\times 4!4!} h_{\vi\vj\vk\vl\va\vb\vc\vd}\Big)\\
    &\approx \frac 1 {4!}\mathcal T_{\vi\vj\vk\vl} h_{\vi\vj\vk\vl} + \frac{\mathcal T_{\vi\vj\vk\vl}\mathcal T_{\va\vb\vc\vd}} {2\times 4!4!} h_{\vi\vj\vk\vl,\va\vb\vc\vd},
\end{align}
where we defined
\begin{equation}
    h_{\vi\vj\vk\vl,\va\vb\vc\vd} \equiv h_{\vi\vj\vk\vl\va\vb\vc\vd} - h_{\vi\vj\vk\vl}h_{\va\vb\vc\vd}.
\end{equation}
Asserting $\partial \mathcal L / \partial \mathcal T_{\vi\vj\vk\vl}=0$ to get the ML estimate for the connected trispectrum yields
\begin{equation}
    \hat{\mathcal T}_{\vi\vj\vk\vl} = -4! h_{\va\vb\vc\vd}\times  (h^{-1})_{\va\vb\vc\vd,\vi\vj\vk\vl}.\label{eq:That}
\end{equation}
From here to find a useful form of $h_{\vi\vj\vk\vl,\va\vb\vc\vd}$, the generalization of \Eq{h8h4}, we make the generalization of the approximation we made in \Eq{appro1} and replace products of $T$ with the appropriate Wick contracted power spectra:
\begin{equation}
    T_{\vi}T_{\vj}\dots T_{\vk}T_{\vl} = \sum_{\rm Wick} \tilde C_{\va\vb}\times\dots \times\tilde C_{\vc\vd}
    \label{eq:Wick}
\end{equation}
where $\{\va,\vb,\dots,\vc,\vd\}$ is a permutation of $\{\vi,\vj,\dots,\vk,\vl\}$ and the summation in \Eq{Wick} is over all permutations.
This approximation allows us to derive\footnote{with the heavy reliance on Mathematica to make this a tractable calculation} a generalization of the relation found in \Eq{h8h4}:
\begin{equation}
    h_{{\color{red}\va\vb\vc\vd},{\color{blue}\vi\vj\vk\vl}} = -[\tilde D_{{\color{red}\va}{\color{blue}\vi}} \tilde D_{{\color{red}\vb}{\color{blue}\vj}}\tilde D_{{\color{red}\vc}{\color{blue}\vk}}\tilde D_{{\color{red}\vd}{\color{blue}\vl}} + ({\rm 23\ perms.})],\label{eq:habcdijkl}
\end{equation}
where the permutations come from pairings of $\color{red}\{\va,\vb,\vc,\vd\}$ with $\color{blue}\{\vi,\vj,\vk,\vl\}$. Now note in the context of a summation, such as in \Eq{That}, all the terms in \Eq{habcdijkl} are the same under relabelling of dummy indices. So \Eq{That} turns into
\begin{align}
\nonumber    \hat {\mathcal T}_{\vi\vj\vk\vl}&= -4!h_{\va\vb\vc\vd}\times [-24\tilde D_{\va\vi}\tilde D_{\vb\vj} \tilde D_{\vc\vk}\tilde D_{\vd\vl}]^{-1}\\
    &= h_{\va\vb\vc\vd} \times \tilde C_{\va\vi}\tilde C_{\vb\vj}\tilde C_{\vc\vk}\tilde C_{\vd\vl}
\end{align}
Plugging in \Eq{hijkl} then leads to the connected four-point function estimator:
\begin{align}
\nonumber\hat {\mathcal T}_{\vi\vj\vk\vl}&= T_\vi T_\vj T_\vk T_\vl \\
\nonumber &- [T_\vi T_\vj \tilde C_{\vk\vl} +({\rm 5\ perms.})]\\
&+ [\tilde C_{\vi\vj} \tilde C_{\vj\vk} + ({\rm 3\ perms.})].\label{eq:tri-est}
\end{align}
This is the generalization of \Eq{Kopt}. 

Now we can specialize this to CMB lensing power spectrum estimation to derive ${\rm RDN}^{(0)}$. Note from \Eq{kk} that our estimate of the CMB lensing power spectrum is not from a general connected four-point function but specifically the combination $\int \langle T_{\vl_1}T_{\vL-\vl_1} T_{-\vl_2} T_{-\vL+\vl_2}\rangle_c  = \int \mathcal T_{\vl_1,\vL-\vl_1,-\vl_2,-\vL+\vl_2}$.
So our estimator \Eq{tri-est} for this specific combination yields 
\begin{align}
 \nonumber&\hat {\mathcal T}_{{\vl_1},\vL-{\vl_1},-{\vl_2},-\vL+{\vl_2}}= {T_{\vl_1} T_{\vL-{\vl_1}} T_{-{\vl_2}} T_{-\vL+{\vl_2}}} \\
\nonumber - [&T_{\vl_1} T_{\vL-{\vl_1}} \tilde C_{-{\vl_2},-\vL+{\vl_2}} + T_{{\vl_1}}T_{-{\vl_2}} \tilde C_{\vL-{\vl_1}, -\vL+{\vl_2}}\\
\nonumber+&T_{\vl_1} T_{-\vL+{\vl_2}}\tilde C_{\vL-{\vl_1}, -{\vl_2} }+ T_{\vL-{\vl_1}}T_{-{\vl_2}}\tilde C_{{\vl_1},-\vL+{\vl_2}}  \\
\nonumber+&T_{\vL-{\vl_1}}T_{-\vL+{\vl_2}}\tilde C_{{\vl_1}, -{\vl_2}} + T_{-{\vl_2}}T_{-\vL+{\vl_2}} \tilde C_{{\vl_1}, \vL-{\vl_1}}]\\
\nonumber+ &\tilde C_{{\vl_1},\vL-{\vl_1}}\tilde C_{-{\vl_2},-\vL+{\vl_2}} + \tilde C_{{\vl_1},-{\vl_2}}C_{\vL-{\vl_1}, -\vL+{\vl_2}}\\
+ &\tilde C_{{\vl_1},-\vL+{\vl_2}} \tilde C_{\vL-{\vl_1},-{\vl_2}}.\label{eq:kcmcb_test_1}
\end{align}

Each temperature map $T$ corresponds to the actual measured data $d$ whereas the $\tilde C$ come from the two sets of simulations $\{s\}, \{s'\}$ needed to compute ${\rm RDN}^{(0)}$ that we discussed in \Sec{RDN0}. 
Note that the covariance matrix $\tilde C_{\vi\vj}$ is defined in terms of the two-point function of Gaussian random fields which have the same total power spectrum as the lensed CMB, $\tilde C_{\vl}^{TT}$, not a lensed temperature map. 
In this case, we can rewrite the covariance matrix in terms of the power spectrum:
\begin{equation}
\tilde C_{\vi\vj} = \langle T_{\vl_{i}}T_{\vl_{j}} \rangle = (2\pi)^2 \delta^{(D)} (\vl_i + \vl_j)\tilde C_{\vl_i}^{TT}.
\end{equation}
So if we assume $\vL\ne 0$, then 
\begin{equation}
 \tilde C_{\vl_1,\vL-\vl_1} = \tilde C_{-\vl_2,-\vL+\vl_2}\propto \delta^{(D)}(\vL)=0.
\end{equation}
So, the connected four-point function estimator \Eq{tri-est} specialized to CMB lensing power spectrum estimation, \Eq{kcmcb_test_1}, reduces to
\begin{align}
\nonumber&\hat {\mathcal T}_{{\vl_1},\vL-{\vl_1},-{\vl_2},-\vL+{\vl_2}}= \underbrace{T_{\vl_1} T_{\vL-{\vl_1}} T_{-{\vl_2}} T_{-\vL+{\vl_2}}}_{\subset C_\vL(\hat\kappa_\vL, \hat\kappa_\vL),{\rm\ \Eq{kk}}} \\
 \nonumber - [&\underbrace{{\color{red}T_{{\vl_1}}}{\color{darkgreen}T_{-{\vl_2}}} \tilde C_{{\color{blue}\vL-{\vl_1}}, {\color{violet}-\vL+{\vl_2}}}}_{\subset C_\vL(\hat\kappa^{{\color{red}d}{\color{blue}s}}, \hat\kappa^{{\color{darkgreen}d}{\color{violet}s}})}+ 
 \underbrace{{\color{red}T_{\vl_1}}{\color{violet} T_{-\vL+{\vl_2}}}\tilde C_{{\color{blue}\vL-{\vl_1}},{\color{darkgreen}-{\vl_2}} }}_{\subset C_\vL(\hat\kappa^{{\color{red}d}{\color{blue}s}},\hat\kappa^{{\color{darkgreen}s}{\color{violet}d}})}\\
 \nonumber+ &\underbrace{{\color{blue}T_{\vL-{\vl_1}}}{\color{darkgreen}T_{-{\vl_2}}}\tilde C_{{\color{red}{\vl_1}},{\color{violet}-\vL+{\vl_2}}}}_{\subset C_\vL(\hat\kappa^{{\color{red}s}{\color{blue}d}},\hat\kappa^{{\color{darkgreen}d}{\color{violet}s}})} 
 + \underbrace{{\color{blue}T_{\vL-{\vl_1}}}{\color{violet}T_{-\vL+{\vl_2}}}\tilde C_{{\color{red}{\vl_1}}, {\color{darkgreen}-{\vl_2}}} }_{\subset C_\vL(\hat\kappa^{{\color{red}s}{\color{blue}d}},\hat\kappa^{{\color{darkgreen}s}{\color{violet}d}})}]\\
+ &\underbrace{\tilde C_{{\color{red}{\vl_1}},{\color{darkgreen}-{\vl_2}}}C_{{\color{blue}\vL-{\vl_1}}, {\color{violet}-\vL+{\vl_2}}}}_{\subset C_\vL(\hat\kappa^{{\color{red}s}{\color{blue}s'}},\hat\kappa^{{\color{darkgreen}s}{\color{violet}s'}})}
+ \underbrace{\tilde C_{{\color{red}{\vl_1}},{\color{violet}-\vL+{\vl_2}}} \tilde C_{{\color{blue}\vL-{\vl_1}},{\color{darkgreen}-{\vl_2}}}}_{\subset C_\vL(\hat\kappa^{{\color{red}s}{\color{blue}s'}},\hat\kappa^{{\color{darkgreen}s'}{\color{violet}s}})}.
\end{align}
{From} this we can read off the terms in \Eq{RDN0} which define ${\rm RDN}^{(0)}$. So we have shown how ${\rm RDN}^{(0)}$ naturally arises from a Edgeworth expansion of the CMB likelihood that we claimed in \Sec{RDN0}. 
We have also established the correspondence between ${\rm RDN}^{(0)}$ and the optimal Kurtosis estimator \Eq{Kopt} for the toy model discussed in \Sec{toy} and derived in \App{Kopt_deriv}.

\section{Computinig $N^{(1)}$ from Simulations}
\label{app:N1}
In addition to the analytic expression for $N^{(1)}$ we derived in \Eq{n1} one can also compute $N^{(1)}$ from simulation as we mentioned in \Sec{Ni}. This makes use of 4 sets of simulations. First we have two sets of simulations $\{s_\phi\}$ and $\{s_{\phi }'\}$ which share a common lensing potential $\phi$ but different realizations of the unlensed Gaussian CMB. The last two sets of simulations $\{s\},\{s'\}$ have different realization of both unlensed Gaussian CMB and lensing potential. From these sets of simulations one can form a estimator based on simulation for $N^{(1)}$ (\Refs{ACT:2023dou, Story_2015}):
\begin{align}
    \nonumber N^{(1),{\rm MC}}_\vL &= \big< C_{\mathbf L}(\hat\kappa^{s_\phi s_\phi'}, \hat\kappa^{s_\phi s_\phi'}) + C_{\mathbf L}(\hat\kappa^{s_\phi s_\phi'},\hat\kappa^{s_\phi' s_\phi})\\
    &- C_{\mathbf L}(\hat\kappa^{ss'},\hat\kappa^{ss'}) - C_{\mathbf{L}} (\hat\kappa^{ss'},\hat\kappa^{s's})\big>_{s,s',s_\phi,s_\phi'}.
    \label{eq:N1}
\end{align}
{To show why this works, lets recall the language of Feynman diagrams that we introduced in \App{Ni}. Because of how we construct $\{s_\phi\}$ and $\{s_\phi'\}$, the only diagrams\footnote{\label{fn:neg}neglecting the same diagrams we neglected in Footnote.~\ref{footnote:negdiag} for the same reasons} contributing to the four-point functions in the first line are the diagrams that contribute to $N^{(0)}$ and $N^{(1)}$:
\begin{equation}
   \vcenter{\hbox{\includegraphics{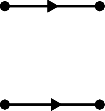}}}  + \vcenter{\hbox{\includegraphics{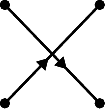}}}  +
   \vcenter{\hbox{\includegraphics{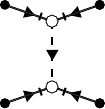}}}  + \vcenter{\hbox{\includegraphics{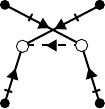}}}  .
\end{equation}
We have no diagrams that contribute to $\langle\kappa\kappa\rangle$ since those diagrams would require the unlensed temperature map in $s_\phi$ to be correlated with the unlensed temperature map in $s_\phi'$. 
So, to $O(\kappa^2)$ the first line yields an estimate for $N^{(0)}+N^{(1)}$. 
The terms on the second line yields only diagrams\textsuperscript{\ref{fn:neg}} contributing $N^{(0)}$ at $O(\kappa)$ and no diagrams at $O(\kappa^2)$ since the lensing potential $\kappa$ is no longer shared.
So, the second line shares only a $N^{(0)}$ contribution with the first line. 
Thus the difference of the two lines yields an simulation based estimate of $N^{(1)}$.
}

We make use of a hybrid approach to estimate $N^{(1)}$ in our numerical studies throughout this paper. 
For a full-sky analysis, \Eq{N1} is sufficient to estimate $N^{(1)}$ for all scales. 
However in our numerical studies we work on small patches where the flat-sky approximation holds. 
In this case \Eq{N1} cannot easily converge to a estimate of $N^{(1)}$ for very large scales where the number of modes is small. 
However, one can see from e.g. \Fig{Nhat_unmasked} that large scales are also the regime where the $N^{(1)}$ is order of magnitudes smaller than the signal $\langle\kappa\kappa\rangle$. 
So, our estimate of $N^{(1)}$ does not need to be perfect for these regimes.
Because of this, we use the analytical result \Eq{n1} to compute $N^{(1)}_L$ for $L < 512$. 
For $L>512$ we use the simulation based computation \Eq{N1} to compute $N^{(1)}_L$ since on small scales we have sufficient statistics to estimate $N^{(1)}$ from simulations.


\section{Fast algorithm to compute $\hat N^\times$}
\label{app:fastNhatX}

In \Sec{splits} Eqs.~(\ref{eq:Nijkl}-\ref{eq:Cnbcross}) we described how our proposed estimator could be combined with split-based methods recently proposed in \Reff{Madhavacheril:2020ido} to build estimators insensitive to modelling of instrument noise. Suppose we have $m$ splits of the CMB map with independent instrument noise. To combine our proposed noise avoidance with the split-based method we must compute \Eq{Nhatcross}. However, \Eq{Nhatcross} is naively a $O(m^4)$ computation. In this appendix we will construct a fast $O(m^2)$ algorithm to compute the $\hat N_\vL^\times$ described in \Eq{Nhatcross} similar to the fast algorithm discussed in \Reff{Madhavacheril:2020ido}.

To start we use an identity stated by \Reff{Madhavacheril:2020ido}
\begin{align}
\nonumber\gamma_{ijkl} &= \gamma_i \gamma_j \gamma_k \gamma_l \\
\nonumber  &- \Big[ \delta^{(K)}_{ij} \gamma_k \gamma_l + (\mbox{5 perm.}) \Big]\\
\nonumber  &+ 2 \Big[ \delta^{(K)}_{ijk} \gamma_l + (\mbox{3 perm.}) \Big]\\
\nonumber  &+ \Big[ \delta^{(K)}_{ij} \delta^{(K)}_{kl} + (\mbox{2 perm.}) \Big]\\
  &- 6 \delta^{(K)}_{ijkl}  \label{eq:gamma4}.
\end{align}
Here $\delta^{(K)}$ is a Kronecker delta:
\begin{equation}
\delta^{(K)}_{i_1\cdots i_n} = \begin{cases}
    1& i_1=\dots=i_n\\
    0&{\rm otherwise}
\end{cases}.
\label{eq:delta_def}
\end{equation}
For future convenience let us define a few things:
\begin{align}
   \label{eq:Qijl} \mathcal Q^{ij}_\vl &= \left|T^{(i)}_\vl T^{(j)}_{-\vl}\right|, \\\
    \mathcal Q^{i+} &= \frac 1 m \sum_{j=1}^m \mathcal Q^{ij}, \\
    \mathcal Q^{i\times} &= \mathcal Q^{i+} - \frac 1 m \mathcal Q^{ii},\\
    \mathcal Q^{++} &= \frac 1 {m^2} \sum_{i,j=1}^m \mathcal Q^{ij},\\
   \label{eq:Qxxl} \mathcal Q^{\times\times} &= \mathcal Q^{++} - \frac 1 {m^2} \sum_i\mathcal Q^{ii}.
\end{align}
This allows us to define
\begin{equation}
    \hat N^{\alpha, \beta}_\vL = 2(N^\kappa_\vL)^2 \int_\vl F^\kappa_{\vl,\vL-\vl} F^{\kappa}_{-\vl,-\vL+\vl}\times \mathcal Q^{\alpha}_\vl \mathcal Q^{\beta}_{\vL-\vl}.
\end{equation}
Here, $\alpha$ and $\beta$ {can be any pair of the superscripts $i, +,\times$ appearing in Eqs.~(\ref{eq:Qijl})-(\ref{eq:Qxxl})}. Note that we already introduced a special case of this, $\hat N^{ij,kl}$, in \Eq{Nijkl}.

To create a fast algorithm to compute \Eq{Nhatcross} we can apply the identity in \Eq{gamma4}:
\begin{align}
\nonumber    \sum_{ijkl} \gamma_i\gamma_j\gamma_k\gamma_l \hat{ N}^{ij,kl}_\vL&= m^4 \hat N^{++,++}_\vL \\
\nonumber    -\sum_{ijkl} \left[\substack{\delta^{(K)}_{ij}\gamma_k\gamma_l\\+ (5\ {\rm perm.})}\right] \fN^{ij,kl}_\vL&= -2m^2 \sum_i \left[2 \fN^{i+,i+}_\vL + \fN^{ii, ++}_\vL \right]\\
\nonumber    2\sum_{ijkl} \left[\substack{\delta^{(K)}_{ijk}\gamma_l\\ + (3\ {\rm perm.})} \right]\fN^{ij,kl}_\vL &= 8m \sum_i  \fN^{+i,ii}_\vL\\
\nonumber    \sum_{ijkl} \left[\substack{\delta^{(K)}_{ij}\delta^{(K)}_{kl}\\+(2\ {\rm perm.})} \right]\fN^{ij,kl}_\vL &= \sum_{ij} \left[\fN^{ii, jj}_\vL + 2 \fN^{ij,ij}\right]\\
    -6\sum_{ijkl}\delta^{(K)}_{ijkl} \fN^{ik,jl} &= - 6 \sum_i \fN^{ii,ii}\label{eq:identity_terms}
\end{align}
We can simplify things tremendously by combining terms:
\begin{align}
    \nonumber \hat N^\times_\vL &=  \frac 1 {m(m-1)(m-2)(m-3)}\\
    &\times \Big[ m^4 \fN_\vL^{\times\times,\times\times}-4m^2 \sum_i \fN^{i\times , i\times }_\vL+4\sum_{i<j} \fN^{ij,ij}_\vL\Big].
    \label{eq:Nhatcross_fast}
\end{align}
Using this we are able to compute $\hat N^\times_\vL$ (\Eq{Nhatcross}) in $O(m^2)$.

\bibliography{bib}

\begin{thebibliography}{51}%
\makeatletter
\providecommand \@ifxundefined [1]{%
 \@ifx{#1\undefined}
}%
\providecommand \@ifnum [1]{%
 \ifnum #1\expandafter \@firstoftwo
 \else \expandafter \@secondoftwo
 \fi
}%
\providecommand \@ifx [1]{%
 \ifx #1\expandafter \@firstoftwo
 \else \expandafter \@secondoftwo
 \fi
}%
\providecommand \natexlab [1]{#1}%
\providecommand \enquote  [1]{``#1''}%
\providecommand \bibnamefont  [1]{#1}%
\providecommand \bibfnamefont [1]{#1}%
\providecommand \citenamefont [1]{#1}%
\providecommand \href@noop [0]{\@secondoftwo}%
\providecommand \href [0]{\begingroup \@sanitize@url \@href}%
\providecommand \@href[1]{\@@startlink{#1}\@@href}%
\providecommand \@@href[1]{\endgroup#1\@@endlink}%
\providecommand \@sanitize@url [0]{\catcode `\\12\catcode `\$12\catcode `\&12\catcode `\#12\catcode `\^12\catcode `\_12\catcode `\%12\relax}%
\providecommand \@@startlink[1]{}%
\providecommand \@@endlink[0]{}%
\providecommand \url  [0]{\begingroup\@sanitize@url \@url }%
\providecommand \@url [1]{\endgroup\@href {#1}{\urlprefix }}%
\providecommand \urlprefix  [0]{URL }%
\providecommand \Eprint [0]{\href }%
\providecommand \doibase [0]{http://dx.doi.org/}%
\providecommand \selectlanguage [0]{\@gobble}%
\providecommand \bibinfo  [0]{\@secondoftwo}%
\providecommand \bibfield  [0]{\@secondoftwo}%
\providecommand \translation [1]{[#1]}%
\providecommand \BibitemOpen [0]{}%
\providecommand \bibitemStop [0]{}%
\providecommand \bibitemNoStop [0]{.\EOS\space}%
\providecommand \EOS [0]{\spacefactor3000\relax}%
\providecommand \BibitemShut  [1]{\csname bibitem#1\endcsname}%
\let\auto@bib@innerbib\@empty
\bibitem [{\citenamefont {Lewis}\ and\ \citenamefont {Challinor}(2006)}]{Lewis:2006fu}%
  \BibitemOpen
  \bibfield  {author} {\bibinfo {author} {\bibfnamefont {Antony}\ \bibnamefont {Lewis}}\ and\ \bibinfo {author} {\bibfnamefont {Anthony}\ \bibnamefont {Challinor}},\ }\bibfield  {title} {\enquote {\bibinfo {title} {{Weak gravitational lensing of the CMB}},}\ }\href {\doibase 10.1016/j.physrep.2006.03.002} {\bibfield  {journal} {\bibinfo  {journal} {Phys. Rept.}\ }\textbf {\bibinfo {volume} {429}},\ \bibinfo {pages} {1--65} (\bibinfo {year} {2006})},\ \Eprint {http://arxiv.org/abs/astro-ph/0601594} {arXiv:astro-ph/0601594} \BibitemShut {NoStop}%
\bibitem [{\citenamefont {Allison}\ \emph {et~al.}(2015)\citenamefont {Allison}, \citenamefont {Caucal}, \citenamefont {Calabrese}, \citenamefont {Dunkley},\ and\ \citenamefont {Louis}}]{Allison:2015qca}%
  \BibitemOpen
  \bibfield  {author} {\bibinfo {author} {\bibfnamefont {R.}~\bibnamefont {Allison}}, \bibinfo {author} {\bibfnamefont {P.}~\bibnamefont {Caucal}}, \bibinfo {author} {\bibfnamefont {E.}~\bibnamefont {Calabrese}}, \bibinfo {author} {\bibfnamefont {J.}~\bibnamefont {Dunkley}}, \ and\ \bibinfo {author} {\bibfnamefont {T.}~\bibnamefont {Louis}},\ }\bibfield  {title} {\enquote {\bibinfo {title} {{Towards a cosmological neutrino mass detection}},}\ }\href {\doibase 10.1103/PhysRevD.92.123535} {\bibfield  {journal} {\bibinfo  {journal} {Phys. Rev. D}\ }\textbf {\bibinfo {volume} {92}},\ \bibinfo {pages} {123535} (\bibinfo {year} {2015})},\ \Eprint {http://arxiv.org/abs/1509.07471} {arXiv:1509.07471 [astro-ph.CO]} \BibitemShut {NoStop}%
\bibitem [{\citenamefont {Schmittfull}\ and\ \citenamefont {Seljak}(2018)}]{Schmittfull:2017ffw}%
  \BibitemOpen
  \bibfield  {author} {\bibinfo {author} {\bibfnamefont {Marcel}\ \bibnamefont {Schmittfull}}\ and\ \bibinfo {author} {\bibfnamefont {Uros}\ \bibnamefont {Seljak}},\ }\bibfield  {title} {\enquote {\bibinfo {title} {{Parameter constraints from cross-correlation of CMB lensing with galaxy clustering}},}\ }\href {\doibase 10.1103/PhysRevD.97.123540} {\bibfield  {journal} {\bibinfo  {journal} {Phys. Rev. D}\ }\textbf {\bibinfo {volume} {97}},\ \bibinfo {pages} {123540} (\bibinfo {year} {2018})},\ \Eprint {http://arxiv.org/abs/1710.09465} {arXiv:1710.09465 [astro-ph.CO]} \BibitemShut {NoStop}%
\bibitem [{\citenamefont {Li}\ \emph {et~al.}(2018)\citenamefont {Li}, \citenamefont {Gluscevic}, \citenamefont {Boddy},\ and\ \citenamefont {Madhavacheril}}]{Li:2018zdm}%
  \BibitemOpen
  \bibfield  {author} {\bibinfo {author} {\bibfnamefont {Zack}\ \bibnamefont {Li}}, \bibinfo {author} {\bibfnamefont {Vera}\ \bibnamefont {Gluscevic}}, \bibinfo {author} {\bibfnamefont {Kimberly~K.}\ \bibnamefont {Boddy}}, \ and\ \bibinfo {author} {\bibfnamefont {Mathew~S.}\ \bibnamefont {Madhavacheril}},\ }\bibfield  {title} {\enquote {\bibinfo {title} {{Disentangling Dark Physics with Cosmic Microwave Background Experiments}},}\ }\href {\doibase 10.1103/PhysRevD.98.123524} {\bibfield  {journal} {\bibinfo  {journal} {Phys. Rev. D}\ }\textbf {\bibinfo {volume} {98}},\ \bibinfo {pages} {123524} (\bibinfo {year} {2018})},\ \Eprint {http://arxiv.org/abs/1806.10165} {arXiv:1806.10165 [astro-ph.CO]} \BibitemShut {NoStop}%
\bibitem [{\citenamefont {{Sherwin}}\ \emph {et~al.}(2011)\citenamefont {{Sherwin}} \emph {et~al.}}]{2011PhRvL.107b1302S}%
  \BibitemOpen
  \bibfield  {author} {\bibinfo {author} {\bibfnamefont {Blake~D.}\ \bibnamefont {{Sherwin}}} \emph {et~al.},\ }\bibfield  {title} {\enquote {\bibinfo {title} {{Evidence for Dark Energy from the Cosmic Microwave Background Alone Using the Atacama Cosmology Telescope Lensing Measurements}},}\ }\href {\doibase 10.1103/PhysRevLett.107.021302} {\bibfield  {journal} {\bibinfo  {journal} {\prl}\ }\textbf {\bibinfo {volume} {107}},\ \bibinfo {eid} {021302} (\bibinfo {year} {2011})},\ \Eprint {http://arxiv.org/abs/1105.0419} {arXiv:1105.0419 [astro-ph.CO]} \BibitemShut {NoStop}%
\bibitem [{\citenamefont {Smith}\ \emph {et~al.}(2007)\citenamefont {Smith}, \citenamefont {Zahn},\ and\ \citenamefont {Dore}}]{Smith:2007rg}%
  \BibitemOpen
  \bibfield  {author} {\bibinfo {author} {\bibfnamefont {Kendrick~M.}\ \bibnamefont {Smith}}, \bibinfo {author} {\bibfnamefont {Oliver}\ \bibnamefont {Zahn}}, \ and\ \bibinfo {author} {\bibfnamefont {Olivier}\ \bibnamefont {Dore}},\ }\bibfield  {title} {\enquote {\bibinfo {title} {{Detection of Gravitational Lensing in the Cosmic Microwave Background}},}\ }\href {\doibase 10.1103/PhysRevD.76.043510} {\bibfield  {journal} {\bibinfo  {journal} {Phys. Rev. D}\ }\textbf {\bibinfo {volume} {76}},\ \bibinfo {pages} {043510} (\bibinfo {year} {2007})},\ \Eprint {http://arxiv.org/abs/0705.3980} {arXiv:0705.3980 [astro-ph]} \BibitemShut {NoStop}%
\bibitem [{\citenamefont {{Das}}\ \emph {et~al.}(2011)\citenamefont {{Das}} \emph {et~al.}}]{2011PhRvL.107b1301D}%
  \BibitemOpen
  \bibfield  {author} {\bibinfo {author} {\bibfnamefont {Sudeep}\ \bibnamefont {{Das}}} \emph {et~al.},\ }\bibfield  {title} {\enquote {\bibinfo {title} {{Detection of the Power Spectrum of Cosmic Microwave Background Lensing by the Atacama Cosmology Telescope}},}\ }\href {\doibase 10.1103/PhysRevLett.107.021301} {\bibfield  {journal} {\bibinfo  {journal} {\prl}\ }\textbf {\bibinfo {volume} {107}},\ \bibinfo {eid} {021301} (\bibinfo {year} {2011})},\ \Eprint {http://arxiv.org/abs/1103.2124} {arXiv:1103.2124 [astro-ph.CO]} \BibitemShut {NoStop}%
\bibitem [{\citenamefont {{van Engelen}}\ \emph {et~al.}(2012)\citenamefont {{van Engelen}} \emph {et~al.}}]{van_Engelen_2012}%
  \BibitemOpen
  \bibfield  {author} {\bibinfo {author} {\bibfnamefont {A.}~\bibnamefont {{van Engelen}}} \emph {et~al.},\ }\bibfield  {title} {\enquote {\bibinfo {title} {{A Measurement of Gravitational Lensing of the Microwave Background Using South Pole Telescope Data}},}\ }\href {\doibase 10.1088/0004-637X/756/2/142} {\bibfield  {journal} {\bibinfo  {journal} {\apj}\ }\textbf {\bibinfo {volume} {756}},\ \bibinfo {eid} {142} (\bibinfo {year} {2012})},\ \Eprint {http://arxiv.org/abs/1202.0546} {arXiv:1202.0546 [astro-ph.CO]} \BibitemShut {NoStop}%
\bibitem [{\citenamefont {Ade}\ \emph {et~al.}(2014)\citenamefont {Ade} \emph {et~al.}}]{polarbear}%
  \BibitemOpen
  \bibfield  {author} {\bibinfo {author} {\bibfnamefont {P.~A.~R.}\ \bibnamefont {Ade}} \emph {et~al.} (\bibinfo {collaboration} {POLARBEAR}),\ }\bibfield  {title} {\enquote {\bibinfo {title} {{Measurement of the Cosmic Microwave Background Polarization Lensing Power Spectrum with the POLARBEAR experiment}},}\ }\href {\doibase 10.1103/PhysRevLett.113.021301} {\bibfield  {journal} {\bibinfo  {journal} {Phys. Rev. Lett.}\ }\textbf {\bibinfo {volume} {113}},\ \bibinfo {pages} {021301} (\bibinfo {year} {2014})},\ \Eprint {http://arxiv.org/abs/1312.6646} {arXiv:1312.6646 [astro-ph.CO]} \BibitemShut {NoStop}%
\bibitem [{\citenamefont {Story}\ \emph {et~al.}(2015)\citenamefont {Story} \emph {et~al.}}]{Story_2015}%
  \BibitemOpen
  \bibfield  {author} {\bibinfo {author} {\bibfnamefont {K.~T.}\ \bibnamefont {Story}} \emph {et~al.} (\bibinfo {collaboration} {SPT}),\ }\bibfield  {title} {\enquote {\bibinfo {title} {{A Measurement of the Cosmic Microwave Background Gravitational Lensing Potential from 100 Square Degrees of SPTpol Data}},}\ }\href {\doibase 10.1088/0004-637X/810/1/50} {\bibfield  {journal} {\bibinfo  {journal} {Astrophys. J.}\ }\textbf {\bibinfo {volume} {810}},\ \bibinfo {pages} {50} (\bibinfo {year} {2015})},\ \Eprint {http://arxiv.org/abs/1412.4760} {arXiv:1412.4760 [astro-ph.CO]} \BibitemShut {NoStop}%
\bibitem [{\citenamefont {Ade}\ \emph {et~al.}(2016{\natexlab{a}})\citenamefont {Ade} \emph {et~al.}}]{BICEP2016}%
  \BibitemOpen
  \bibfield  {author} {\bibinfo {author} {\bibfnamefont {P.~A.~R.}\ \bibnamefont {Ade}} \emph {et~al.} (\bibinfo {collaboration} {BICEP2, Keck Array}),\ }\bibfield  {title} {\enquote {\bibinfo {title} {{BICEP2 / Keck Array VIII: Measurement of gravitational lensing from large-scale B-mode polarization}},}\ }\href {\doibase 10.3847/1538-4357/833/2/228} {\bibfield  {journal} {\bibinfo  {journal} {Astrophys. J.}\ }\textbf {\bibinfo {volume} {833}},\ \bibinfo {pages} {228} (\bibinfo {year} {2016}{\natexlab{a}})},\ \Eprint {http://arxiv.org/abs/1606.01968} {arXiv:1606.01968 [astro-ph.CO]} \BibitemShut {NoStop}%
\bibitem [{\citenamefont {Sherwin}\ \emph {et~al.}(2017)\citenamefont {Sherwin} \emph {et~al.}}]{PhysRevD.95.123529}%
  \BibitemOpen
  \bibfield  {author} {\bibinfo {author} {\bibfnamefont {Blake~D.}\ \bibnamefont {Sherwin}} \emph {et~al.},\ }\bibfield  {title} {\enquote {\bibinfo {title} {{Two-season Atacama Cosmology Telescope polarimeter lensing power spectrum}},}\ }\href {\doibase 10.1103/PhysRevD.95.123529} {\bibfield  {journal} {\bibinfo  {journal} {Phys. Rev. D}\ }\textbf {\bibinfo {volume} {95}},\ \bibinfo {pages} {123529} (\bibinfo {year} {2017})},\ \Eprint {http://arxiv.org/abs/1611.09753} {arXiv:1611.09753 [astro-ph.CO]} \BibitemShut {NoStop}%
\bibitem [{\citenamefont {Omori}\ \emph {et~al.}(2017)\citenamefont {Omori} \emph {et~al.}}]{Omori_2017}%
  \BibitemOpen
  \bibfield  {author} {\bibinfo {author} {\bibfnamefont {Y.}~\bibnamefont {Omori}} \emph {et~al.},\ }\bibfield  {title} {\enquote {\bibinfo {title} {{A 2500 deg$^2$ CMB Lensing Map from Combined South Pole Telescope and Planck Data}},}\ }\href {\doibase 10.3847/1538-4357/aa8d1d} {\bibfield  {journal} {\bibinfo  {journal} {Astrophys. J.}\ }\textbf {\bibinfo {volume} {849}},\ \bibinfo {pages} {124} (\bibinfo {year} {2017})},\ \Eprint {http://arxiv.org/abs/1705.00743} {arXiv:1705.00743 [astro-ph.CO]} \BibitemShut {NoStop}%
\bibitem [{\citenamefont {Wu}\ \emph {et~al.}(2019)\citenamefont {Wu} \emph {et~al.}}]{Wu_2019}%
  \BibitemOpen
  \bibfield  {author} {\bibinfo {author} {\bibfnamefont {W.~L.~K.}\ \bibnamefont {Wu}} \emph {et~al.},\ }\bibfield  {title} {\enquote {\bibinfo {title} {{A Measurement of the Cosmic Microwave Background Lensing Potential and Power Spectrum from 500 deg$^2$ of SPTpol Temperature and Polarization Data}},}\ }\href {\doibase 10.3847/1538-4357/ab4186} {\bibfield  {journal} {\bibinfo  {journal} {Astrophys. J.}\ }\textbf {\bibinfo {volume} {884}},\ \bibinfo {pages} {70} (\bibinfo {year} {2019})},\ \Eprint {http://arxiv.org/abs/1905.05777} {arXiv:1905.05777 [astro-ph.CO]} \BibitemShut {NoStop}%
\bibitem [{\citenamefont {Bianchini}\ \emph {et~al.}(2020)\citenamefont {Bianchini} \emph {et~al.}}]{SPT:2019fqo}%
  \BibitemOpen
  \bibfield  {author} {\bibinfo {author} {\bibfnamefont {F.}~\bibnamefont {Bianchini}} \emph {et~al.} (\bibinfo {collaboration} {SPT}),\ }\bibfield  {title} {\enquote {\bibinfo {title} {{Constraints on Cosmological Parameters from the 500 deg$^2$ SPTpol Lensing Power Spectrum}},}\ }\href {\doibase 10.3847/1538-4357/ab6082} {\bibfield  {journal} {\bibinfo  {journal} {Astrophys. J.}\ }\textbf {\bibinfo {volume} {888}},\ \bibinfo {pages} {119} (\bibinfo {year} {2020})},\ \Eprint {http://arxiv.org/abs/1910.07157} {arXiv:1910.07157 [astro-ph.CO]} \BibitemShut {NoStop}%
\bibitem [{\citenamefont {Qu}\ \emph {et~al.}(2023)\citenamefont {Qu} \emph {et~al.}}]{ACT:2023dou}%
  \BibitemOpen
  \bibfield  {author} {\bibinfo {author} {\bibfnamefont {Frank~J.}\ \bibnamefont {Qu}} \emph {et~al.} (\bibinfo {collaboration} {ACT}),\ }\bibfield  {title} {\enquote {\bibinfo {title} {{The Atacama Cosmology Telescope: A Measurement of the DR6 CMB Lensing Power Spectrum and its Implications for Structure Growth}},}\ }\href@noop {} {\  (\bibinfo {year} {2023})},\ \Eprint {http://arxiv.org/abs/2304.05202} {arXiv:2304.05202 [astro-ph.CO]} \BibitemShut {NoStop}%
\bibitem [{\citenamefont {Pan}\ \emph {et~al.}(2023)\citenamefont {Pan} \emph {et~al.}}]{SPT:2023jql}%
  \BibitemOpen
  \bibfield  {author} {\bibinfo {author} {\bibfnamefont {Z.}~\bibnamefont {Pan}} \emph {et~al.} (\bibinfo {collaboration} {SPT}),\ }\bibfield  {title} {\enquote {\bibinfo {title} {{Measurement of gravitational lensing of the cosmic microwave background using SPT-3G 2018 data}},}\ }\href {\doibase 10.1103/PhysRevD.108.122005} {\bibfield  {journal} {\bibinfo  {journal} {Phys. Rev. D}\ }\textbf {\bibinfo {volume} {108}},\ \bibinfo {pages} {122005} (\bibinfo {year} {2023})},\ \Eprint {http://arxiv.org/abs/2308.11608} {arXiv:2308.11608 [astro-ph.CO]} \BibitemShut {NoStop}%
\bibitem [{\citenamefont {Madhavacheril}\ \emph {et~al.}(2020)\citenamefont {Madhavacheril}, \citenamefont {Smith}, \citenamefont {Sherwin},\ and\ \citenamefont {Naess}}]{Madhavacheril:2020ido}%
  \BibitemOpen
  \bibfield  {author} {\bibinfo {author} {\bibfnamefont {Mathew~S.}\ \bibnamefont {Madhavacheril}}, \bibinfo {author} {\bibfnamefont {Kendrick~M.}\ \bibnamefont {Smith}}, \bibinfo {author} {\bibfnamefont {Blake~D.}\ \bibnamefont {Sherwin}}, \ and\ \bibinfo {author} {\bibfnamefont {Sigurd}\ \bibnamefont {Naess}},\ }\bibfield  {title} {\enquote {\bibinfo {title} {{CMB lensing power spectrum estimation without instrument noise bias}},}\ }\href {\doibase 10.1088/1475-7516/2021/05/028} {\  (\bibinfo {year} {2020}),\ 10.1088/1475-7516/2021/05/028},\ \Eprint {http://arxiv.org/abs/2011.02475} {arXiv:2011.02475 [astro-ph.CO]} \BibitemShut {NoStop}%
\bibitem [{\citenamefont {{Sherwin}}\ and\ \citenamefont {{Das}}(2010)}]{2010arXiv1011.4510S}%
  \BibitemOpen
  \bibfield  {author} {\bibinfo {author} {\bibfnamefont {Blake~D.}\ \bibnamefont {{Sherwin}}}\ and\ \bibinfo {author} {\bibfnamefont {Sudeep}\ \bibnamefont {{Das}}},\ }\bibfield  {title} {\enquote {\bibinfo {title} {{CMB Lensing - Power Without Bias}},}\ }\href {\doibase 10.48550/arXiv.1011.4510} {\bibfield  {journal} {\bibinfo  {journal} {arXiv e-prints}\ ,\ \bibinfo {eid} {arXiv:1011.4510}} (\bibinfo {year} {2010})},\ \Eprint {http://arxiv.org/abs/1011.4510} {arXiv:1011.4510 [astro-ph.CO]} \BibitemShut {NoStop}%
\bibitem [{\citenamefont {Smith}\ \emph {et~al.}(2015)\citenamefont {Smith}, \citenamefont {Senatore},\ and\ \citenamefont {Zaldarriaga}}]{Smith:2015uia}%
  \BibitemOpen
  \bibfield  {author} {\bibinfo {author} {\bibfnamefont {Kendrick~M.}\ \bibnamefont {Smith}}, \bibinfo {author} {\bibfnamefont {Leonardo}\ \bibnamefont {Senatore}}, \ and\ \bibinfo {author} {\bibfnamefont {Matias}\ \bibnamefont {Zaldarriaga}},\ }\bibfield  {title} {\enquote {\bibinfo {title} {{Optimal analysis of the CMB trispectrum}},}\ }\href@noop {} {\  (\bibinfo {year} {2015})},\ \Eprint {http://arxiv.org/abs/1502.00635} {arXiv:1502.00635 [astro-ph.CO]} \BibitemShut {NoStop}%
\bibitem [{\citenamefont {Hirata}\ and\ \citenamefont {Seljak}(2003)}]{Hirata:2003ka}%
  \BibitemOpen
  \bibfield  {author} {\bibinfo {author} {\bibfnamefont {Christopher~M.}\ \bibnamefont {Hirata}}\ and\ \bibinfo {author} {\bibfnamefont {Uros}\ \bibnamefont {Seljak}},\ }\bibfield  {title} {\enquote {\bibinfo {title} {{Reconstruction of lensing from the cosmic microwave background polarization}},}\ }\href {\doibase 10.1103/PhysRevD.68.083002} {\bibfield  {journal} {\bibinfo  {journal} {Phys. Rev. D}\ }\textbf {\bibinfo {volume} {68}},\ \bibinfo {pages} {083002} (\bibinfo {year} {2003})},\ \Eprint {http://arxiv.org/abs/astro-ph/0306354} {arXiv:astro-ph/0306354} \BibitemShut {NoStop}%
\bibitem [{\citenamefont {B\"ohm}\ \emph {et~al.}(2016)\citenamefont {B\"ohm}, \citenamefont {Schmittfull},\ and\ \citenamefont {Sherwin}}]{Bohm:2016gzt}%
  \BibitemOpen
  \bibfield  {author} {\bibinfo {author} {\bibfnamefont {Vanessa}\ \bibnamefont {B\"ohm}}, \bibinfo {author} {\bibfnamefont {Marcel}\ \bibnamefont {Schmittfull}}, \ and\ \bibinfo {author} {\bibfnamefont {Blake~D.}\ \bibnamefont {Sherwin}},\ }\bibfield  {title} {\enquote {\bibinfo {title} {{Bias to CMB lensing measurements from the bispectrum of large-scale structure}},}\ }\href {\doibase 10.1103/PhysRevD.94.043519} {\bibfield  {journal} {\bibinfo  {journal} {Phys. Rev. D}\ }\textbf {\bibinfo {volume} {94}},\ \bibinfo {pages} {043519} (\bibinfo {year} {2016})},\ \Eprint {http://arxiv.org/abs/1605.01392} {arXiv:1605.01392 [astro-ph.CO]} \BibitemShut {NoStop}%
\bibitem [{\citenamefont {B\"ohm}\ \emph {et~al.}(2018)\citenamefont {B\"ohm}, \citenamefont {Sherwin}, \citenamefont {Liu}, \citenamefont {Hill}, \citenamefont {Schmittfull},\ and\ \citenamefont {Namikawa}}]{Bohm:2018omn}%
  \BibitemOpen
  \bibfield  {author} {\bibinfo {author} {\bibfnamefont {Vanessa}\ \bibnamefont {B\"ohm}}, \bibinfo {author} {\bibfnamefont {Blake~D.}\ \bibnamefont {Sherwin}}, \bibinfo {author} {\bibfnamefont {Jia}\ \bibnamefont {Liu}}, \bibinfo {author} {\bibfnamefont {J.~Colin}\ \bibnamefont {Hill}}, \bibinfo {author} {\bibfnamefont {Marcel}\ \bibnamefont {Schmittfull}}, \ and\ \bibinfo {author} {\bibfnamefont {Toshiya}\ \bibnamefont {Namikawa}},\ }\bibfield  {title} {\enquote {\bibinfo {title} {{Effect of non-Gaussian lensing deflections on CMB lensing measurements}},}\ }\href {\doibase 10.1103/PhysRevD.98.123510} {\bibfield  {journal} {\bibinfo  {journal} {Phys. Rev. D}\ }\textbf {\bibinfo {volume} {98}},\ \bibinfo {pages} {123510} (\bibinfo {year} {2018})},\ \Eprint {http://arxiv.org/abs/1806.01157} {arXiv:1806.01157 [astro-ph.CO]} \BibitemShut {NoStop}%
\bibitem [{\citenamefont {Beck}\ \emph {et~al.}(2018)\citenamefont {Beck}, \citenamefont {Fabbian},\ and\ \citenamefont {Errard}}]{Beck:2018wud}%
  \BibitemOpen
  \bibfield  {author} {\bibinfo {author} {\bibfnamefont {Dominic}\ \bibnamefont {Beck}}, \bibinfo {author} {\bibfnamefont {Giulio}\ \bibnamefont {Fabbian}}, \ and\ \bibinfo {author} {\bibfnamefont {Josquin}\ \bibnamefont {Errard}},\ }\bibfield  {title} {\enquote {\bibinfo {title} {{Lensing Reconstruction in Post-Born Cosmic Microwave Background Weak Lensing}},}\ }\href {\doibase 10.1103/PhysRevD.98.043512} {\bibfield  {journal} {\bibinfo  {journal} {Phys. Rev. D}\ }\textbf {\bibinfo {volume} {98}},\ \bibinfo {pages} {043512} (\bibinfo {year} {2018})},\ \Eprint {http://arxiv.org/abs/1806.01216} {arXiv:1806.01216 [astro-ph.CO]} \BibitemShut {NoStop}%
\bibitem [{\citenamefont {Fabbian}\ \emph {et~al.}(2019)\citenamefont {Fabbian}, \citenamefont {Lewis},\ and\ \citenamefont {Beck}}]{Fabbian:2019tik}%
  \BibitemOpen
  \bibfield  {author} {\bibinfo {author} {\bibfnamefont {Giulio}\ \bibnamefont {Fabbian}}, \bibinfo {author} {\bibfnamefont {Antony}\ \bibnamefont {Lewis}}, \ and\ \bibinfo {author} {\bibfnamefont {Dominic}\ \bibnamefont {Beck}},\ }\bibfield  {title} {\enquote {\bibinfo {title} {{CMB lensing reconstruction biases in cross-correlation with large-scale structure probes}},}\ }\href {\doibase 10.1088/1475-7516/2019/10/057} {\bibfield  {journal} {\bibinfo  {journal} {JCAP}\ }\textbf {\bibinfo {volume} {10}},\ \bibinfo {pages} {057} (\bibinfo {year} {2019})},\ \Eprint {http://arxiv.org/abs/1906.08760} {arXiv:1906.08760 [astro-ph.CO]} \BibitemShut {NoStop}%
\bibitem [{\citenamefont {Namikawa}\ \emph {et~al.}(2013)\citenamefont {Namikawa}, \citenamefont {Hanson},\ and\ \citenamefont {Takahashi}}]{Namikawa:2012pe}%
  \BibitemOpen
  \bibfield  {author} {\bibinfo {author} {\bibfnamefont {Toshiya}\ \bibnamefont {Namikawa}}, \bibinfo {author} {\bibfnamefont {Duncan}\ \bibnamefont {Hanson}}, \ and\ \bibinfo {author} {\bibfnamefont {Ryuichi}\ \bibnamefont {Takahashi}},\ }\bibfield  {title} {\enquote {\bibinfo {title} {{Bias-Hardened CMB Lensing}},}\ }\href {\doibase 10.1093/mnras/stt195} {\bibfield  {journal} {\bibinfo  {journal} {Mon. Not. Roy. Astron. Soc.}\ }\textbf {\bibinfo {volume} {431}},\ \bibinfo {pages} {609--620} (\bibinfo {year} {2013})},\ \Eprint {http://arxiv.org/abs/1209.0091} {arXiv:1209.0091 [astro-ph.CO]} \BibitemShut {NoStop}%
\bibitem [{\citenamefont {{Planck Collaboration}}(2020)}]{Planck:2018}%
  \BibitemOpen
  \bibfield  {author} {\bibinfo {author} {\bibnamefont {{Planck Collaboration}}},\ }\bibfield  {title} {\enquote {\bibinfo {title} {{Planck 2018 results. VIII. Gravitational lensing}},}\ }\href {\doibase 10.1051/0004-6361/201833886} {\bibfield  {journal} {\bibinfo  {journal} {\aap}\ }\textbf {\bibinfo {volume} {641}},\ \bibinfo {eid} {A8} (\bibinfo {year} {2020})},\ \Eprint {http://arxiv.org/abs/1807.06210} {arXiv:1807.06210 [astro-ph.CO]} \BibitemShut {NoStop}%
\bibitem [{\citenamefont {Kesden}\ \emph {et~al.}(2003)\citenamefont {Kesden}, \citenamefont {Cooray},\ and\ \citenamefont {Kamionkowski}}]{Kesden:2003cc}%
  \BibitemOpen
  \bibfield  {author} {\bibinfo {author} {\bibfnamefont {Michael~H.}\ \bibnamefont {Kesden}}, \bibinfo {author} {\bibfnamefont {Asantha}\ \bibnamefont {Cooray}}, \ and\ \bibinfo {author} {\bibfnamefont {Marc}\ \bibnamefont {Kamionkowski}},\ }\bibfield  {title} {\enquote {\bibinfo {title} {{Lensing reconstruction with CMB temperature and polarization}},}\ }\href {\doibase 10.1103/PhysRevD.67.123507} {\bibfield  {journal} {\bibinfo  {journal} {Phys. Rev. D}\ }\textbf {\bibinfo {volume} {67}},\ \bibinfo {pages} {123507} (\bibinfo {year} {2003})},\ \Eprint {http://arxiv.org/abs/astro-ph/0302536} {arXiv:astro-ph/0302536} \BibitemShut {NoStop}%
\bibitem [{\citenamefont {{Hanson}}\ \emph {et~al.}(2011)\citenamefont {{Hanson}}, \citenamefont {{Challinor}}, \citenamefont {{Efstathiou}},\ and\ \citenamefont {{Bielewicz}}}]{2011PhRvD..83d3005H}%
  \BibitemOpen
  \bibfield  {author} {\bibinfo {author} {\bibfnamefont {Duncan}\ \bibnamefont {{Hanson}}}, \bibinfo {author} {\bibfnamefont {Anthony}\ \bibnamefont {{Challinor}}}, \bibinfo {author} {\bibfnamefont {George}\ \bibnamefont {{Efstathiou}}}, \ and\ \bibinfo {author} {\bibfnamefont {Pawel}\ \bibnamefont {{Bielewicz}}},\ }\bibfield  {title} {\enquote {\bibinfo {title} {{CMB temperature lensing power reconstruction}},}\ }\href {\doibase 10.1103/PhysRevD.83.043005} {\bibfield  {journal} {\bibinfo  {journal} {\prd}\ }\textbf {\bibinfo {volume} {83}},\ \bibinfo {eid} {043005} (\bibinfo {year} {2011})},\ \Eprint {http://arxiv.org/abs/1008.4403} {arXiv:1008.4403 [astro-ph.CO]} \BibitemShut {NoStop}%
\bibitem [{\citenamefont {Jenkins}\ \emph {et~al.}(2014{\natexlab{a}})\citenamefont {Jenkins}, \citenamefont {Manohar}, \citenamefont {Waalewijn},\ and\ \citenamefont {Yadav}}]{Jenkins:2014hza}%
  \BibitemOpen
  \bibfield  {author} {\bibinfo {author} {\bibfnamefont {Elizabeth~E.}\ \bibnamefont {Jenkins}}, \bibinfo {author} {\bibfnamefont {Aneesh~V.}\ \bibnamefont {Manohar}}, \bibinfo {author} {\bibfnamefont {Wouter~J.}\ \bibnamefont {Waalewijn}}, \ and\ \bibinfo {author} {\bibfnamefont {Amit P.~S.}\ \bibnamefont {Yadav}},\ }\bibfield  {title} {\enquote {\bibinfo {title} {{Higher-Order Gravitational Lensing Reconstruction using Feynman Diagrams}},}\ }\href {\doibase 10.1088/1475-7516/2014/09/024} {\bibfield  {journal} {\bibinfo  {journal} {JCAP}\ }\textbf {\bibinfo {volume} {09}},\ \bibinfo {pages} {024} (\bibinfo {year} {2014}{\natexlab{a}})},\ \Eprint {http://arxiv.org/abs/1403.4607} {arXiv:1403.4607 [astro-ph.CO]} \BibitemShut {NoStop}%
\bibitem [{\citenamefont {Jenkins}\ \emph {et~al.}(2014{\natexlab{b}})\citenamefont {Jenkins}, \citenamefont {Manohar}, \citenamefont {Waalewijn},\ and\ \citenamefont {Yadav}}]{Jenkins:2014oja}%
  \BibitemOpen
  \bibfield  {author} {\bibinfo {author} {\bibfnamefont {Elizabeth~E.}\ \bibnamefont {Jenkins}}, \bibinfo {author} {\bibfnamefont {Aneesh~V.}\ \bibnamefont {Manohar}}, \bibinfo {author} {\bibfnamefont {Wouter~J.}\ \bibnamefont {Waalewijn}}, \ and\ \bibinfo {author} {\bibfnamefont {Amit P.~S.}\ \bibnamefont {Yadav}},\ }\bibfield  {title} {\enquote {\bibinfo {title} {{Gravitational Lensing of the CMB: a Feynman Diagram Approach}},}\ }\href {\doibase 10.1016/j.physletb.2014.07.002} {\bibfield  {journal} {\bibinfo  {journal} {Phys. Lett. B}\ }\textbf {\bibinfo {volume} {736}},\ \bibinfo {pages} {6--10} (\bibinfo {year} {2014}{\natexlab{b}})},\ \Eprint {http://arxiv.org/abs/1403.2386} {arXiv:1403.2386 [astro-ph.CO]} \BibitemShut {NoStop}%
\bibitem [{\citenamefont {{Lewis}}\ \emph {et~al.}(2011)\citenamefont {{Lewis}}, \citenamefont {{Challinor}},\ and\ \citenamefont {{Hanson}}}]{2011JCAP...03..018L}%
  \BibitemOpen
  \bibfield  {author} {\bibinfo {author} {\bibfnamefont {Antony}\ \bibnamefont {{Lewis}}}, \bibinfo {author} {\bibfnamefont {Anthony}\ \bibnamefont {{Challinor}}}, \ and\ \bibinfo {author} {\bibfnamefont {Duncan}\ \bibnamefont {{Hanson}}},\ }\bibfield  {title} {\enquote {\bibinfo {title} {{The shape of the CMB lensing bispectrum}},}\ }\href {\doibase 10.1088/1475-7516/2011/03/018} {\bibfield  {journal} {\bibinfo  {journal} {\jcap}\ }\textbf {\bibinfo {volume} {2011}},\ \bibinfo {eid} {018} (\bibinfo {year} {2011})},\ \Eprint {http://arxiv.org/abs/1101.2234} {arXiv:1101.2234 [astro-ph.CO]} \BibitemShut {NoStop}%
\bibitem [{\citenamefont {Ade}\ \emph {et~al.}(2016{\natexlab{b}})\citenamefont {Ade} \emph {et~al.}}]{Planck:2015mym}%
  \BibitemOpen
  \bibfield  {author} {\bibinfo {author} {\bibfnamefont {P.~A.~R.}\ \bibnamefont {Ade}} \emph {et~al.} (\bibinfo {collaboration} {Planck}),\ }\bibfield  {title} {\enquote {\bibinfo {title} {{Planck 2015 results. XV. Gravitational lensing}},}\ }\href {\doibase 10.1051/0004-6361/201525941} {\bibfield  {journal} {\bibinfo  {journal} {Astron. Astrophys.}\ }\textbf {\bibinfo {volume} {594}},\ \bibinfo {pages} {A15} (\bibinfo {year} {2016}{\natexlab{b}})},\ \Eprint {http://arxiv.org/abs/1502.01591} {arXiv:1502.01591 [astro-ph.CO]} \BibitemShut {NoStop}%
\bibitem [{\citenamefont {Aghanim}\ \emph {et~al.}(2020)\citenamefont {Aghanim} \emph {et~al.}}]{Planck:2018lbu}%
  \BibitemOpen
  \bibfield  {author} {\bibinfo {author} {\bibfnamefont {N.}~\bibnamefont {Aghanim}} \emph {et~al.} (\bibinfo {collaboration} {Planck}),\ }\bibfield  {title} {\enquote {\bibinfo {title} {{Planck 2018 results. VIII. Gravitational lensing}},}\ }\href {\doibase 10.1051/0004-6361/201833886} {\bibfield  {journal} {\bibinfo  {journal} {Astron. Astrophys.}\ }\textbf {\bibinfo {volume} {641}},\ \bibinfo {pages} {A8} (\bibinfo {year} {2020})},\ \Eprint {http://arxiv.org/abs/1807.06210} {arXiv:1807.06210 [astro-ph.CO]} \BibitemShut {NoStop}%
\bibitem [{\citenamefont {Mirmelstein}\ \emph {et~al.}(2019)\citenamefont {Mirmelstein}, \citenamefont {Carron},\ and\ \citenamefont {Lewis}}]{Mirmelstein:2019sxi}%
  \BibitemOpen
  \bibfield  {author} {\bibinfo {author} {\bibfnamefont {Mark}\ \bibnamefont {Mirmelstein}}, \bibinfo {author} {\bibfnamefont {Julien}\ \bibnamefont {Carron}}, \ and\ \bibinfo {author} {\bibfnamefont {Antony}\ \bibnamefont {Lewis}},\ }\bibfield  {title} {\enquote {\bibinfo {title} {{Optimal filtering for CMB lensing reconstruction}},}\ }\href {\doibase 10.1103/PhysRevD.100.123509} {\bibfield  {journal} {\bibinfo  {journal} {Phys. Rev. D}\ }\textbf {\bibinfo {volume} {100}},\ \bibinfo {pages} {123509} (\bibinfo {year} {2019})},\ \Eprint {http://arxiv.org/abs/1909.02653} {arXiv:1909.02653 [astro-ph.CO]} \BibitemShut {NoStop}%
\bibitem [{\citenamefont {Alonso}\ \emph {et~al.}(2019)\citenamefont {Alonso}, \citenamefont {Sanchez},\ and\ \citenamefont {Slosar}}]{Alonso:2018jzx}%
  \BibitemOpen
  \bibfield  {author} {\bibinfo {author} {\bibfnamefont {David}\ \bibnamefont {Alonso}}, \bibinfo {author} {\bibfnamefont {Javier}\ \bibnamefont {Sanchez}}, \ and\ \bibinfo {author} {\bibfnamefont {An\v{z}e}\ \bibnamefont {Slosar}} (\bibinfo {collaboration} {LSST Dark Energy Science}),\ }\bibfield  {title} {\enquote {\bibinfo {title} {{A unified pseudo-$C_\ell$ framework}},}\ }\href {\doibase 10.1093/mnras/stz093} {\bibfield  {journal} {\bibinfo  {journal} {Mon. Not. Roy. Astron. Soc.}\ }\textbf {\bibinfo {volume} {484}},\ \bibinfo {pages} {4127--4151} (\bibinfo {year} {2019})},\ \Eprint {http://arxiv.org/abs/1809.09603} {arXiv:1809.09603 [astro-ph.CO]} \BibitemShut {NoStop}%
\bibitem [{\citenamefont {Hivon}\ \emph {et~al.}(2002)\citenamefont {Hivon}, \citenamefont {Gorski}, \citenamefont {Netterfield}, \citenamefont {Crill}, \citenamefont {Prunet},\ and\ \citenamefont {Hansen}}]{Hivon:2001jp}%
  \BibitemOpen
  \bibfield  {author} {\bibinfo {author} {\bibfnamefont {E.}~\bibnamefont {Hivon}}, \bibinfo {author} {\bibfnamefont {K.~M.}\ \bibnamefont {Gorski}}, \bibinfo {author} {\bibfnamefont {C.~B.}\ \bibnamefont {Netterfield}}, \bibinfo {author} {\bibfnamefont {B.~P.}\ \bibnamefont {Crill}}, \bibinfo {author} {\bibfnamefont {S.}~\bibnamefont {Prunet}}, \ and\ \bibinfo {author} {\bibfnamefont {F.}~\bibnamefont {Hansen}},\ }\bibfield  {title} {\enquote {\bibinfo {title} {{Master of the cosmic microwave background anisotropy power spectrum: a fast method for statistical analysis of large and complex cosmic microwave background data sets}},}\ }\href {\doibase 10.1086/338126} {\bibfield  {journal} {\bibinfo  {journal} {Astrophys. J.}\ }\textbf {\bibinfo {volume} {567}},\ \bibinfo {pages} {2} (\bibinfo {year} {2002})},\ \Eprint {http://arxiv.org/abs/astro-ph/0105302} {arXiv:astro-ph/0105302} \BibitemShut {NoStop}%
\bibitem [{\citenamefont {Sailer}\ \emph {et~al.}(in prep.)\citenamefont {Sailer}, \citenamefont {Farren}, \citenamefont {Ferraro},\ and\ \citenamefont {White}}]{noah}%
  \BibitemOpen
  \bibfield  {author} {\bibinfo {author} {\bibfnamefont {Noah}\ \bibnamefont {Sailer}}, \bibinfo {author} {\bibfnamefont {Gerrit}\ \bibnamefont {Farren}}, \bibinfo {author} {\bibfnamefont {Simone}\ \bibnamefont {Ferraro}}, \ and\ \bibinfo {author} {\bibfnamefont {Martin}\ \bibnamefont {White}},\ }\href@noop {} {\  (\bibinfo {year} {in prep.})}\BibitemShut {NoStop}%
\bibitem [{\citenamefont {Millea}\ \emph {et~al.}(2021)\citenamefont {Millea} \emph {et~al.}}]{Millea:2020iuw}%
  \BibitemOpen
  \bibfield  {author} {\bibinfo {author} {\bibfnamefont {M.}~\bibnamefont {Millea}} \emph {et~al.},\ }\bibfield  {title} {\enquote {\bibinfo {title} {{Optimal Cosmic Microwave Background Lensing Reconstruction and Parameter Estimation with SPTpol Data}},}\ }\href {\doibase 10.3847/1538-4357/ac02bb} {\bibfield  {journal} {\bibinfo  {journal} {Astrophys. J.}\ }\textbf {\bibinfo {volume} {922}},\ \bibinfo {pages} {259} (\bibinfo {year} {2021})},\ \Eprint {http://arxiv.org/abs/2012.01709} {arXiv:2012.01709 [astro-ph.CO]} \BibitemShut {NoStop}%
\bibitem [{\citenamefont {Hu}\ and\ \citenamefont {Okamoto}(2002)}]{Hu:2001kj}%
  \BibitemOpen
  \bibfield  {author} {\bibinfo {author} {\bibfnamefont {Wayne}\ \bibnamefont {Hu}}\ and\ \bibinfo {author} {\bibfnamefont {Takemi}\ \bibnamefont {Okamoto}},\ }\bibfield  {title} {\enquote {\bibinfo {title} {{Mass reconstruction with cmb polarization}},}\ }\href {\doibase 10.1086/341110} {\bibfield  {journal} {\bibinfo  {journal} {Astrophys. J.}\ }\textbf {\bibinfo {volume} {574}},\ \bibinfo {pages} {566--574} (\bibinfo {year} {2002})},\ \Eprint {http://arxiv.org/abs/astro-ph/0111606} {arXiv:astro-ph/0111606} \BibitemShut {NoStop}%
\bibitem [{\citenamefont {Sailer}\ \emph {et~al.}(2023)\citenamefont {Sailer}, \citenamefont {Ferraro},\ and\ \citenamefont {Schaan}}]{Sailer:2022jwt}%
  \BibitemOpen
  \bibfield  {author} {\bibinfo {author} {\bibfnamefont {Noah}\ \bibnamefont {Sailer}}, \bibinfo {author} {\bibfnamefont {Simone}\ \bibnamefont {Ferraro}}, \ and\ \bibinfo {author} {\bibfnamefont {Emmanuel}\ \bibnamefont {Schaan}},\ }\bibfield  {title} {\enquote {\bibinfo {title} {{Foreground-immune CMB lensing reconstruction with polarization}},}\ }\href {\doibase 10.1103/PhysRevD.107.023504} {\bibfield  {journal} {\bibinfo  {journal} {Phys. Rev. D}\ }\textbf {\bibinfo {volume} {107}},\ \bibinfo {pages} {023504} (\bibinfo {year} {2023})},\ \Eprint {http://arxiv.org/abs/2211.03786} {arXiv:2211.03786 [astro-ph.CO]} \BibitemShut {NoStop}%
\bibitem [{\citenamefont {Sailer}\ \emph {et~al.}(2020)\citenamefont {Sailer}, \citenamefont {Schaan},\ and\ \citenamefont {Ferraro}}]{Sailer:2020lal}%
  \BibitemOpen
  \bibfield  {author} {\bibinfo {author} {\bibfnamefont {Noah}\ \bibnamefont {Sailer}}, \bibinfo {author} {\bibfnamefont {Emmanuel}\ \bibnamefont {Schaan}}, \ and\ \bibinfo {author} {\bibfnamefont {Simone}\ \bibnamefont {Ferraro}},\ }\bibfield  {title} {\enquote {\bibinfo {title} {{Lower bias, lower noise CMB lensing with foreground-hardened estimators}},}\ }\href {\doibase 10.1103/PhysRevD.102.063517} {\bibfield  {journal} {\bibinfo  {journal} {Phys. Rev. D}\ }\textbf {\bibinfo {volume} {102}},\ \bibinfo {pages} {063517} (\bibinfo {year} {2020})},\ \Eprint {http://arxiv.org/abs/2007.04325} {arXiv:2007.04325 [astro-ph.CO]} \BibitemShut {NoStop}%
\bibitem [{\citenamefont {Maniyar}\ \emph {et~al.}(2021)\citenamefont {Maniyar}, \citenamefont {Ali-Ha\"\i{}moud}, \citenamefont {Carron}, \citenamefont {Lewis},\ and\ \citenamefont {Madhavacheril}}]{Maniyar:2021msb}%
  \BibitemOpen
  \bibfield  {author} {\bibinfo {author} {\bibfnamefont {Abhishek~S.}\ \bibnamefont {Maniyar}}, \bibinfo {author} {\bibfnamefont {Yacine}\ \bibnamefont {Ali-Ha\"\i{}moud}}, \bibinfo {author} {\bibfnamefont {Julien}\ \bibnamefont {Carron}}, \bibinfo {author} {\bibfnamefont {Antony}\ \bibnamefont {Lewis}}, \ and\ \bibinfo {author} {\bibfnamefont {Mathew~S.}\ \bibnamefont {Madhavacheril}},\ }\bibfield  {title} {\enquote {\bibinfo {title} {{Quadratic estimators for CMB weak lensing}},}\ }\href {\doibase 10.1103/PhysRevD.103.083524} {\bibfield  {journal} {\bibinfo  {journal} {Phys. Rev. D}\ }\textbf {\bibinfo {volume} {103}},\ \bibinfo {pages} {083524} (\bibinfo {year} {2021})},\ \Eprint {http://arxiv.org/abs/2101.12193} {arXiv:2101.12193 [astro-ph.CO]} \BibitemShut {NoStop}%
\bibitem [{\citenamefont {Kalaja}\ \emph {et~al.}(2023)\citenamefont {Kalaja}, \citenamefont {Orlando}, \citenamefont {Bowkis}, \citenamefont {Challinor}, \citenamefont {Meerburg},\ and\ \citenamefont {Namikawa}}]{Kalaja:2022xhi}%
  \BibitemOpen
  \bibfield  {author} {\bibinfo {author} {\bibfnamefont {Alba}\ \bibnamefont {Kalaja}}, \bibinfo {author} {\bibfnamefont {Giorgio}\ \bibnamefont {Orlando}}, \bibinfo {author} {\bibfnamefont {Aleksandr}\ \bibnamefont {Bowkis}}, \bibinfo {author} {\bibfnamefont {Anthony}\ \bibnamefont {Challinor}}, \bibinfo {author} {\bibfnamefont {P.~Daniel}\ \bibnamefont {Meerburg}}, \ and\ \bibinfo {author} {\bibfnamefont {Toshiya}\ \bibnamefont {Namikawa}},\ }\bibfield  {title} {\enquote {\bibinfo {title} {{The reconstructed CMB lensing bispectrum}},}\ }\href {\doibase 10.1088/1475-7516/2023/04/041} {\bibfield  {journal} {\bibinfo  {journal} {JCAP}\ }\textbf {\bibinfo {volume} {04}},\ \bibinfo {pages} {041} (\bibinfo {year} {2023})},\ \Eprint {http://arxiv.org/abs/2210.16203} {arXiv:2210.16203 [astro-ph.CO]} \BibitemShut {NoStop}%
\bibitem [{\citenamefont {Lewis}\ \emph {et~al.}(2000)\citenamefont {Lewis}, \citenamefont {Challinor},\ and\ \citenamefont {Lasenby}}]{Lewis:1999bs}%
  \BibitemOpen
  \bibfield  {author} {\bibinfo {author} {\bibfnamefont {Antony}\ \bibnamefont {Lewis}}, \bibinfo {author} {\bibfnamefont {Anthony}\ \bibnamefont {Challinor}}, \ and\ \bibinfo {author} {\bibfnamefont {Anthony}\ \bibnamefont {Lasenby}},\ }\bibfield  {title} {\enquote {\bibinfo {title} {{Efficient computation of CMB anisotropies in closed FRW models}},}\ }\href {\doibase 10.1086/309179} {\bibfield  {journal} {\bibinfo  {journal} {Astrophys. J.}\ }\textbf {\bibinfo {volume} {538}},\ \bibinfo {pages} {473--476} (\bibinfo {year} {2000})},\ \Eprint {http://arxiv.org/abs/astro-ph/9911177} {arXiv:astro-ph/9911177} \BibitemShut {NoStop}%
\bibitem [{\citenamefont {{Lewis}}\ and\ \citenamefont {{Challinor}}(2011)}]{2011ascl.soft02026L}%
  \BibitemOpen
  \bibfield  {author} {\bibinfo {author} {\bibfnamefont {Antony}\ \bibnamefont {{Lewis}}}\ and\ \bibinfo {author} {\bibfnamefont {Anthony}\ \bibnamefont {{Challinor}}},\ }\href@noop {} {\enquote {\bibinfo {title} {{CAMB: Code for Anisotropies in the Microwave Background}},}\ }\bibinfo {howpublished} {Astrophysics Source Code Library, record ascl:1102.026} (\bibinfo {year} {2011}),\ \Eprint {http://arxiv.org/abs/1102.026} {ascl:1102.026} \BibitemShut {NoStop}%
\bibitem [{\citenamefont {{Howlett}}\ \emph {et~al.}(2012)\citenamefont {{Howlett}}, \citenamefont {{Lewis}}, \citenamefont {{Hall}},\ and\ \citenamefont {{Challinor}}}]{2012JCAP...04..027H}%
  \BibitemOpen
  \bibfield  {author} {\bibinfo {author} {\bibfnamefont {Cullan}\ \bibnamefont {{Howlett}}}, \bibinfo {author} {\bibfnamefont {Antony}\ \bibnamefont {{Lewis}}}, \bibinfo {author} {\bibfnamefont {Alex}\ \bibnamefont {{Hall}}}, \ and\ \bibinfo {author} {\bibfnamefont {Anthony}\ \bibnamefont {{Challinor}}},\ }\bibfield  {title} {\enquote {\bibinfo {title} {{CMB power spectrum parameter degeneracies in the era of precision cosmology}},}\ }\href {\doibase 10.1088/1475-7516/2012/04/027} {\bibfield  {journal} {\bibinfo  {journal} {\jcap}\ }\textbf {\bibinfo {volume} {2012}},\ \bibinfo {eid} {027} (\bibinfo {year} {2012})},\ \Eprint {http://arxiv.org/abs/1201.3654} {arXiv:1201.3654 [astro-ph.CO]} \BibitemShut {NoStop}%
\bibitem [{\citenamefont {Schaan}\ and\ \citenamefont {Ferraro}(2019)}]{Schaan:2018tup}%
  \BibitemOpen
  \bibfield  {author} {\bibinfo {author} {\bibfnamefont {Emmanuel}\ \bibnamefont {Schaan}}\ and\ \bibinfo {author} {\bibfnamefont {Simone}\ \bibnamefont {Ferraro}},\ }\bibfield  {title} {\enquote {\bibinfo {title} {{Foreground-Immune Cosmic Microwave Background Lensing with Shear-Only Reconstruction}},}\ }\href {\doibase 10.1103/PhysRevLett.122.181301} {\bibfield  {journal} {\bibinfo  {journal} {Phys. Rev. Lett.}\ }\textbf {\bibinfo {volume} {122}},\ \bibinfo {pages} {181301} (\bibinfo {year} {2019})},\ \Eprint {http://arxiv.org/abs/1804.06403} {arXiv:1804.06403 [astro-ph.CO]} \BibitemShut {NoStop}%
\bibitem [{\citenamefont {{Amendola}}(1996)}]{1996MNRAS.283..983A}%
  \BibitemOpen
  \bibfield  {author} {\bibinfo {author} {\bibfnamefont {Luca}\ \bibnamefont {{Amendola}}},\ }\bibfield  {title} {\enquote {\bibinfo {title} {{Non-Gaussian likelihood function and COBE data}},}\ }\href {\doibase 10.1093/mnras/283.3.983} {\bibfield  {journal} {\bibinfo  {journal} {\mnras}\ }\textbf {\bibinfo {volume} {283}},\ \bibinfo {pages} {983--989} (\bibinfo {year} {1996})}\BibitemShut {NoStop}%
\bibitem [{\citenamefont {Regan}\ \emph {et~al.}(2010)\citenamefont {Regan}, \citenamefont {Shellard},\ and\ \citenamefont {Fergusson}}]{Regan:2010cn}%
  \BibitemOpen
  \bibfield  {author} {\bibinfo {author} {\bibfnamefont {D.~M.}\ \bibnamefont {Regan}}, \bibinfo {author} {\bibfnamefont {E.~P.~S.}\ \bibnamefont {Shellard}}, \ and\ \bibinfo {author} {\bibfnamefont {J.~R.}\ \bibnamefont {Fergusson}},\ }\bibfield  {title} {\enquote {\bibinfo {title} {{General CMB and Primordial Trispectrum Estimation}},}\ }\href {\doibase 10.1103/PhysRevD.82.023520} {\bibfield  {journal} {\bibinfo  {journal} {Phys. Rev. D}\ }\textbf {\bibinfo {volume} {82}},\ \bibinfo {pages} {023520} (\bibinfo {year} {2010})},\ \Eprint {http://arxiv.org/abs/1004.2915} {arXiv:1004.2915 [astro-ph.CO]} \BibitemShut {NoStop}%
\bibitem [{\citenamefont {Schmittfull}\ \emph {et~al.}(2013)\citenamefont {Schmittfull}, \citenamefont {Challinor}, \citenamefont {Hanson},\ and\ \citenamefont {Lewis}}]{Schmittfull:2013uea}%
  \BibitemOpen
  \bibfield  {author} {\bibinfo {author} {\bibfnamefont {Marcel~M.}\ \bibnamefont {Schmittfull}}, \bibinfo {author} {\bibfnamefont {Anthony}\ \bibnamefont {Challinor}}, \bibinfo {author} {\bibfnamefont {Duncan}\ \bibnamefont {Hanson}}, \ and\ \bibinfo {author} {\bibfnamefont {Antony}\ \bibnamefont {Lewis}},\ }\bibfield  {title} {\enquote {\bibinfo {title} {{Joint analysis of CMB temperature and lensing-reconstruction power spectra}},}\ }\href {\doibase 10.1103/PhysRevD.88.063012} {\bibfield  {journal} {\bibinfo  {journal} {Phys. Rev. D}\ }\textbf {\bibinfo {volume} {88}},\ \bibinfo {pages} {063012} (\bibinfo {year} {2013})},\ \Eprint {http://arxiv.org/abs/1308.0286} {arXiv:1308.0286 [astro-ph.CO]} \BibitemShut {NoStop}%
\end{thebibliography}%

\end{document}